\documentclass[mathpazo]{cicp}

\usepackage{amsmath}
\usepackage{amssymb}
\usepackage{amsthm}
\usepackage{url}
\usepackage{array}

\newcommand{\en}[1]{(\ref{#1})}
\newcommand{\leavethisout}[1]{}

\newcommand{\dt}{\Delta t}
\newcommand{\dx}{h_x}
\newcommand{\dy}{h_y}
\newcommand{\Reynolds}{\text{\emph{Re}}}
\newcommand{\units}[1]{\mbox{$\mathrm{#1}$}}
\newcommand{\bunits}[1]{[\mbox{$\mathrm{#1}$}]}
\newcommand{\starone}{(1)}
\newcommand{\startwo}{(2)}
\newcommand{\starthree}{(3)}
\newcommand{\fsub}[2]{#1_{\mbox{}\!\text{\scriptsize\emph{#2}}}}

\newcommand{\myvec}[1]{\vec{#1}}
\renewcommand{\myvec}[1]{\boldsymbol{#1}}
\newcommand{\vd}{\myvec{d}}
\newcommand{\vX}{\myvec{X}}

\newcommand{\vF}{\myvec{F}}
\newcommand{\vu}{\myvec{u}}
\newcommand{\vx}{\myvec{x}}
\newcommand{\vf}{\myvec{f}}
\newcommand{\vq}{\myvec{q}}

\begin{document}
\title{Numerical simulations of particle sedimentation using the
  immersed boundary method}

\author[S.~Ghosh and J.~M.~Stockie]{Sudeshna Ghosh\affil{1} and 
  John M. Stockie\affil{1}\comma\corrauth} 
\address{\affilnum{1}\ Department of Mathematics, 
  Simon Fraser University, Burnaby, British Columbia, V5A~1S6, Canada}
\emails{{\tt sga33@sfu.ca} (S.~Ghosh), {\tt jstockie@sfu.ca}
  (J.~M.~Stockie)} 


\begin{abstract}
  We study the settling of solid particles within a viscous
  incompressible fluid contained in a two-dimensional channel, where the
  mass density of the particles is slightly greater than that of the
  fluid.  The fluid-structure interaction problem is simulated
  numerically using the immersed boundary method, with an added mass
  term that is incorporated using a Boussinesq approximation.
  Simulations are performed with a single circular particle, and also
  with two particles in various initial configurations.  The terminal
  settling velocities for the particles correspond closely with both
  theoretical and experimental results, and the single-particle dynamics
  reproduce expected behavior qualitatively.  The two-particle
  simulations exhibit drafting-kissing-tumbling dynamics that is similar
  to what is observed in other experimental and numerical studies.
\end{abstract}



\maketitle


\section{Introduction}
\label{sec:intro}

Particulate flows involve a dynamically evolving fluid that interacts
with solid suspended particles, and arise in a wide range of
applications in natural and industrial processes
\cite{burger-wendland-2001}.  We are particularly interested in the
gravitational settling or sedimentation problem, in which the suspended
solid particles have large enough mass that they settle under their own
weight.  Sedimentation is observed in many applications, including flow
of pollutants in rivers and the atmosphere, tea leaves settling to the
bottom of a teacup, industrial crystal precipitation, mineral ore
processing, and hail formation in thunderclouds, to name just a few.

There is an extensive literature on experimental, theoretical and
computational studies of particulate flows involving sedimentation.  We
make no attempt here to perform a comprehensive review, but will rather
highlight a few of the more important results.  Experimental studies of
sedimentation have had a long history including the earlier work of
Richardson and Zaki~\cite{richardson-zaki-1954b} and extending to more
recent
years~\cite{difelice-1999,fortes-joseph-lundgren-1987,hu-joseph-fortes-1997,jayaweera-mason-1965}.
Many analytical and approximate solutions have been developed to explain
the behavior of settling suspensions, especially in the dilute limit
where there are only a small number of particles.  Back in 1851,
Stokes~\cite{stokes-1966} derived an analytical solution for a single
particle settling within an unbounded fluid, and many other authors have
since extended these results to other more practical sedimentation
problems~\cite{brenner-1966,davis-acrivos-1985,guazzelli-morris-2012,vasseur-cox-1977}.
More recently, many numerical approaches have been applied to simulate
settling particles, including the finite element method
\cite{feng-hu-joseph-1994,glowinski-etal-2001,hu-1996,munster-mierka-turek-2012},
lattice-Boltzmann method
\cite{dupuis-etal-2008,ladd-1994II,qi-1999}, and
boundary element method
\cite{hernandez-ortiz-phdthesis-2004,phanthien-fan-2002}.  The
underlying feature of these numerical methods is that the fluid flow is
governed by the Navier-Stokes equations whereas the particles are
governed by Newton's equations of motion.  The hydrodynamic forces
between the particle and fluid are obtained from the solution of this
coupled system, which typically requires either complex interfacial
matching conditions at the fluid-particle interface, or else some form
of dynamic boundary-fitted meshing.  In any case, these methods tend to
be complex and extremely CPU-intensive, especially for three-dimensional
flows.

One numerical approach that has proven to be especially effective for
solving complex fluid-structure interaction problems involving dynamic
moving structures is the \emph{immersed boundary} (or IB) method.  This
approach has been used extensively to simulate deformable structures
arising in problems in biofluid mechanics~\cite{peskin-2002}.  Wang and
Layton~\cite{wang-layton-2009} have recently used the IB method to
simulate sedimentation of multiple rigid 1D fibers suspended in a
viscous incompressible fluid, and several other authors have applied the
the IB approach to solve related sedimentation
problems~\cite{breugem-2012,dupuis-etal-2008,hopkins-fauci-2002,uhlmann-2005,wang-fan-luo-2008}.

The IB method is a mixed Eulerian-Lagrangian approach, in which the
fluid equations are solved on an equally-spaced rectangular mesh, while
the moving solid boundaries are approximated at a set of points that
moves relative to the underlying fluid grid.  In the original IB method,
the effect of these immersed boundaries is represented as a singular
force that is computed from the IB configuration and which is then
spread onto fluid grid points by means of a regularized delta function.
The added mass due to a sedimenting particle can also be distributed
onto the fluid in a similar manner.  With the exception of the papers by
Wang and Layton~\cite{wang-layton-2009} and Hopkins and
Fauci~\cite{hopkins-fauci-2002}, the other authors mentioned above have
employed a modification of this IB approach known as the ``direct
forcing IB method,'' wherein the force is an artificial quantity that is
calculated directly from the governing equations so as to satisfy the
velocity boundary conditions exactly on the immersed boundary
(see~\cite{mittal-iaccarino-2005} for more details).

Our aim in this paper is to apply the original IB method to solving
sedimentation problems, rather than the direct forcing approach.  We
restrict ourselves to a two-dimensional geometry, in which one or two
particles with a circular cross-section settle under the influence of
gravity within a rectangular channel that has vertical bounding walls.
Although the IB approach has been applied to solve certain sedimentation
problems, there has not yet been an extensive comparison to other
results in the literature.  Our primary aim is therefore to perform
such a comparison to a number of
experimental~\cite{sucker-brauer-1975,white-1946},
theoretical~\cite{faxen-1946,takaisi-1955}, and
numerical~\cite{feng-hu-joseph-1994} studies, in order to ascertain the
validity of the IB approach in simulating sedimentation problems.
Although we focus here on solid particles, the long-term goal of our
work is to develop a numerical framework that can be used to 
investigate the settling of highly deformable particles.

We begin in Section~\ref{sec:ib-method} by describing the IB method and
defining the forces used to simulate the presence of both settling
particles and channel walls.  Section~\ref{sec:vs-approx} contains a
review of previous analytical and experimental results on the settling
velocity for a single particle in both unbounded and wall-bounded
domains.  We then perform a series of numerical simulations of
sedimentation at small to moderate Reynolds numbers, and report the
results in Sections~\ref{sec:num-one} and \ref{sec:num-two}.  Most of
the results appearing in this article are contained in the PhD thesis of
the first author~\cite{ghosh-2013}.

\section{Immersed boundary method}
\label{sec:ib-method}

The immersed boundary method is both a mathematical formulation and a
numerical scheme.  We begin in this section by describing the model
equations that underlie the IB formulation for fluid-structure
interaction.  Following that, we discretize the equations and describe
the numerical algorithm used to determine an approximate solution.
Finally, we provide details on the specification of the discrete IB
force density representing the channel walls and sedimenting particles.

\subsection{Model formulation}
\label{sec:model}

In this section we describe a two-dimensional IB model that is capable
of capturing solid (and potentially deformable) elastic bodies with
general shape and that move within a surrounding incompressible,
Newtonian fluid under the action of gravitational force. The details of
the IB force density used to handle a solid circular object in the
presence of two parallel bounding walls are left for
section~\ref{sec:force-density}.  All variables and parameters in this
paper are stated in CGS units, unless otherwise indicated.

\begin{figure}[hbtp]
  \centering
  \includegraphics[width=0.7\columnwidth]{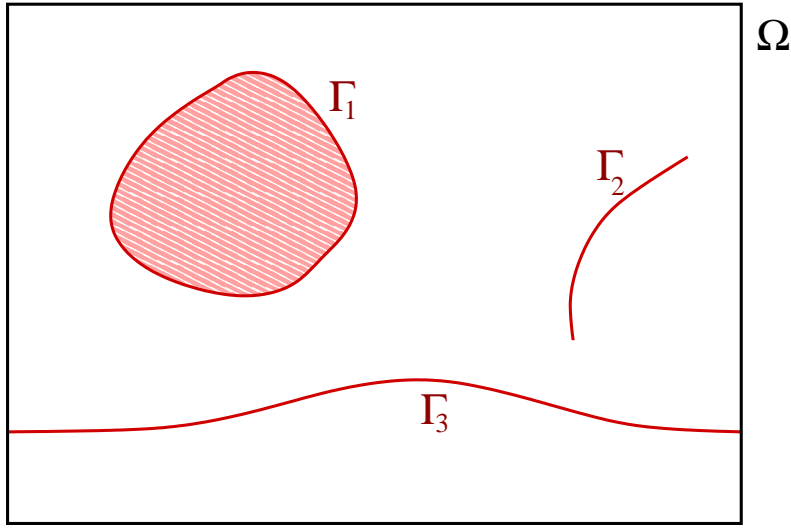}
  \caption{A general immersed boundary configuration  
    $\Gamma = \mathop{\bigcup}_{i=1}^{3} \Gamma_i$
    %
    \ consisting of several disconnected components immersed
    within a doubly-periodic fluid domain $\Omega$.} 
  \label{fig:domain}
\end{figure}

Suppose that a moving elastic solid body $\Gamma$ is contained within a
fluid domain $\Omega$ as pictured in Figure~\ref{fig:domain}.
In general, $\Gamma$ may consist of several disconnected components,
$\Gamma=\mathop{\bigcup}_i \Gamma_i$, where each $\Gamma_i$ can be a
one-dimensional elastic membrane (parameterized by a single real
parameter $s$) or an elastic solid region (whose specification requires
two parameters, $r$ and $s$).  We denote the location or
``configuration'' of the immersed boundary by $\vX(\vq,t)$ \bunits{cm},
where $\vq$ is a dimensionless IB parameterization that is used to
represent either a scalar $s$ or a vector $(r,s)$, depending on the
context.  For simplicity, we assume that $\Omega=[0,L_x]\times[0,L_y]$
is rectangular in shape and that periodic boundary conditions are
applied in both the $x$- and $y$-directions.

The effect of the elastic body on the fluid is to impose a force
$\fsub{\vf}{IB}$ \bunits{g/cm^2\,s^2} onto the adjacent fluid particle
at location $\vx=\vX(\vq,t)$, which is incorporated into the
incompressible Navier-Stokes equations as follows:
\begin{gather}
  \rho \frac{\partial\vu}{\partial t} + \rho \vu\cdot \nabla \vu = 
  \mu \nabla^2 \vu  - \nabla p + \fsub{\vf}{IB}, 
  \label{eq:ns-mom}
  \\
  \nabla \cdot \vu = 0. \label{eq:ns-inc}
\end{gather}
Here, $\vu(\vx,t)$ is the fluid velocity \bunits{cm/s}, $p(\vx,t)$ is
the pressure \bunits{g/cm\,s^2}, $\vx=(x,y)$ are the Eulerian
coordinates \bunits{cm} for the fluid domain $\Omega$, $\rho$ is density
\bunits{g/cm^3} and $\mu$ is dynamic viscosity \bunits{g/cm\,s}.  The IB
forcing term in the momentum equations \en{eq:ns-mom} is represented by
a force density $\fsub{\vF}{IB}(\vq,t)$ \bunits{g/s^2} that is spread
onto the surrounding fluid by means of a delta-function convolution
\begin{gather}
  \fsub{\vf}{IB}(\vx,t) = \int_\Gamma
  \fsub{\vF}{IB}(\vq,t) \, \delta(\vx-\vX(\vq,t) )\, 
  d\vq,
  \label{eq:ib-force}
\end{gather}
where $\delta(\vx)=\delta(x)\delta(y)$ is the Cartesian product of two
one-dimensional Dirac delta functions.  The consistency of the above
equations with the dynamics of an actual incompressible elastic material
interacting with an incompressible fluid is demonstrated under very
general conditions in the review paper by Peskin~\cite{peskin-2002}.

Most papers in the immersed boundary literature assume that $\Gamma$ has
the same constant density $\rho_f$ as the surrounding fluid, and hence
$\Gamma$ is neutrally buoyant. However, for the particle sedimentation
application considered here, we must take $\Gamma$ (or at least
portions of it) to have density $\rho_s>\rho_f$ that is greater than
that of the fluid. Consequently, the density of the fluid-solid
composite material $\rho(\vx,t)$ is a variable quantity that may also be
written in terms of a delta function convolution
as~\cite{zhu-peskin-2002}
\begin{gather*}
  \rho(\vx,t) = \rho_f + \Delta \rho(\vx,t)
  \intertext{where}
  \Delta \rho(\vx,t) = \int_\Gamma M(\vq) \, \delta(\vx-\vX(\vq,t)) 
  \, d\vq.
\end{gather*}
The quantity $M(\vq)\geqslant 0$\ 
is the added Lagrangian mass density due to $\Gamma$, with $M=0$ only
for those components that are neutrally buoyant.

In all examples in this paper, we will take $M\equiv M_o$ (constant),
and we also assume that the solid density is close to that of the fluid
so that $\Delta\rho \ll \rho_f$.  Consequently, it is reasonable to
apply a Boussinesq approximation as in~\cite{hopkins-fauci-2002} so that
the extra intertial term involving $\Delta\rho$ is neglected and the
density on the left hand side of the momentum equations \en{eq:ns-mom}
is taken equal to the constant $\rho_f$:
\begin{gather}
  \rho_f \frac{\partial\vu}{\partial t} + \rho_f \vu\cdot \nabla \vu = 
  \mu \nabla^2 \vu - \nabla p + \fsub{\vf}{IB} + \fsub{\vf}{G}.
  \label{eq:ns-mom2}
\end{gather}
The extra forcing term $\fsub{\vf}{G}$ derives from the force of gravity
acting on the immersed boundary and can be written
as~\cite{hopkins-fauci-2002}
\begin{gather}
  \fsub{\vf}{G}(\vx,t) = - g \hat{k}\, \Delta\rho 
  = - g \hat{k} \int_\Gamma M(\vq)\, \delta(\vx-\vX(\vq,t))\, d\vq,
  \label{eq:gravity}
\end{gather}
where $g=980~\units{cm/s^2}$ is the gravitational acceleration and
$\hat{k}=(0,1)$ is the unit vector in the vertical direction.  

Finally, the immersed boundary is assumed to move with the fluid so that
\begin{gather}
  \frac{\partial \vX}{\partial t} = \int_{\Omega} \vu(\vx,t) \,
  \delta(\vx-\vX(\vq,t))\, d\vx, 
  \label{eq:ib-velocity}
\end{gather}
which is simply the ``no-slip'' condition for fluid particles
located adjacent to the immersed boundary.

In summary, the governing equations consist of \en{eq:ns-inc},
\en{eq:ns-mom2}--\en{eq:ib-velocity}, with the IB force density being
the only component that remains to be specified.  Since it is easiest to
write $\fsub{\vf}{IB}$ in discrete form, we will first derive the
discretized governing equations, after which we will provide a
specification for the IB force.

\subsection{Numerical algorithm}
\label{sec:algo}

The algorithm we describe next is a semi-implicit scheme that is closely
related to the method outlined in \cite{stockie-thesis-1997}.  The fluid
domain $\Omega$ is divided into an equally-spaced grid of points denoted
by $\vx_{i,j}=(x_i,y_j)=(i\dx,j\dy)$, with $\dx=L_x/N_x$, $\dy=L_y/N_y$,
$i=1,2,\dots,N_x$, and $j=1,2,\dots,N_y$.  We consider a time interval
$[0,T]$ divided into equally-spaced points $t_n=n \dt$ with time step
$\dt=T/N_t$ and $n=0,1,2,\dots,N_t$.  We may then define discrete
approximations of the velocity and pressure $\vu_{i,j}^n$ and
$p_{i,j}^n$ at points $(x_i, y_j, t_n)$.  The immersed boundary $\Gamma$
is similarly discretized at points $\vX_\ell$ for $\ell=1,2,\dots,N_b$,
and the IB configuration and force density are approximated by
$\vX_\ell^n$ and $\vF_\ell^n$ respectively.

Using the above notation, we introduce finite difference operators that
approximate the spatial derivatives appearing in the governing
equations.  In particular, we define two one-sided difference
approximations of the $x$--derivative of a grid quantity $w_{i,j}$
\begin{gather}
  D_x^+ w_{i,j} = \frac{w_{i+1,j}-w_{i,j}}{\dx} 
  \quad \text{and} \quad
  D_x^- w_{i,j} = \frac{w_{i,j}-w_{i-1,j}}{\dx} ,
  \label{eq:deriv-pm}
\end{gather}
as well as the centered approximation
\begin{gather}
  D_x^0 w_{i,j} = \frac{w_{i+1,j}-w_{i-1,j}}{2\dx} .
  \label{eq:deriv-0}
\end{gather}
Analogous definitions apply for the $y$--derivative approximations
$D_y^+$, $D_y^-$ and $D_y^0$, and the gradient is replaced by the
centered approximation $\nabla_h = (D_x^0, D_y^0)$.  Finally, the delta
function appearing in the integral terms is replaced by the 
regularized function
\begin{gather}
  \delta_h(\vx) = \frac{1}{\dx \dy} \phi \left( \frac{x}{\dx} \right) 
  \phi \left( \frac{y}{\dy} \right),\\
  \intertext{where}
  \phi(r) = \begin{cases}
    \displaystyle 
    \frac{1}{4} \left( 1 + \cos\left( \frac{\pi r}{2} \right) \right), 
    & \text{if $|r| \leqslant 2$}, \\
    0, & \text{otherwise}.  
  \end{cases}
  \label{eq:delta}
\end{gather}

We are now prepared to state the immersed boundary algorithm.  In any
given time step, we assume that values of the velocity $\vu_{i,j}^{n-1}$
and IB configuration $\vX_{\ell}^{n-1}$ are known from the previous
step.  These quantities are evolved to time $t_n$ using the following
procedure:
\begin{enumerate}
\item Compute the IB force density $\fsub{\vF}{IB,$\ell$}^{n-1}$ based on
  the configuration $\vX_\ell^{n-1}$ as described in
  section~\ref{sec:force-density}.
\item Spread the IB force to the fluid grid points using a
  discretization of the integral in \en{eq:ib-force}
  \begin{gather}
    \fsub{\vf}{IB,i,j}^{n-1} = \sum_{\ell=1}^{N_b}
    \fsub{\vF}{IB,$\ell$}^{n-1} \, \delta_h (\vx_{i,j} - \vX_\ell^{n-1})
    \, A_b, 
    \label{eq:ib-force2}
  \end{gather}
  and a similar approximation of the integral in \en{eq:gravity} yields
  a formula for $\fsub{\vf}{G,i,j}^{n-1}$.  The scaling factor $A_b$ in
  both cases is inversely proportional to the number of IB points
  ($N_b$) and has a different interpretation depending on whether the
  immersed boundary is a 1D fiber (channel wall) or a 2D solid block
  (circular particle).  In the case of a fiber $A_b$ is a length, while
  for a solid region $A_b$ is an area; in both cases, the factor $A_b$
  ensures that the formula \en{eq:ib-force2} scales properly with the
  number of IB points and that it is a consistent approximation of the
  corresponding integral.  More details on the precise form of
  \en{eq:ib-force2} and the specification of $A_b$ are provided in
  section~\ref{sec:force-density}.

\item Integrate the incompressible Navier-Stokes equations using a
  split-step projection scheme:
  \begin{enumerate}

  \item Compute an intermediate velocity
    $\vu^{\starone}_{i,j}$ by applying the elastic and gravitational
    forces on the immersed boundary:
    \begin{gather}
      \rho_f \left( \frac{\vu^{\starone}_{i,j} - \vu^{n-1}_{i,j}}{\dt} \right) =
      \fsub{\vf}{IB,i,j}^{n-1} + \fsub{\vf}{G,i,j}^{n-1}
      \label{eq:project1}
    \end{gather}

  \item Apply an ADI discretization of the advection and diffusion
    terms:
    \begin{gather}
      \rho_f \left(\frac{\vu^{\startwo}_{i,j}-\vu^{\starone}_{i,j}}{\dt} +
        u^{n-1}_{i,j} D^0_x\vu^{\startwo}_{i,j} \right) = 
      \mu D^+_x D^-_x \vu^{\startwo}_{i,j}, 
      \label{eq:project2}
    \end{gather}
    \begin{gather}
      \rho_f \left( \frac{\vu^{\starthree}_{i,j}-\vu^{\startwo}_{i,j}}{\dt} +
        v^{n-1}_{i,j} D^0_y \vu^{\starthree}_{i,j} \right) =
      \mu D^+_y D^-_y \vu^{\starthree}_{i,j}.
      \label{eq:project3}
    \end{gather}
    These equations represent a sequence of tridiagonal solves for
    $\vu_{i,j}^{\startwo}$ and $\vu_{i,j}^{\starthree}$.

  \item Project the intermediate velocity $\vu_{i,j}^{\starthree}$ onto the
    space of divergence-free vector fields by:
    \begin{enumerate}
    \item Solving the pressure Poisson equation
      \begin{gather}
        \nabla_h \cdot \nabla_h p_{i,j} = \frac{\rho_f}{\dt} \, \nabla_h
        \vu_{i,j}^{\starthree}. 
        \label{eq:poisson}
      \end{gather}
      Note that $\nabla_h\cdot\nabla_h$ represents a wide finite
      difference stencil for the Laplacian involving the pressure values
      $p_{i,j}$, $p_{i-2,j}$, $p_{i+2,j}$, $p_{i,j-2}$ and $p_{i,j+2}$.
      Owing to the periodic boundary conditions on $\Omega$, the
      resulting system of linear equations is solved most easily by
      means of the discrete Fourier transform, which is calculated using
      the Fast Fourier Transform (FFT) algorithm~\cite{numrecipes}.  The
      discrete Fourier transform fully decouples the system and reduces
      the solution to a single linear equation for each wave number in
      Fourier space.  The pressure variables may then be obtained by
      applying the inverse FFT.  Details of this approach are described
      in \cite{stockie-thesis-1997,tu-peskin-1992}.
    \item Updating the velocity according to 
      \begin{gather}
        \vu_{i,j}^n = \vu_{i,j}^{\starthree} - \frac{\dt}{\rho_f}\,
        \nabla_h p_{i,j}. 
        \label{eq:project4}
      \end{gather}
    \end{enumerate}
  \end{enumerate}

\item Evolve the immersed boundary using
  \begin{gather}
    \vX_{\ell}^n = \vX_{\ell}^{n-1} + \dt\sum_{i,j} \vu_{i,j}^n \,
    \delta_h(\vx_{i,j} - \vX_{\ell}^{n-1}) \, \dx \dy.
    \label{eq:ib-evolve}
  \end{gather}
\end{enumerate}
This simple semi-implicit time discretization described above introduces
a CFL-like time-step restriction on the numerical scheme that depends on
the Reynolds number as well as the elastic IB force.  The dependence of
the stable time step on parameters can be characterized in certain
idealized
cases~\cite{boffi-gastaldi-heltai-2007,hou-shi-2008a,lai-thesis-1998},
and these results can be used as a guide to selecting a value for
$\Delta t$, but in practical computations the time step must be
determined manually.

This algorithm yields a solution that is first-order accurate in time,
and although all spatial derivatives are approximated using second-order
finite differences, the method is also first-order accurate in space
owing to errors in velocity interpolation near the immersed boundary
that arise from the use of the regularized delta function.  It is
straightforward to increase the temporal accuracy to second order using
an algorithm such as that proposed by Lai and
Peskin~\cite{lai-peskin-2000}, but it is much more difficult to increase
the spatial accuracy~\cite{griffith-etal-2007}.  Since the focus of the
current study is to validate the general IB approach in the study of
particle sedimentation, we have chosen to employ the simple scheme
above, and leave for future work the implementation of higher order
extensions to the algorithm.

\subsection{Discrete IB force density for particle and channel walls}
\label{sec:force-density}

\begin{figure}[htbp]
  \centering
  \includegraphics[trim=3in 0in 3in 0in,clip,width=0.55\columnwidth]{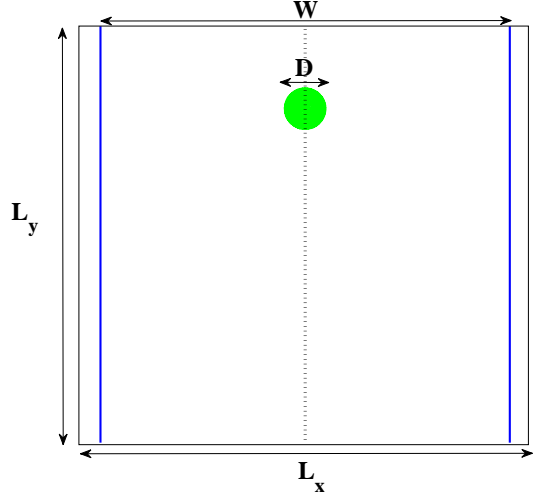}
  \caption{Initial geometry for the gravitational settling problem. The
    parallel channel walls are denoted by dashed vertical lines,
    separated by a distance $W$.  The first test case has a single solid
    particle of diameter $D$ that is released at time $t=0$ along the
    channel centerline (which is indicated by a dotted line).}
  \label{fig:channel} 
\end{figure}

We begin by describing the geometry for the particle sedimentation
problem.  Referring to Figure~\ref{fig:channel}, we take a rectangular
fluid domain of size $L_x \times L_y$ and place two vertical immersed
fibers representing the channel walls a distance $W<L_x$ apart,
symmetric relative to the channel centerline, and separated from the
domain boundary by a narrow strip of fluid.  With periodic boundary
conditions applied on all sides of the domain, the channel walls
naturally connect to each other across the top and bottom boundaries.  A
single, solid, circular particle of diameter $D$ is initially located at
the center of the channel.  Later on, we will consider other initial
configurations with one and two particles, but for now this will suffice
to illustrate the calculation of the IB force density.  This circular
particle in 2D may be thought of as corresponding in 3D to a
cross-section of a solid cylinder with infinite length.

In our sedimentation model, the IB force density $\fsub{\vF}{IB}$ is the
sum of two terms, $\fsub{\vF}{IB}=\vF^w + \vF^c$, where $\vF^w$
represents the force density generated by the channel walls and $\vF^c$
is that generated by the circular particle.  These forces are discussed
separately in the next two sections.

\subsubsection{Elastic force from the channel walls, $\vF^w$}
\label{sec:wall-force}

The vertical walls are discretized using an equally-spaced array of IB
points that are initially at locations $\vX^{w,L}_\ell = \left(
  \frac{1}{2}(L_x-W), \, \ell h_w \right)$ for the left wall, and
$\vX^{w,R}_\ell = \left(\frac{1}{2}(L_x+W), \, \ell h_w \right)$ for the
right wall, where the wall point spacing is $h_w=L_y/N_w$ and $\ell=1,
2, \dots, N_w$.  Each IB point is connected to a fixed ``tether point''
(at the same initial location) by a very stiff spring that exerts a
force of the form
\begin{gather}
  \vF^{w,L}_\ell = \sigma_w (\vX^{w,L}_\ell - \vX_\ell), 
  \label{eq:fwall-l}
\end{gather}
where $\sigma_w$ \bunits{g/cm\,s^2} is the spring stiffness and
$\vX_\ell$ is the moving IB point location.  Any motion of the IB point
away from corresponding the tether point location generates a spring
force that drives it back towards the target, so that as long as
$\sigma_w$ is chosen large enough the wall points can be made to mimic a
rigid structure.  We emphasize that tether points neither move with the
fluid nor generate any force themselves.  A similar expression is
developed for the force density at the right wall points,
$\vF^{w,R}_\ell$ so that the total wall force density may be written as
\begin{gather}
  \vF^w = \sum_{\ell=1}^{N_w} (\vF^{w,L}_\ell + \vF^{w,R}_\ell).
\end{gather}
The natural choice of scaling factor in the force spreading step
\en{eq:ib-force2} is the wall point spacing, $A_b=h_w$. 

\subsubsection{Elastic force from the particle, $\vF^c$}
\label{sec:particle-force}

\begin{figure}[htbp]
  \centering
  \includegraphics[width=0.45\columnwidth]{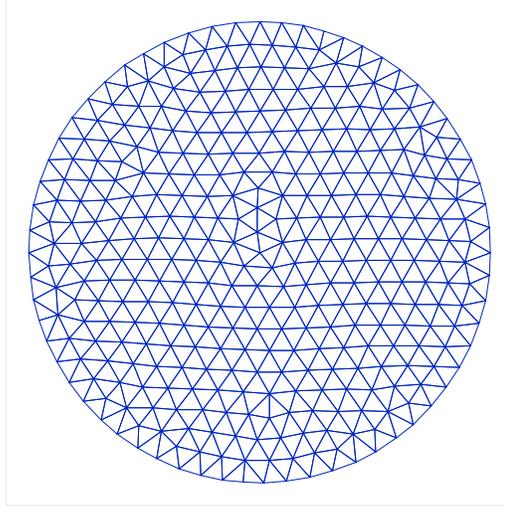}
  \caption{Uniform triangular mesh generated by {\tt distmesh2d}.}
  \label{fig:dist_mesh}
\end{figure}

The circular particle is represented by a collection of $N_c$ Lagrangian
points that lie on its circumference and throughout its interior.  We
make use of the unstructured triangular mesh generator
DistMesh~\cite{persson-strang-2004} that generates a nearly uniform
triangulation such as that shown in Figure~\ref{fig:dist_mesh}.  The
nodes of the triangulation are the IB points $\vX_\ell$, for
$\ell=1,2,\dots, N_c$, while the edges of the triangles define a network
of springs that maintains the shape of the particle.  In addition to
bearing IB spring forces, the network nodes are also employed in
equation \en{eq:gravity} to distribute added mass throughout the
particle.
In practice, we generate the triangulation by calling the Matlab
function {\tt distmesh2d} with the ``scaled edge length function'' {\tt
  huniform} (a function provided by the authors that attempts to find a
mesh that is as uniform as possible).  We also set the ``initial edge
length'' parameter equal to $\frac{1}{3}\,\min(\dx,\dy)$, which ensures
that the mesh obeys
\begin{gather*}
  \max_{k,\ell} |\vX_k-\vX_\ell| < \frac{1}{2}\,\min(\dx,\dy),
\end{gather*}
which is a standard ``rule of thumb'' that avoids leakage of fluid
between IB points~\cite{peskin-2002}.

This form of particle discretization should be compared with the more
common IB approach that uses an open circular ring of points with a
freely-moving fluid inside, such as
in~\cite{lai-peskin-2000,wang-fan-luo-2008}.  This approach has been
criticised~\cite{haeri-shrimpton-2012} for generating non-physical fluid
motions inside the particle and in some cases leading to significant
deviations in the shape of the particle.  In contrast, our
discretization of the particle interior with a network of IB springs
suppresses this spurious fluid motion and also helps to maintain the
rigidity of the particle boundary.

We now define the spring forces that act on the network, following the
development of Alpkvist and Klapper for viscoelastic biofilm
structures~\cite{alpkvist-klapper-2007}.  Let $\vd_{\ell,m}(t) =
\vX_\ell(t) - \vX_m(t)$ be the vector joining two IB points labeled
$\ell$ and $m$, and let $d_{\ell,m}(t) = |\vd_{\ell,m}(t)|$ be the
corresponding distance.  We assume that the spring network is initially
in equilibrium (i.e., zero force) so that all springs have a resting length
equal to their initial length, $d_{\ell,m}(0)$.  Let $\mathbb{I}$ be an
incidence matrix whose entries $\mathbb{I}_{\ell,m}$ are either 1 or 0
depending on whether or not points $\ell$ and $m$ are connected,
respectively. Then the force density acting on the $\ell^{th}$ IB point
in the network is
\begin{gather}
  \vF^c_\ell = \sigma_c 
  \sum_{\substack{m=1 \\ \mathbb{I}_{\ell,m}\neq 0}}^{N_b} 
  \mathbb{I}_{\ell,m} \frac{\vd_{\ell,m}}{d_{\ell,m}}
  (d_{\ell,m}(0) - d_{\ell,m}),
  \label{eq:fcircle}
\end{gather}
where the sum is taken only over those $m$ for which $\vX_m$ is
connected to $\vX_\ell$ in the network.  We have also assumed that the
spring stiffness $\sigma_c$ \bunits{g/cm\,s^2} is constant for all
network connections.  The total elastic force density generated by all
IB points making up the circular particle is then given by
\begin{gather}
  \vF^c = \sum_{\ell=1}^{N_c}\vF^c_\ell.
\end{gather} 
The appropriate scaling factor for the force integral \en{eq:ib-force2}
is the average area of a triangular mesh cell, $A_b=\pi(D/2)^2/N_c$.  A
similar approach was employed by Hopkins and
Fauci~\cite{hopkins-fauci-2002} to simulate a suspension of microbial
cells that they treated as point particles.

\section{Approximate formulas for settling velocity} 
\label{sec:vs-approx}

We next review some of the existing analytical and experimental results
on the settling of a single particle falling under the action of
gravity.  The study of a spherical particle in an unbounded fluid medium
in 3D is a classical problem that was considered by
Stokes~\cite{stokes-1966}, who obtained a formula for the settling
velocity that is now known as Stokes' law.  We will first state Stokes'
result and then modify it for a circular particle in 2D, which
corresponds to an idealized ``infinite cylinder'' in 3D.  We then
consider the case of a circular particle falling in a bounded fluid
domain between two vertical walls and then review several of the most
commonly-used formulas for the ``wall-correction factors'' that have
been obtained from either fitting to experimental data or using
approximate analytical techniques.  A fairly extensive overview of
settling for cylindrical particles, including many of the wall
correction formulas reported in the literature, is given by Champmartin
and Ambari~\cite{champmartin-ambari-2007}.

\subsection{Stokes' law for a spherical particle in 3D}
\label{sec:vs-stokes}

There are two main forces acting upon a massive particle settling in a
fluid: the gravitational force $\fsub{F}{g}$, and the drag force $F_d$
due to the ``friction'' between the particle and the fluid.  A particle
that is initially at rest will accelerate under the action of gravity,
and as the particle begins to move through the fluid it experiences a
drag force in the direction opposite to its motion that increases with
the speed of the particle relative to the fluid.  If the drag force
becomes large enough that it equals the gravitational force, then the
two forces are in balance and no further acceleration occurs.  The
particle velocity in this equilibrium state is known as the settling or
terminal velocity.

We take a sphere of diameter $D$ and density $\rho_s$ whose added mass
relative to the fluid is $\frac{4}{3} \pi \left(\frac{D}{2}\right)^3
(\rho_s-\rho_f)$.  The net gravitational force acting on the sphere is
\begin{gather}
  \fsub{F}{g} = \frac{4\pi}{3} \left(\frac{D}{2}\right)^3
  g (\rho_s-\rho_f),  
  \label{sphereg}
\end{gather}
and the corresponding drag force is 
\begin{gather}
  F_d = \frac{1}{2}\, C_d \rho_f V^2 \pi \left(\frac{D}{2}\right)^2, 
  \label{spheredrag}
\end{gather}
where $C_d$ is the drag coefficient for a sphere and $V$ is the velocity
of the sphere relative to the fluid.  The settling velocity $V_s$
corresponds to the long-term steady state in which drag and gravity
forces are in balance, so that $F_d = \fsub{F}{g}$.  By equating
\en{sphereg} and \en{spheredrag}, we can solve for
\begin{gather}
  V_s = \sqrt{\frac{4 g D (\rho_s-\rho_f)}{3 C_d\rho_f}}, 
  \label{terminal_sphere}
\end{gather}
keeping in mind that the drag coefficient on the right hand side also
typically depends on the settling velocity, $V_s$.  Indeed, we know from
\cite{batchelor-1967} that the drag coefficient for a sphere can be
approximated for small Reynolds number by
\begin{gather}
  C_d = \frac{24}{\Reynolds} = \frac{24\mu}{\rho_f V_s D},
\end{gather}
where we have taken 
\begin{gather}
  \Reynolds = \frac{\rho_f D V_s}{\mu}, 
\end{gather}
based on the particle diameter.  Substituting this expression into
\en{terminal_sphere} and solving for $V_s$ we obtain Stokes' law
\begin{gather}
  V_s = \frac{gD^2(\rho_s-\rho_f)}{18\mu}, 
\end{gather}
which is valid for $\Reynolds \lesssim 0.1$.

\subsection{Settling velocity for a circular particle in 2D}
\label{sec:vs-cylinder}

A similar argument may be used to derive the corresponding expression
for a circular particle in 2D.  We begin by considering a cylinder with
diameter $D$ and length $\ell$ and take the limit as
$\ell\rightarrow\infty$ in order to obtain a result that is relevant to
our 2D geometry.  The immersed cylinder has an added mass (relative to
the fluid) of $m = \pi \left(\frac{D}{2}\right)^2 \ell \,
(\rho_s-\rho_f)$ for which the net gravitational force is
\begin{gather}
  \fsub{\widetilde{F}}{g} = \frac{\pi}{4} g \ell D^2 (\rho_s-\rho_f),
  \label{fg}
\end{gather}
and the drag force is 
\begin{gather}
  \widetilde{F}_d = \frac{1}{2} \widetilde{C}_d \rho_f \widetilde{V}^2 D 
  \ell.  
  \label{fd}
\end{gather}
Notice that the cross-sectional area factor $\pi\left({D}/{2}\right)^2$
for the sphere from \en{spheredrag} is replaced by $D\ell$ for the
cylinder, and the tildes are used here to denote cylindrical quantities.
We also make use of the drag coefficient for a cylinder from
\cite{batchelor-1967}
\begin{gather}
  \widetilde{C}_d = \frac{8\pi}{\Reynolds\,
    \ln\left(\frac{7.4}{\Reynolds}\right)},
  \label{drag_cylinder}
\end{gather}
which holds when $\Reynolds \ll 1$.  

The settling velocity for the cylinder is then obtained by equating the
gravitational and drag forces in \en{fg} and \en{fd}, which yields
\begin{gather}
  \widetilde{V}_s = \sqrt{\frac{\pi g D (\rho_s-\rho_f)}{2
      \widetilde{C}_d \rho_f}}. 
  \label{classical_cylinder}
\end{gather}
Observe that the factor of length $\ell$ cancels in the above
expression, so that this same expression is valid also for the 2D
geometry in the $\ell\rightarrow\infty$ limit.  Furthermore, this
expression is the same as that for the sphere in \en{terminal_sphere}
except that the factor $\sqrt{4/3}$ is replaced here with
$\sqrt{\pi/2}$, and of course the cylinder drag coefficient is also
different.  When equations \en{drag_cylinder}--\en{classical_cylinder}
are taken together, they reduce to a nonlinear equation in
$\Reynolds$
\begin{gather}
  f(\Reynolds) = \Reynolds - \frac{\rho_f g D^3 (\rho_s -
    \rho_f)}{16\mu^2}\; \ln \left( \frac{7.4}{\Reynolds} \right) = 0,  
  \label{eq:fRe}
\end{gather}
which can alternatively be written as an equation in $\widetilde{V}_s$.
It is easy to show that the function $f(x)$ is continuous on the
interval $0<x<\infty$ and has the following properties:
\begin{gather*}
  \lim_{x \to 0^+} f(x)    = -\infty, \quad
  \lim_{x \to \infty} f(x) = +\infty 
  \quad \text{and} \quad  
  f^\prime(x) > 0. 
\end{gather*}
Therefore, $f$ is guaranteed to have a unique positive real root by the
intermediate value theorem.

Newton's method may be used to solve \en{eq:fRe} for $\Reynolds$, and we
find that any initial guess for $\Reynolds$ suffices since the
convergence is to rapid.  Table~\ref{asett} lists values of
$\widetilde{V}_s$ from \en{eq:fRe} for parameters $D=0.018$ and $\rho_s$
ranging from 1.01 to 1.05.
\begin{table}
  \begin{center}
    \renewcommand{\extrarowheight}{3pt}
    \begin{tabular}{|c|c|c|}
      \hline
      $\rho_s $ & $\Delta\rho$ & $\widetilde{V}_{s}$ \\
      \hline\hline
      1.01 & 0.01 & 0.0024\\
      1.02 & 0.02 & 0.0045\\
      1.03 & 0.03 & 0.0066\\
      1.04 & 0.04 & 0.0086\\
      1.05 & 0.05 & 0.0105\\
      \hline
    \end{tabular}
  \end{center}
  \caption{Settling velocities $\widetilde{V}_s$ for a cylindrical
    particle, obtained by solving equation \en{eq:fRe} with $\rho_f=1$
    and $D=0.018$.} 
  \label{asett}
\end{table}
As expected, the settling velocity increases with particle density as in
the Stokes case.

\subsection{Wall-corrected settling velocities}
\label{sec:vs-walls}

In this section, we summarize a number of formulas that approximate
settling velocity for a particle in a bounded fluid domain that consists
of a channel with two parallel, vertical walls separated by a distance
$W$.  It is well-known that the bounding walls exert an additional
retarding effect on a sedimenting
particle~\cite{benrichou-etal-2005,chhabra-etal-2003,faxen-1946,happel-brenner-1983,pianet-arquis-2008,takaisi-1955,white-1946}
so that the settling velocity is lower in a channel than in an unbounded
domain under the same conditions. The effect of these wall interactions
may be approximated by means of a ``wall correction factor'' $\lambda$,
that is usually expressed in the
form~\cite{benrichou-etal-2005,benrichou-etal-2004}
\begin{gather}
  \lambda(k, \Reynolds) = \frac{\widetilde{F}_d(k)}{\mu V_c},
  \label{rich}
\end{gather}
where $\widetilde{F}_d(k)$ represents a drag force per unit length,
$V_c$ is the terminal settling velocity in the channel, and $k = D/W$ is
the dimensionless particle size with $0<k<1$.  

Note that the factor $\lambda(k, \Reynolds)$ depends on both Reynolds
number and particle size; however, it is well
known~\cite{chhabra-etal-2003} that at either very low or very high
values of $\Reynolds$, $\lambda$ is nearly independent of the Reynolds number.
In this study, we are concerned with the low $\Reynolds$ regime and so
it is reasonable to assume that $\lambda$ depends on $k$ only.  Various
formulas for the wall correction factor have been reported in the
literature (see \cite{benrichou-etal-2005}, for example) all of which
have the same channel geometry as pictured in Figure~\ref{fig:channel}.
Some of the more common wall correction factors are listed below:
\begin{itemize}
\item White~\cite{white-1946}: carried out experiments with various
  wires and ebonite rods in a channel containing viscous liquids such as
  glycerin and paraffin.  He obtained the following experimental fit for
  the drag force on a cylinder
  \begin{gather}
    \lambda(k) = \frac{-6.4}{\ln(k)}, 
    \label{white_wall}
  \end{gather}
  whose domain of validity is restricted to $0 < k < 0.2$.

\item Fax\'{e}n~\cite{faxen-1946,happel-brenner-1983}: derived an
  approximate analytical solution of the Stokes equations, from which he
  obtained
  \begin{gather}
    \lambda(k) = \frac{-4\pi}{0.9157 + \ln(k) -
      1.724k^2 + 1.730k^4 - 2.406k^6 + 4.591k^8}. 
    \label{faxen_wall}
  \end{gather}
  Some authors claim that this approximation is valid for $k$ as large
  as 0.5~\cite{benrichou-etal-2005}, while others cite an upper bound of
  $k=0.3$ or even lower~\cite{pianet-arquis-2008} which is more in line
  with our numerical simulations (see Figure~\ref{fig:our_sett} in
  Section~\ref{sec:num-one}).

\item Takaisi~\cite{takaisi-1955}: used an analytical solution of 
  Oseen's equations to obtain the approximation
  \begin{gather}
    \lambda(k) = \frac{-4\pi}{0.9156 + \ln(k)},
    \label{takaisi_wall}
  \end{gather}
  which is restricted to $0 < k < 0.2$.  He also performed a comparison
  with White's experimental fit and showed that the two
  expressions match reasonably well when $k < 0.05$.
\end{itemize}

If we now consider $\lambda(k)$ to be a known function of the
dimensionless particle size $k$, then equation \en{rich} can be solved
for the drag force per unit length as
\begin{gather*}
  \widetilde{F}_d(k) = V_c \mu \lambda(k).
\end{gather*}
Equating this expression with the gravitational force
\begin{gather*}
  \fsub{\widetilde{F}}{g} = \frac{\pi}{4} g D^2 (\rho_s -\rho_f), 
\end{gather*}
we find the following formula for the confined (or
wall-corrected) terminal settling velocity of a cylinder
\begin{gather}
  V_c = \frac{\pi g D^2 (\rho_s -\rho_f)}{4\mu \lambda(k)}.
  \label{V_D}
\end{gather}
In the next section, this expression will be compared with numerically
simulated values for the three choices of $\lambda(k)$ listed above.

\section{Numerical results: Single particle case}
\label{sec:num-one}

In this section, we concentrate on a single particle that settles under
the influence of gravity.  Two initial configurations are investigated:
one a symmetric case in which the particle is released along the
centerline, and a second asymmetric case where the particle is released
from an off-center location.

We restrict ourselves to a low Reynolds number regime corresponding to
$\Reynolds\lesssim 7$, where $\Reynolds$ refers to a ``final'' Reynolds
number that is based on the vertical velocity after a particle has
achieved its terminal settling velocity.  Unless otherwise indicated, we
choose physical parameters $\rho_f=1$, $\rho_s=1.01$ and $D=0.08$.  The
wall and particle IB spring stiffness values are taken large enough that
the walls and particle boundary do not deform ``too much'' from their
initial shapes -- taking $\sigma_w=\sigma_c=3\times 10^4$ keeps the
relative error in these boundary shapes to within approximately 0.2\%.

Except for the convergence study in the next section, most of our
simulations are performed at the same grid resolution of
$\dx=\dy=0.0083$ and with a time step of $\dt=10^{-5}$.  We also select
the number of IB points for the two channel walls ($N_w$) such that the
ratio of spacing between wall points to fluid grid size is
$h_w/\min(\dx,\dy) \approx \frac{1}{3}$ -- this ratio is well within the
factor of $\frac{1}{2}$ that is recommended to avoid leakage of fluid
between IB points~\cite{peskin-2002}.  We also adjust the number of IB
points for the particle ($N_c$) until the initial mesh computed by
DistMesh satisfies the same criterion.

\subsection{Convergence study}
\label{sec:converge}

\begin{figure}
  \centering
  \includegraphics[width=\columnwidth]{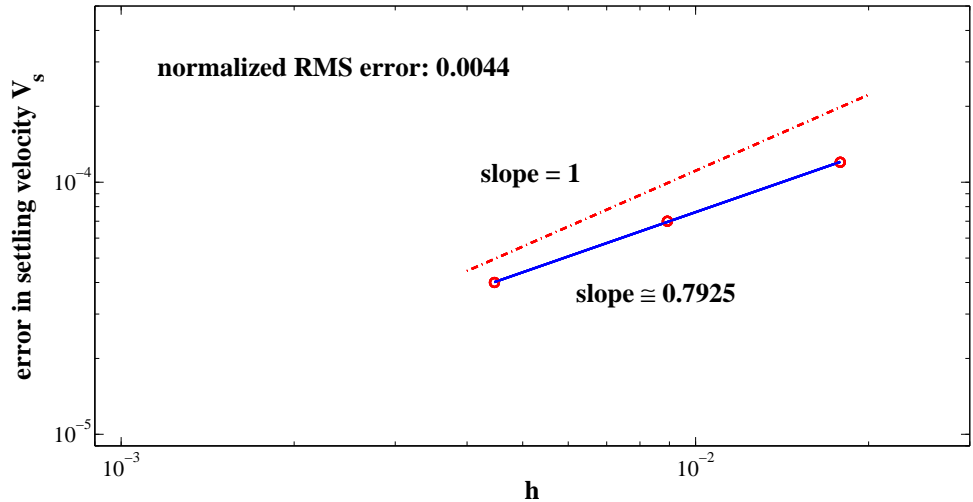}
  \caption{A convergence study in the settling velocity $V_s$ on a
    sequence of grids.  The difference in values of $V_s$ on two
    successively finer grids is plotted versus the grid spacing $h$.  A
    straight line with slope 1 corresponding to a first-order method is
    shown for comparison purposes.}
  \label{fig:converge}
\end{figure}

We begin by performing a convergence study that validates the spatial
accuracy of our numerical method.  As mentioned earlier in
section~\ref{sec:algo}, the IB algorithm being employed here is
well-known to be first order accurate in space.  To verify this result,
we select a sequence of fluid grids with $N_x=N_y=56$, 112, 224 and 448
on a square domain with side length $L_x=L_y=1$, and use the settling
velocity $V_s$ as a representative measure of the solution for each
case.  The difference between values of $V_s$ on successive grids is
calculated and the results are plotted in Figure~\ref{fig:converge},
which demonstrates that our numerical solution converges as the grid
spacing is reduced.  The curve is nearly a straight line on a log-log
scale, and the slope of 0.79 obtained from a least squares fit suggests
that our implementation of the IB method is close to the expected
first-order accuracy.  Similar convergence rates are observed for other
quantities such as fluid velocity, IB position, etc.

\subsection{Comparison with Stokes' law}
\label{sec:domain_size}

We aim next to validate the numerical method against the settling
velocity $\widetilde{V}_s$ for a cylinder in an unbounded medium.
However, we recall that our doubly periodic geometry implies that a
single particle actually corresponds to an infinite array of sedimenting
particles.  Therefore, in the absence of solid boundaries or any other
mechanism for dissipating energy, the net effect of gravity acting on
such an infinite array of mass-bearing particles will be to accelerate
the particles and the surrounding fluid indefinitely.  This situation is
clearly non-physical, and so instead we introduce walls into the domain
that are situated ``far enough'' from the particle so as to minimize
wall-particle interactions and yet still permit the particle to reach
its natural terminal velocity. To this end, we take a square domain with
side length $L_x=L_y=L$ that contains two vertical walls separated by a distance
$W=L-0.04$, and perform a sequence of computations with successively
larger $L$.  In particular, we fix the particle diameter at
$D=0.018~\units{cm}$ and vary $L$ between 1 and 7~\units{cm}.  The fluid
viscosity here and in the next section is $\mu=1~\units{g/cm\,s}$, which
we remark is much larger than in later sections because a meaningful
comparison to $\widetilde{V}_s$ is only possible at low Reynolds number.

The computed values of settling velocity are summarized in
Table~\ref{varydomain}, from which we observe that as $L$ increases $V$
approaches a limiting value of roughly $0.001674~\units{cm/s}$.  The
results have clearly converged on the largest domain size, but the
limiting value of $V$ is significantly less than the Stokes settling
velocity from equation \en{classical_cylinder}, $\widetilde{V}_s =
0.00239~\units{cm/s}$.  We suspect that the discrepancy is due to a
combination of effects arising from grid resolution, including the
first-order dispersive errors in our numerical scheme and the increase
in the effective thickness of the walls and particle owing to the delta
function smoothing width (this last effect is discussed further in
section~\ref{sec:wall_distance}).
\begin{table}
  \centering
  \renewcommand{\extrarowheight}{3pt}
  \begin{tabular}{|c|c|} \hline
    $L$ & computed $V$ \\ \hline\hline
    1 & 0.001230\\
    2 & 0.001462\\
    3 & 0.001565\\
    4 & 0.001635\\
    5 & 0.001673\\
    6 & 0.001674\\
    7 & 0.001674\\ \hline
  \end{tabular}
  \caption{Computed settling velocity as a function of domain size $L$ for
    $D=0.018$.} 
  \label{varydomain}
\end{table}

\subsection{Single particle initially along the centerline} 

We next consider the channel domain pictured in Figure~\ref{fig:channel}
wherein the particle is initially released along the center of the
channel.  In this case, the symmetry suggests that any forces generated
by particle-wall interactions are balanced and so the particle should
fall along the centerline without veering to either side.  We perform a
number of sensitivity studies that investigate the effect of parameters
such as the fluid domain size $[0,L_x] \times [0,L_y]$, density
difference $\Delta\rho$, and relative particle size $k$ on the settling
velocity.  As in the previous section, simulations are restricted to low
Reynolds number by taking $\mu=1$.

\subsubsection{Dependence of settling velocity on density difference
  $\Delta\rho$} 
\label{sec:delta_rho}

\begin{figure}[htbp]
  \centering
  \includegraphics[width=\columnwidth]{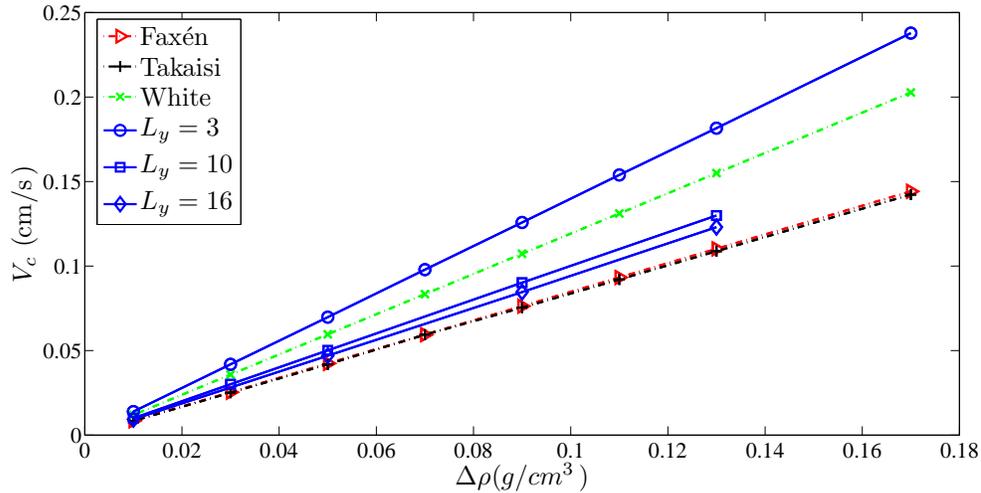}
  \caption{Settling velocity in a channel as a function of $\Delta\rho$.
    parameter values: $L_x=1,\; L_y\in\{3,10,16\}, \; D=0.1, \; W=0.98,
    \; k= 0.102$.}
  \label{banded5}
\end{figure}

For a fixed channel size with $W=0.98$ and $L_x=1$, we vary $\Delta\rho$
between $0.01$ and $0.17$.  The plot of settling velocity in
Figure~\ref{banded5} shows that the wall-corrected settling velocity
$V_c$ increases linearly with $\Delta\rho$, which is consistent with
equation \en{V_D}.  We also consider the effect of changes in the
channel length by taking values of $L_y\in \{3, 10, 16\}$, from which we
find that the influence of periodic copies in the $y$--direction is
most greatest for the shortest channel ($L_y=3$) which also exhibits the
highest settling velocity.  As the channel length is increased, the
settling velocity decreases until by $L_y=16$ the results appear to have
converged and are incidentally closest to the result predicted by
Fax\'en's formula~\en{faxen_wall}.

\subsubsection{Dependence of settling velocity on particle size $k$}
\label{sec:wall_distance}

\begin{figure}[htbp]
  \centering
  \includegraphics[width=0.95\columnwidth]{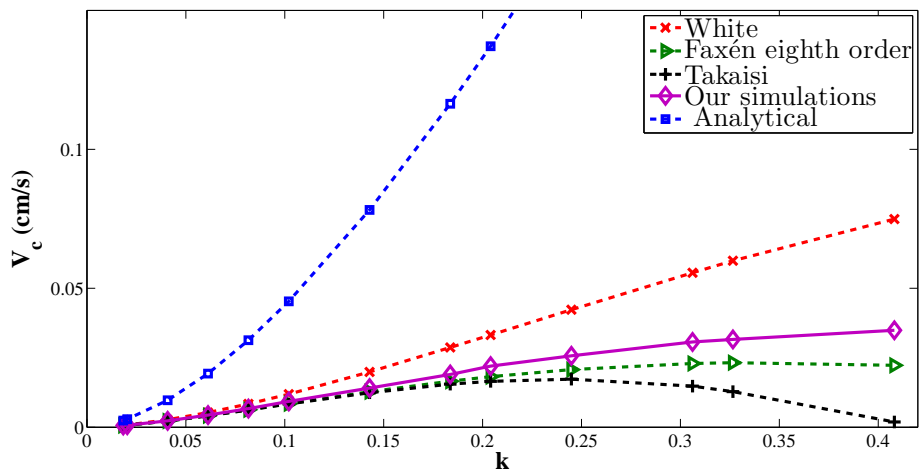}
  \caption{Variation of settling velocity with $k$ for the computed
    results and various analytical expressions. Parameter values:
    $L_x=1,$ $L_y=16$, $\mu=1$.}
  \label{fig:faxen_my_comp} 
\end{figure}

As mentioned earlier in section \ref{sec:vs-walls}, the work of
Fax\'{e}n, White, Takaisi, and others suggests that $\lambda$ depends
only on $k$ at low values of Reynolds number.  This motivates our next
sensitivity study of the effect of dimensionless particle size, for
which we again fix the channel width at $W=L_x-0.04$, and then vary $k$
by choosing values of particle diameter $D\in[0.018, 0.96]$.  Alongside
our computational results in Figure~\ref{fig:faxen_my_comp}, we have
displayed corresponding estimates of the confined settling velocity
$V_c$ calculated using equation \en{V_D} with the three wall correction
factors \en{white_wall}--\en{takaisi_wall}.  The unbounded cylindrical
settling velocity $\widetilde{V}_s$ is also included for comparison
purposes, which clearly diverges from the wall-corrected values away
from $k=0$.  For these simulations, $\Reynolds$ was in the range $[1.2
\times 10^{-5}, 1.8 \times 10^{-2}]$.

\leavethisout{
  \begin{table}
    \begin{center}
      \renewcommand{\extrarowheight}{3pt}
      \begin{tabular}{|c|c|c|c||c|}
        \hline
        $k$ & Fax\'{e}n & Takaisi & White & Computed  \\
        \hline\hline
        $0.0184$& $0.00061$&$0.00061$&$0.00067$&$0.00065$\\
        $0.0204$& $0.00073$&$0.00073$&$0.00081$&$0.00073$\\
        $0.0408$& $0.0022$ &$0.0022$ &$0.0027$ &$0.0023$\\
        $0.0612$& $0.0042$ &$0.0041$ &$0.0053$ &$0.0044$\\
        $0.0816$& $0.0063$ &$0.0062$ &$0.0084$ &$0.0067$\\
        $0.1020$& $0.0085$ &$0.0084$ &$0.0119$ &$0.0093$\\
        $0.1429$& $0.0128$ &$0.0124$ &$0.0199$ &$0.0140$\\
        $0.1837$& $0.0166$ &$0.0155$ &$0.0287$ &$0.0189$\\
        $0.2041$& $0.0182$ &$0.0165$ &$0.0332$ &$0.0220$\\
        $0.2449$& $0.0208$ &$0.0173$ &$0.0423$ &$0.0257$\\
        $0.3061$& $0.0228$ &$0.0148$ &$0.0556$ &$0.0307$\\
        $0.3265$& $0.0232$ &$0.0128$ &$0.0599$ &$0.0316$\\
        $0.4082$& $0.0215$ &$0.0019$ &$0.0749$ &$0.0349$\\
        \hline
      \end{tabular}
    \end{center}
    \caption{Variation of settling velocity with $k$ for 
      three wall-corrected formulas and our computed solution.} 
    \label{tab:comp_a}
  \end{table}
}

\begin{figure}[htbp]
  \centering
  \includegraphics[width=0.95\columnwidth]{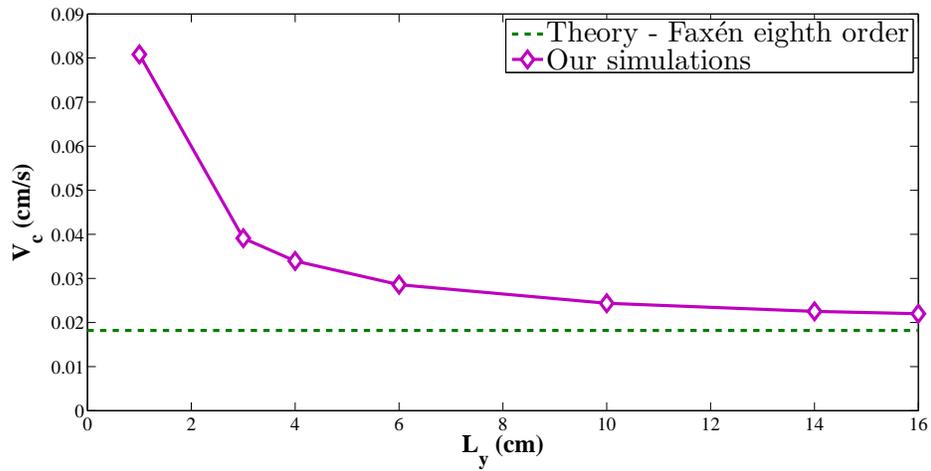}
  \caption{Effect of increasing the channel length $L_y$ and hence
    reducing the impact of the periodic copies in $y$.  Parameter
    values: $\mu=1$, $L_x=1$, $D=0.2$, $W=0.98$, $k=0.2041$.}
  \label{conv_faxen}
\end{figure}

Our computed results match most closely with Fax\'{e}n's formula, which
is most often cited as the most accurate approximation for the
wall-corrected settling velocity.  We also performed a study of the
effect of changes in the channel length $L_y$, in order to determine the
effect of any possible interference between vertical periodic copies of
the particle due to the periodicity assumption.  Results for
$L_y\in[1,16]$ are displayed in Figure~\ref{conv_faxen} which clearly
show that our computed settling velocity converges to a value quite
close to that predicted by Fax\'{e}n's formula when $k\lesssim 0.2$.

\begin{figure}[htbp]
  \centering
  \includegraphics[width=0.95\columnwidth]{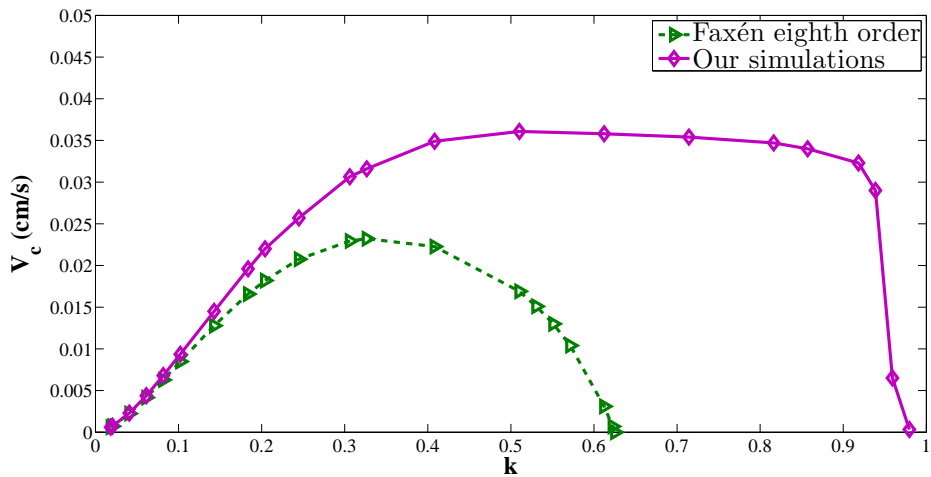}
  \caption{Plot of settling velocity versus $k$, comparing our simulations
    to Fax\'en's approximation.  Parameter values:
    $L_x=1$, $L_y=16$, $\mu=0.01$.} 
  \label{fig:our_sett}
\end{figure}

On the other hand, there remain significant deviations between Fax\'en's
results and our computations for values of $k$ larger than 0.2, as
demonstrated in Figure~\ref{fig:our_sett}.  Fax\'en's settling velocity
levels out and attains a maximum near $k= 0.3$ and then falls to zero
near $k=0.6$. In contrast, our computed settling velocity reaches a
maximum that is roughly 60\%\ larger ($V_c\approx 0.035$) and plateaus
for $k$ roughly in the range $[0.4,0.8]$. 

Our computed settling velocity only drops to zero when $k$ is very close
to 1, which is easily justified since the particle must come to a stop
as it come into direct contact with the stationary walls.  However, our
results in Figure~\ref{fig:our_sett} show that $V_c$ actually tends to
zero not at $k=1$ but rather $k\approx 0.96$.  The reason for this
apparent reduction of 0.04 in the channel width is that the approximate
delta function in our numerical scheme has a finite smoothing width that
has the effect of introducing an extra ``effective thickness'' to both
the walls and the particle.  The numerical simulations in
\cite{stockie-2009} show that when using the cosine delta function, the
effective thickness of an immersed boundary is approximately $1.6h$,
where $h$ is the fluid grid spacing\footnote{Note that the effective
  thickness depends on the choice of regularized delta function.
  Bringley~\cite{bringley-thesis-2008} computed an effective thickness
  closer to $1.25h$ for a different but closely-related approximate
  delta function.}.  Consequently, a particle with diameter $D$ should
have an effective diameter of roughly $\fsub{D}{eff} \approx D + 3.2h$,
while the walls should each extend an additional distance of $3.2h$ into
the channel.  Taken together this suggests a total reduction of $6.4h$
in the effective channel width, which for $h=0.0083$ equals
approximately 0.053.  This is not far away from the observed reduction
of 0.04.

We summarize the behavior from our numerical simulations as follows:
\begin{itemize}
\item For small particle diameters corresponding to $k\in [0,0.2]$, the
  particle is far enough from the channel walls that the retarding
  effects of wall drag are not as prominent.  In this range, the
  dependence of the settling velocity is roughly proportional to $k$,
  which is consistent with Fax\'en's result.
\item For intermediate values of $k$, roughly in the range $[0.4,0.8]$,
  the settling velocity has attained a maximum value and remains
  approximately constant.  For these particle sizes, the interactions
  with the walls are at long range and are mediated by the fluid.
\item For values of $k\in [0.8,1.0]$, the particle is very close to the
  walls, giving rise to close-range interactions that slow the particle
  significantly.
\end{itemize}
Of course, the validity of Fax\'en's approximation is limited to
$k\lesssim 0.2$ and so it is no surprise that our results differ so much
for larger $k$.

\subsection{Single particle initially off-center}

\begin{figure}[htbp]
  \centering
  \includegraphics[trim=6in 0in 4.5in 0in,clip,width=0.28\columnwidth]{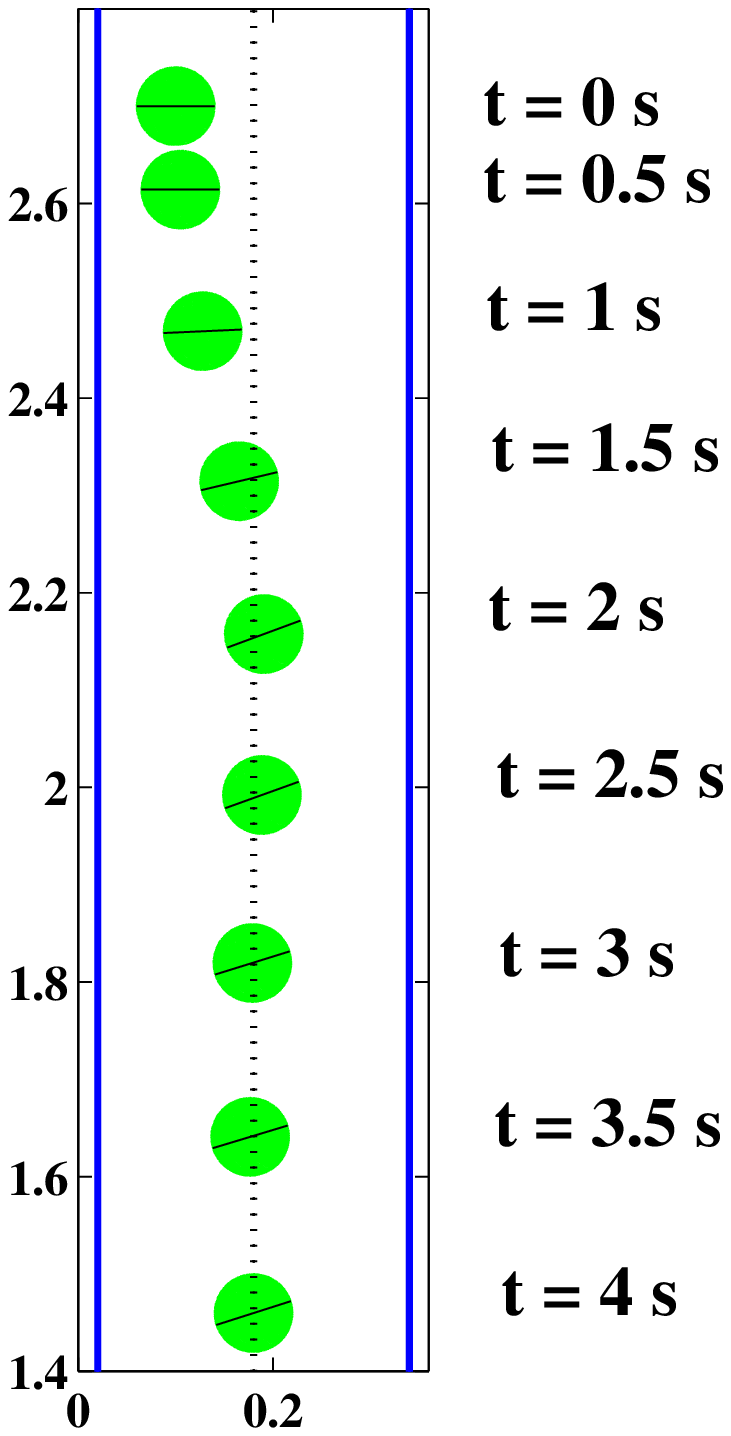} \qquad 
  \includegraphics[trim=5in 0in 4.5in 0in,clip,width=0.36\columnwidth]{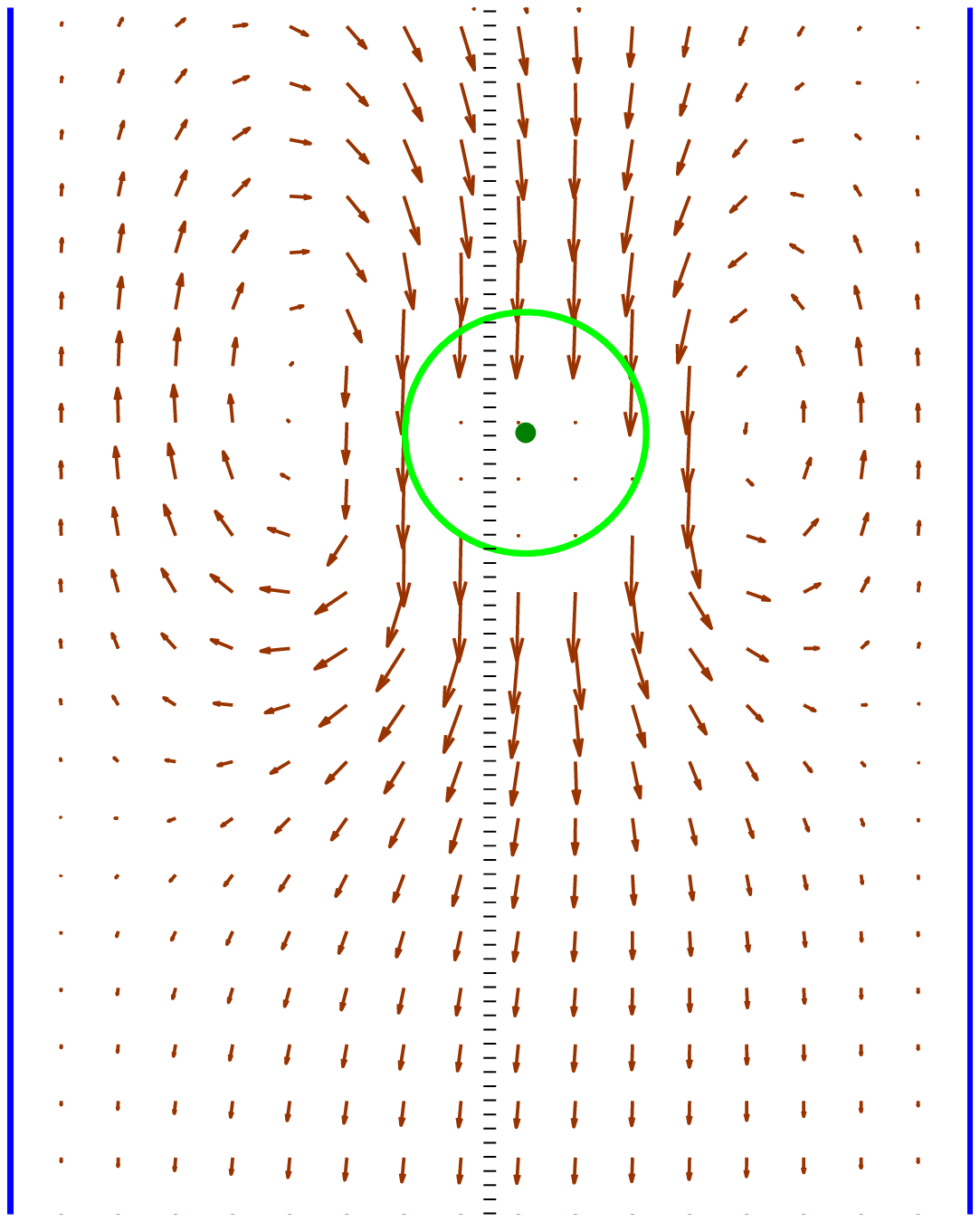}
  \caption{(Left) Settling dynamics for a single particle released
    off-center, in a channel of width $W=4D$ and $\Reynolds=4.9$.  The
    solid line drawn through the center of the particle highlights the
    rotational motion. (Right) Velocity vector plot at time $t\approx
    2~\units{s}$.}
  \label{dynamics_4d_single}
\end{figure}

In this section, we consider an asymmetry initial condition in which the
particle is released from an off-center location.  In
Figure~\ref{dynamics_4d_single}, the initial configuration labeled
``$t=0~\units{s}$'' shows the particle a distance $W/2$ to the left of
center.  The diameter of the particle is taken to be
$D=0.08~\units{cm}$, the channel length is $L_y=3$, and we consider two
different channel widths, $W=4D$ and $8D$.  We also vary Reynolds number
by taking values of the viscosity $\mu\in[0.006,
0.018]~\units{g/cm\,s}$.  This choice of parameters allows us to draw a
comparison with the analytical and experimental results reported by
Sucker and Brauer~\cite{sucker-brauer-1975}, as well as numerical
simulations of Feng et al.~\cite{feng-hu-joseph-1994}.

\begin{figure}[htbp]
  \centering
  \includegraphics[width=0.95\columnwidth]{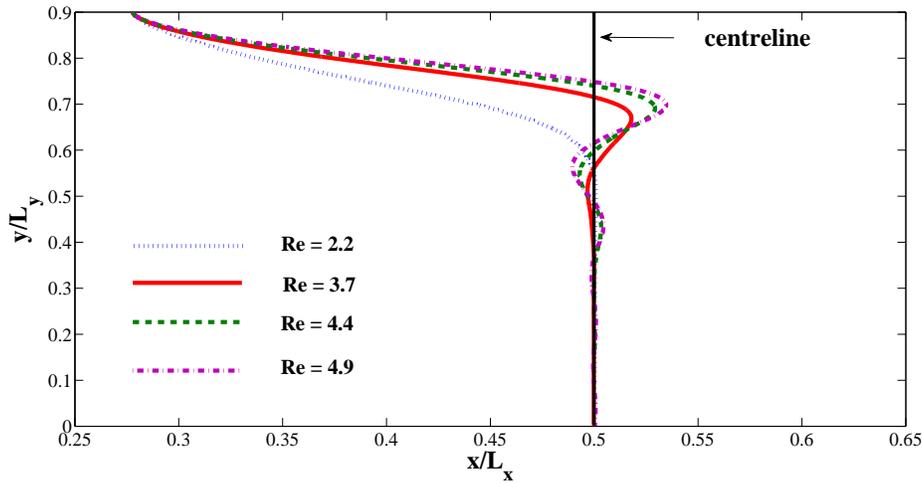}
  \caption{Settling trajectories for an off-center particle at different
    Reynolds number, in a channel of width $W=4D$.}
  \label{4d_traject1} 
\end{figure}

The settling dynamics are pictured in Figure~\ref{dynamics_4d_single}
for $W=4D$ and $\Reynolds=4.9$.  As the particle falls, it initially
drifts to the right toward the channel centerline, eventually attaining
its terminal settling velocity there.  The plot of particle trajectories
in Figure~\ref{4d_traject1} shows that the particle actually undergoes a
damped oscillation about the channel centerline with an initial
overshoot.  Simulations were also performed for three other Reynolds
numbers, $\Reynolds=2.2$, 3.7 and 4.4, and the corresponding
trajectories in Figure~\ref{4d_traject1} show that increasing
$\Reynolds$ leads to larger oscillations about the centerline.  For the
smallest value of $\Reynolds=2.2$, the particle trajectory undergoes a
nearly monotonic approach toward the centerline; in an analogy with
simple harmonic oscillation, this behavior can be described as an
overdamped oscillation.

\begin{figure}[htbp]
  \centering
  (a)\\
  \includegraphics[width=0.95\columnwidth]{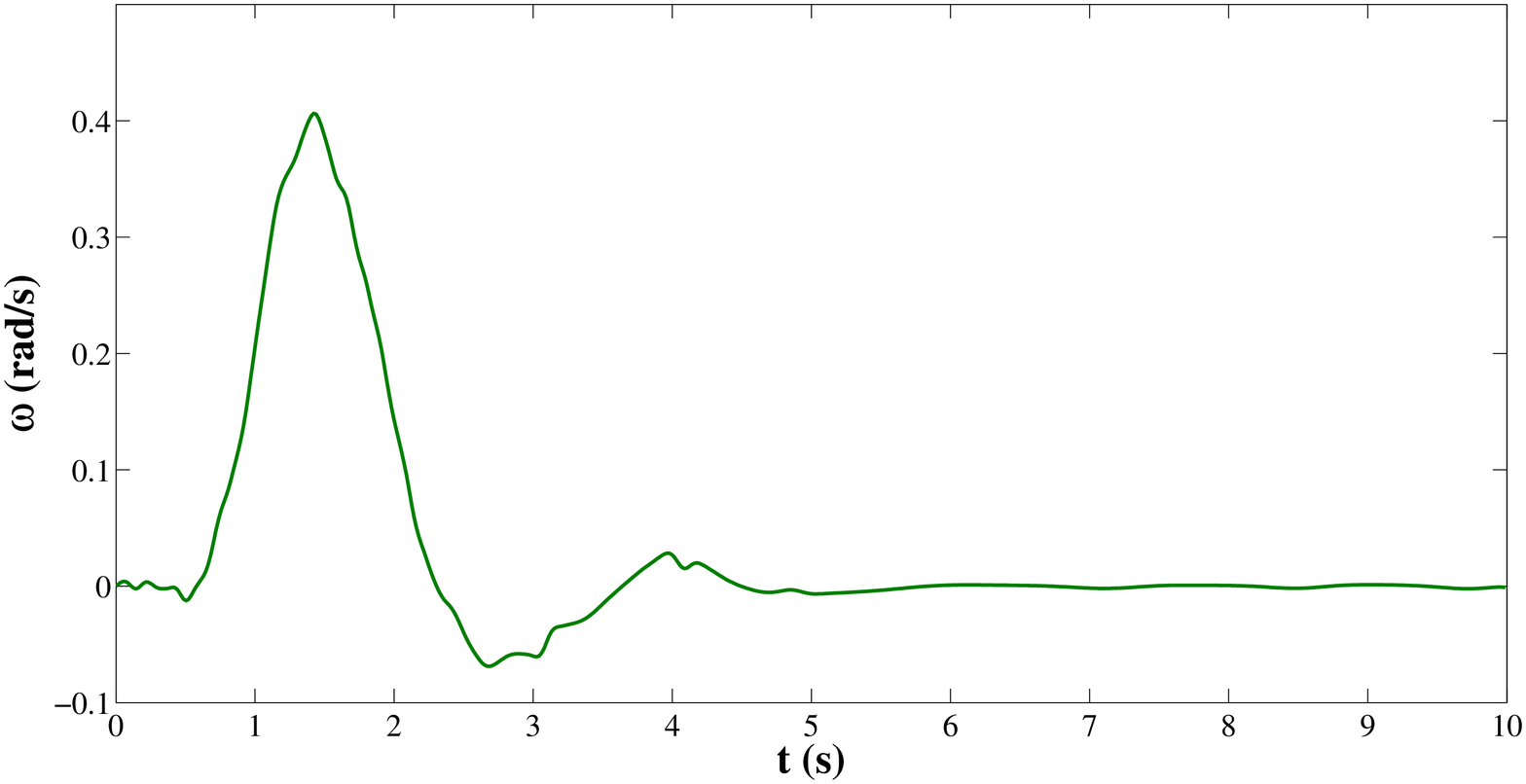}\\
  \bigskip  (b)\\
  \includegraphics[width=0.95\columnwidth]{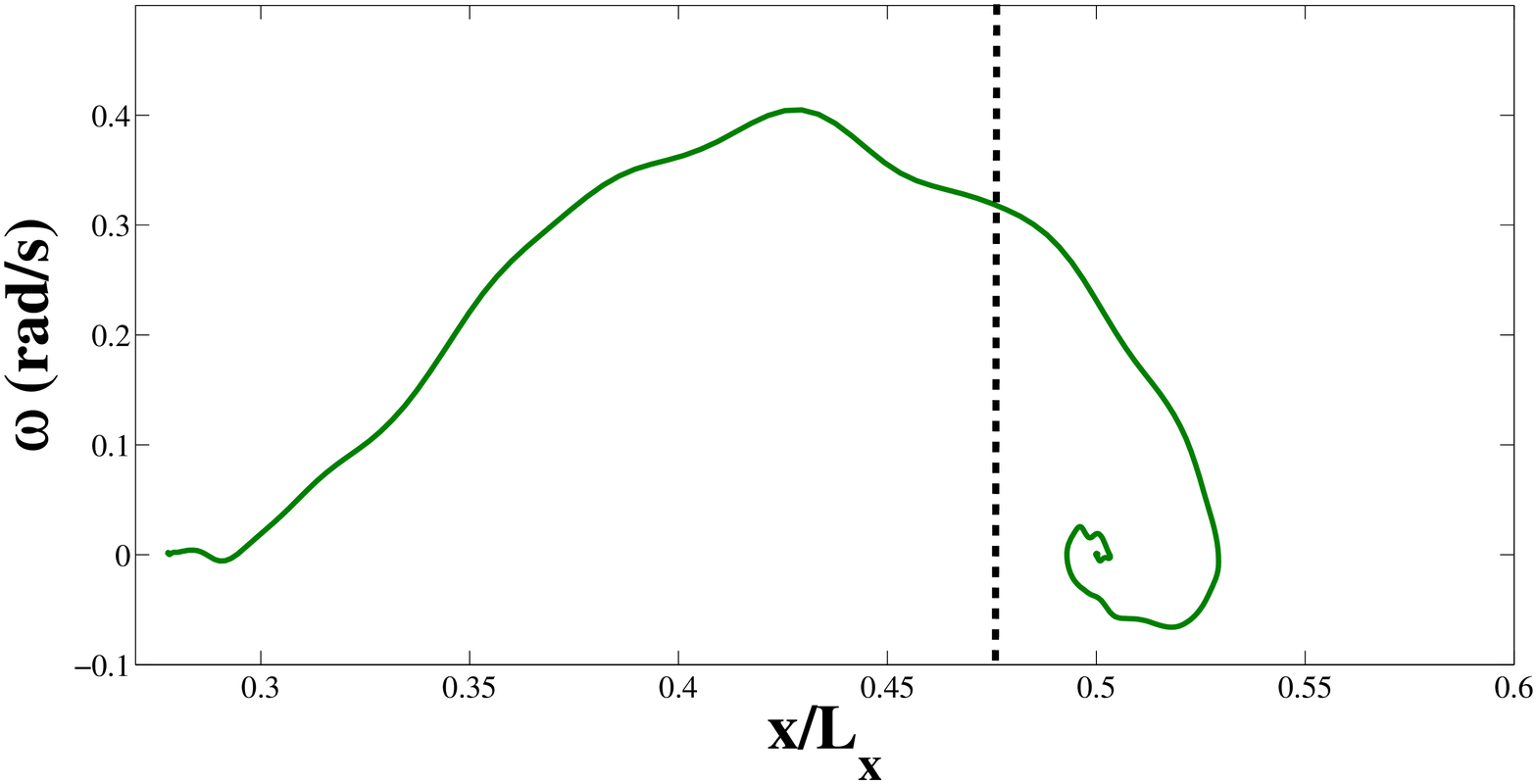}\\
  \caption{Plots of the angular velocity $\omega$ in \units{rad/s},
    for a single particle released from an off-center location, with
    $W=4D$ and $\Reynolds=4.9$.  Positive $\omega$ corresponds to
    counter-clockwise rotation.  (a) Variation of angular velocity with
    time. (b) Angular velocity versus horizontal location, where the
    dashed line represents the channel centerline.}
  \label{omega_4d}
\end{figure}

In addition to the vertical and horizontal translations of the center of
mass, the particle also undergoes a small-amplitude rotational motion as
it settles, which can be seen by tracking the progress of the
straight line drawn through the center of the
particle in Figure~\ref{dynamics_4d_single}.  This rotation can be more
easily seen in the plot of angular velocity in Figure~\ref{omega_4d} for
the $\Reynolds=4.9$ case.
Initially, as the particle drifts from its starting location toward the
centerline, it experiences a slight counter-clockwise rotation.  As the
particle approaches its equilibrium horizontal location, the rotation
slows and the particle ends up with an orientation that is slightly
tilted relative to the initial state.

\begin{figure}[htbp]
  \centering
  \includegraphics[width=0.95\columnwidth]{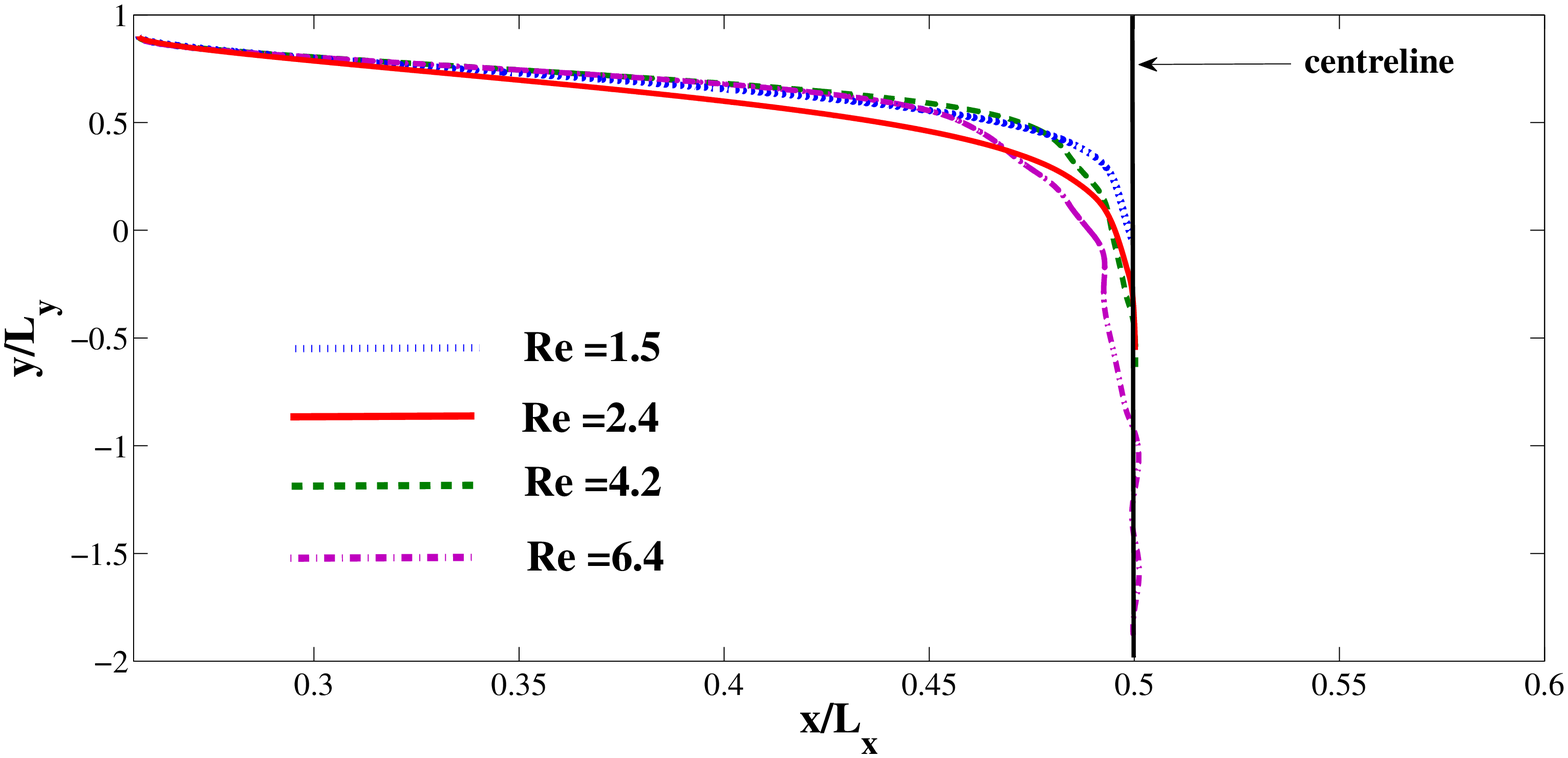}
  \caption{Settling trajectories for an off-center particle at different
    Reynolds number and in a channel of width $W=8D$.}  
  \label{8d_traject} 
\end{figure}

\begin{figure}[htbp]
  \centering
  \includegraphics[width=0.95\columnwidth]{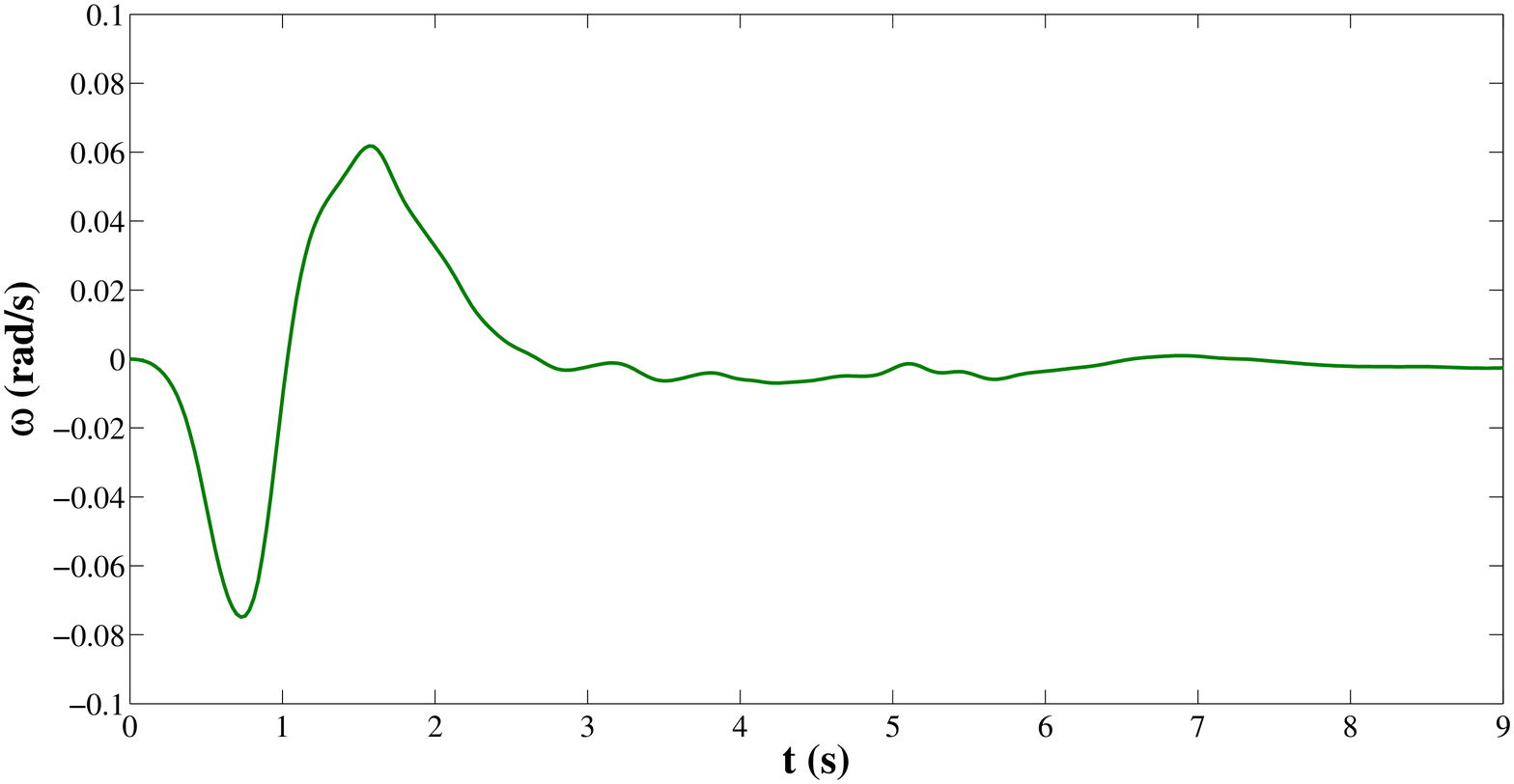}
  \caption{Plot of angular velocity $\omega$ for a single particle
    released from an off-center location, with $W=8D$ and
    $\Reynolds=6.4$.}
  \label{omega_8d_single}
\end{figure}

We next perform simulations on a channel twice as wide ($W=8D$) and take
four different values of Reynolds number, $\Reynolds=1.5$, 2.4, 4.2,
and~6.4.  The particle trajectories are shown in Figure~\ref{8d_traject}
where we observe that in contrast with the $W=4D$ results in
Figure~\ref{4d_traject1}, there is no overshoot of the centerline even
for the highest value of $\Reynolds$.  We attribute this behavior to the
fact that in a wider channel, the hydrodynamic interactions between wall
and particle that drive the horizontal motions are substantially weaker.
The plot of angular velocity in Figure~\ref{omega_8d_single} exhibits
slightly different dynamics than the narrower channel, although the
amplitude of the rotational motion is at least an order of magnitude
smaller.  This is to be expected since the rotational motion is also
driven by the wall-particle interactions which are weaker for the wider
channel case.

\begin{figure}[htbp]
  \centering
  \includegraphics[width=0.95\columnwidth]{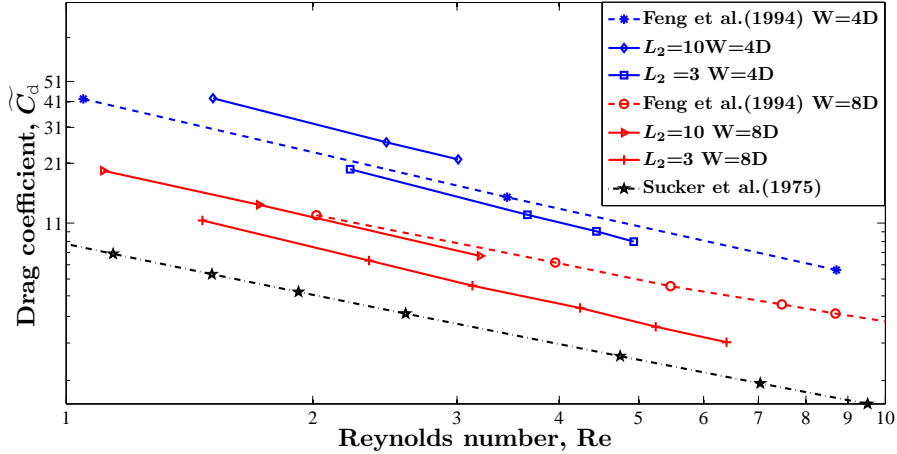}
  \caption{Comparison of drag coefficients for a single particle
    settling in channels of width $4D$ and $8D$, on a log--log scale.
    The experimental correlation of Sucker and
    Brauer~\cite{sucker-brauer-1975} and numerical results from
    FHJ~\cite{feng-hu-joseph-1994} are included for comparison purposes
    (reproduced with permission). Parameter values for our simulations:
    $D=0.08$, $L_y=3$, $L_x=0.36$ (when $W=4D$) and $L_x=0.68$ (when
    $W=8D$).}
  \label{drag_coeff_compare}
\end{figure}

We conclude our examination of the single-particle settling dynamics by
comparing in Figure~\ref{drag_coeff_compare} the drag coefficients for
the two different channel widths considered above, based on the formula
$\widetilde{C}_d=\pi(\rho_s-\rho_f)gD / (2\rho_f V_c^2)$ derived
from equation~\en{classical_cylinder}.  We have also included in this
figure values of the drag coefficient taken from the following two
papers:
\begin{itemize}
\item Feng, Hu and Joseph~\cite{feng-hu-joseph-1994}, who performed
  numerical simulations using a finite element method for a single
  particle settling in channels of width $4D$ and $8D$.  Because we will
  refer to this paper so often, we will refer to it with the
  abbreviation FHJ.
\item Sucker and Brauer~\cite{sucker-brauer-1975}, who developed an
  empirical formula that is a fit to experimental data for the cylinder
  in a very large fluid domain.  They also developed an approximate
  analytical formula for an unbounded domain that matched closely with
  the experimental data.
\end{itemize}
Our simulations match reasonably well with those of FHJ particularly for
the $W=8D$ channel in the large $L_y$ limit.  Sucker and Brauer's
empirical formula clearly deviates from both results because it applies
strictly only to unbounded domains.  While these results are
encouraging, a much more comprehensive comparison is needed in order to
draw any solid conclusions about the accuracy of our numerical approach.

\section{Numerical results: Two particle case}
\label{sec:num-two}

\begin{figure}[htbp]
  \centering
  \begin{tabular}{ccc}
    (a) & & (b) \\
    \includegraphics[trim=5.0in 0in 4.5in 0in,clip,width=0.29\columnwidth]{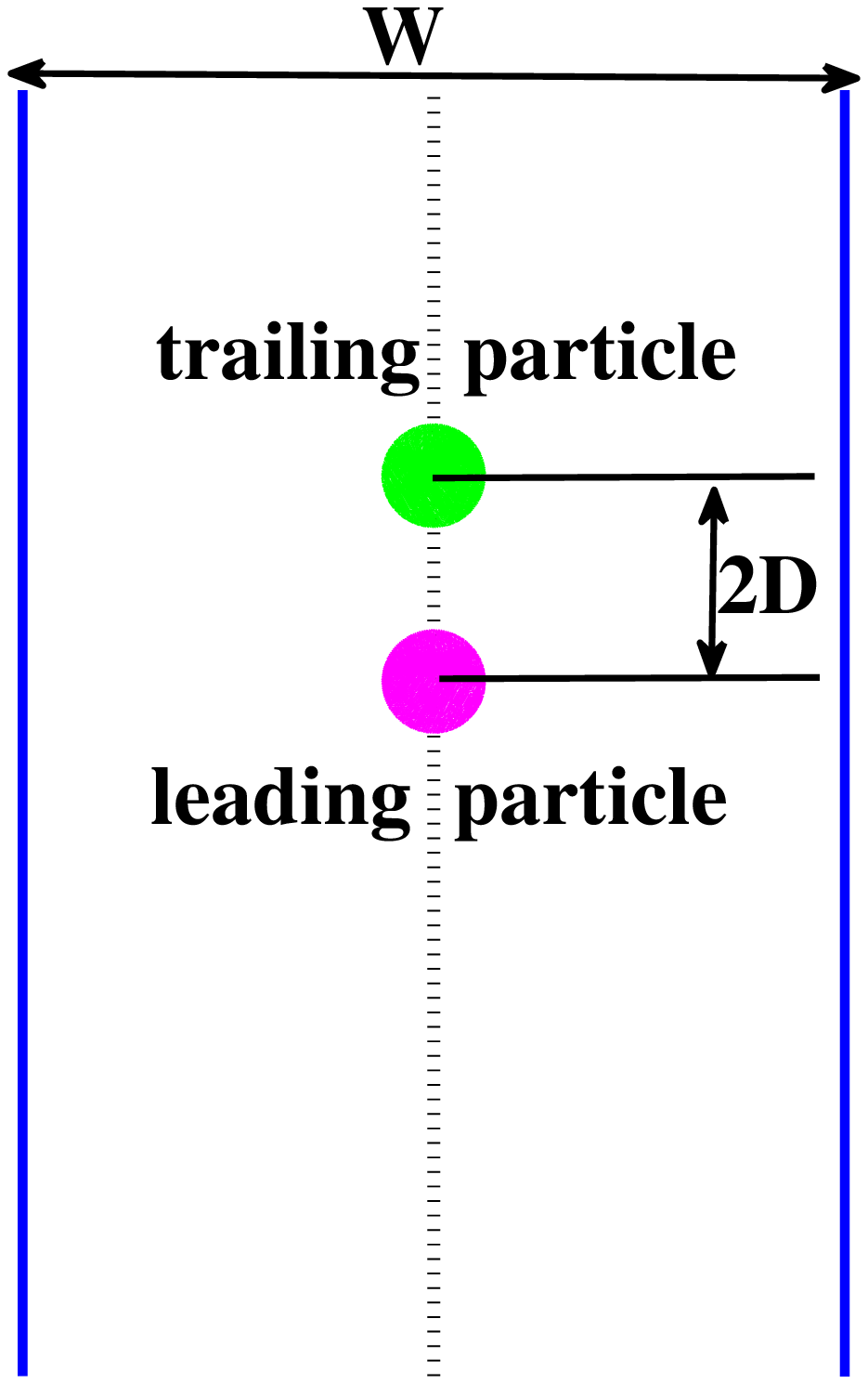} & &
    \includegraphics[trim=5.0in 0in 4.5in 0in,clip,width=0.29\columnwidth]{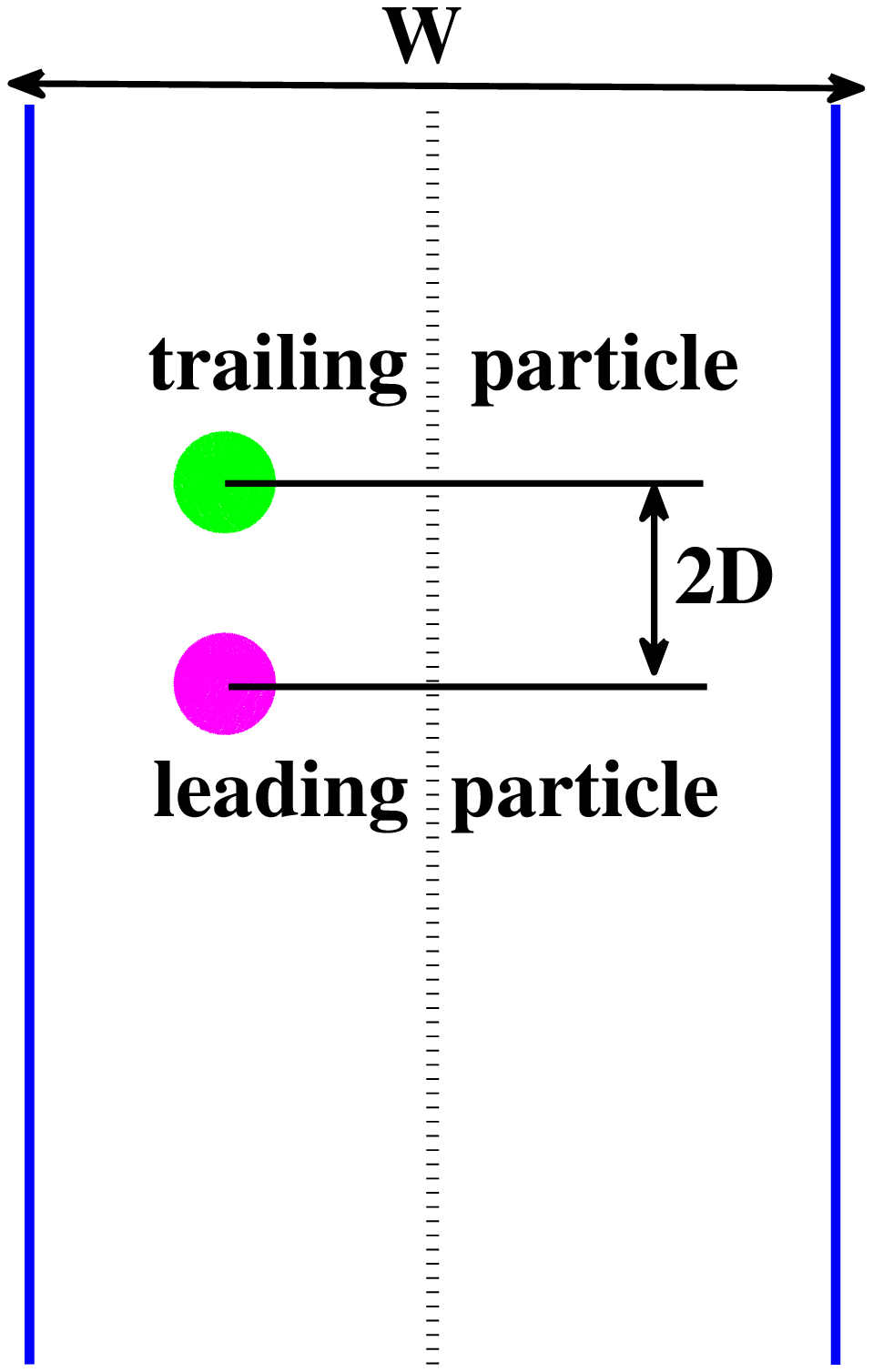}\\
    (c) & & (d) \\
    \includegraphics[trim=5.0in 0in 4.5in 0in,clip,width=0.29\columnwidth]{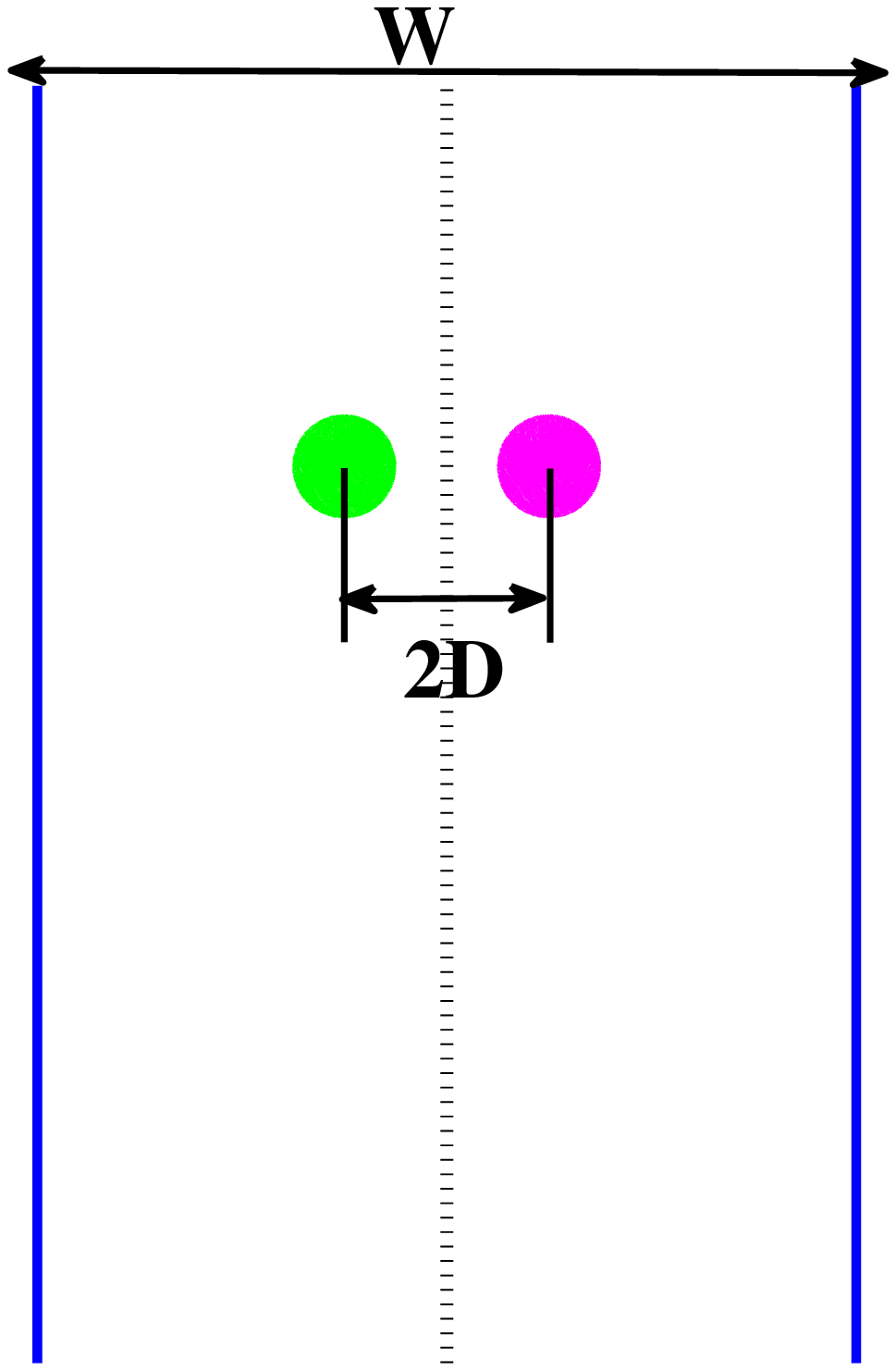} & &
    \includegraphics[trim=5.0in 0in 4.5in 0in,clip,width=0.29\columnwidth]{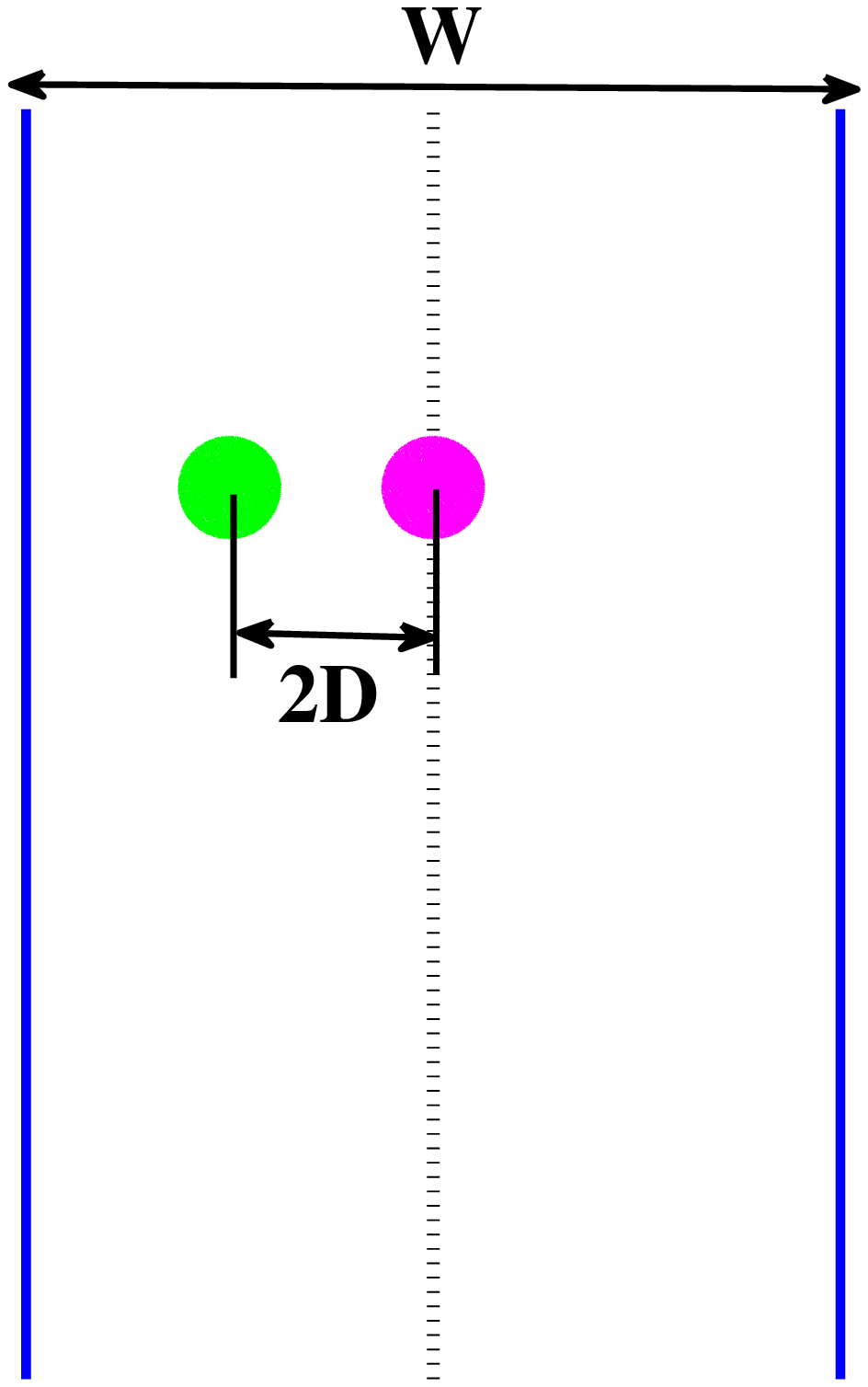}
  \end{tabular}
  \caption{Four initial configurations for the two-particle settling
    problem, where the particle centers are separated by a
    distance $2D$: (a) aligned vertically, along the channel centerline;
    (b) aligned vertically, offset to the left of center; (c) aligned
    horizontally, symmetric about the centerline; (d) aligned
    horizontally, offset to the left of center.}
  \label{fig:init_profile}
\end{figure}

This section investigates the interactions between two circular
particles with identical diameter $D$ that are settling in a channel of
length $L_y=3$.  We consider several initial configurations pictured in
Figure~\ref{fig:init_profile}, and again compare the solution at
different values of Reynolds number by varying viscosity over the range
$\mu\in[0.0015, 0.16]$.  

We begin by describing a well-known phenomenon in particle suspension
flows wherein pairs of particles interact and undergo a ``drafting,
kissing and tumbling'' behavior (which we abbreviate by DKT).  This
phenomenon has been established experimentally in papers such as
\cite{fortes-joseph-lundgren-1987,joseph-etal-1987} and demonstrated
numerically in \cite{feng-hu-joseph-1994}, and can be justified
physically as follows.  The leading particle creates in its wake a
reduction of pressure as it falls under the influence of gravity.
Provided that the trailing particle is close enough to interact with
this wake, it experiences a smaller drag force than the leading
particle.  As a result the trailing particle falls faster and the
particles approach each other -- this is the initiation of the
``drafting phase''.  As the distance between the particles decreases,
they eventually become close enough to nearly touch, which is referred
to as ``kissing''.  The kissing particles momentarily form a single
longer body that is aligned parallel with the flow; however, this
parallel arrangement is unstable and the particles eventually tumble
relative to each other and swap leading/trailing positions -- this is
the ``tumbling phase''.  The particles subsequently separate and one of
two things happens: either the DKT process repeats, or the particles
continue to separate until the interaction force becomes so weak that
they fall independently at their ``natural'' wall-corrected vertical
settling velocity \cite{prosperetti-tryggvason-2007}.

The simulations in this section are performed using the four initial
configurations depicted in Figure~\ref{fig:init_profile}:
\begin{enumerate}
  \renewcommand{\theenumi}{\alph{enumi}}
  \renewcommand{\labelenumi}{(\theenumi)}
\item Aligned vertically, one above the other along the channel
  centerline.
\item Aligned vertically, but shifted to the left to a position midway
  between the channel centerline and the left wall.
\item Aligned horizontally, and placed symmetrically about the
  centerline. 
\item Aligned horizontally, but shifted to the left of center.
\end{enumerate}
In all cases, the particle diameter is $D=0.08~\units{cm}$
and the initial separation distance between the centers of mass of the
two particles is $2D$.  As in the previous section, we also consider two
channel widths, $W=4D$ and $W=8D$. 

\subsection{Two vertically-aligned particles, released along the centerline}
\label{sec:cent}

As a first test of the two-particle case, we use the initial set-up
shown in Figure~\ref{fig:init_profile}(a) wherein the particles are both
released along the centerline with their centers of mass separated
vertically by a distance $2D$.  We perform simulations with channel
width $W=8D$ and three different Reynolds numbers, $\Reynolds=3$, 14 and
80.

\begin{figure}[htbp]
  \centering
  (a) Vertical separation distance\\
  \includegraphics[width=0.95\columnwidth]{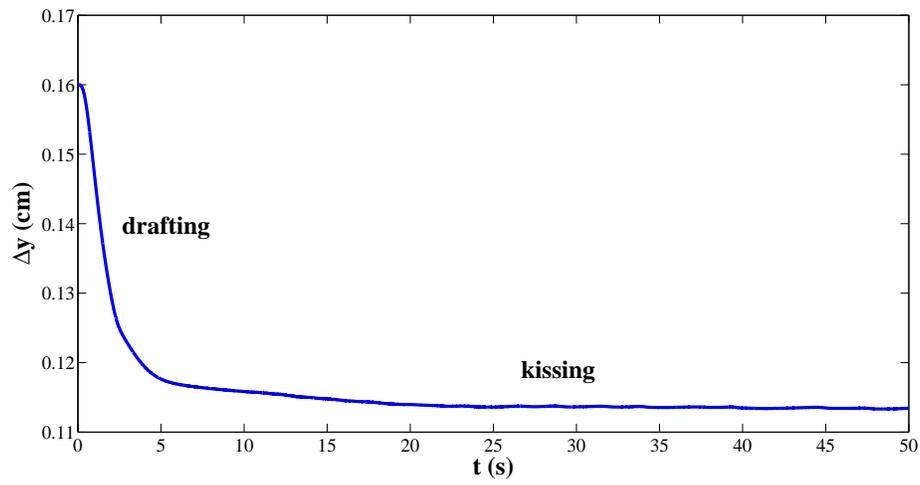}\\
  \bigskip  (b) Angular velocities\\
  \includegraphics[width=0.95\columnwidth]{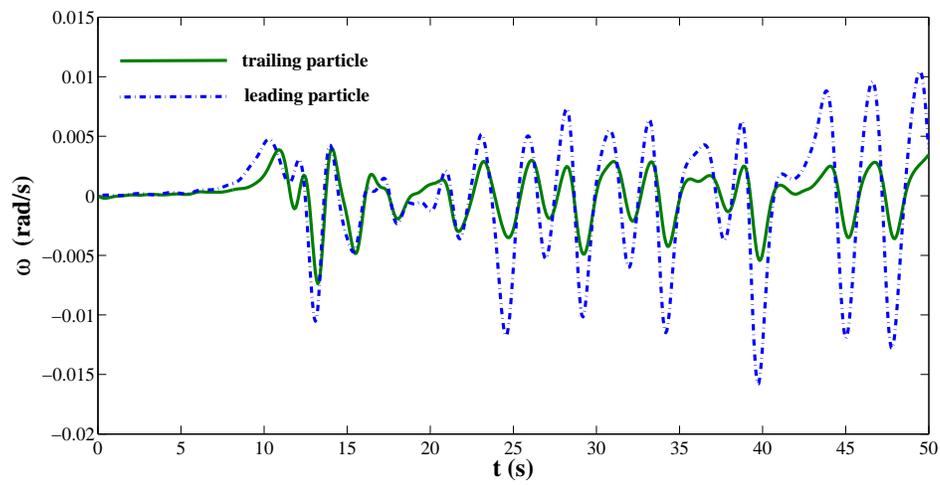}
  \caption{Settling dynamics of two particles initially aligned
    vertically, with parameters $W=8D$ and $\Reynolds=3$. (a) Vertical
    separation distance. (b) Angular velocity $\omega$ (positive $=$
    counter-clockwise rotation).}
  \label{fig:cl1}
\end{figure}

Starting with the smallest value of $\Reynolds=3$, we find that both
particles remain along the channel centerline throughout the simulation,
and while drafting and kissing behavior is observed, no tumbling occurs.
As seen in Figure~\ref{fig:cl1}(a), the trailing particle approaches
quite close to the leading particle, but never touches it.  This is
because of the increase in the effective diameter of the particle owing
to the delta function smoothing radius, as was discussed already at the
end of section~\ref{sec:wall_distance}.  After kissing, the particles
continue to fall as a single body with no significant relative motion,
except for a very slight ``wobble'' that corresponds to a
small-amplitude oscillation in the orientation angle (refer to the
angular velocity plot in Figure~\ref{fig:cl1}(b)).  This motion can be
attributed primarily to oscillations of the IB points making up the
particles that arises from the IB spring forces driving slight
deviations from the equilibrium (stress-free) state; this is an artifact
of the IB method that is not present in actual solid particles.

\begin{figure}[htbp]
  \centering
  (a) Vertical separation distance\\
  \includegraphics[width=0.85\columnwidth]{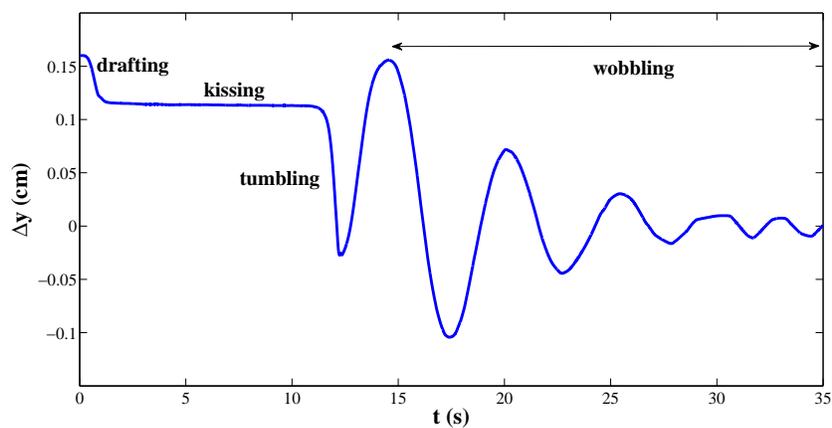}\\
  \bigskip  (b) Horizontal separation distance\\
  \includegraphics[width=0.85\columnwidth]{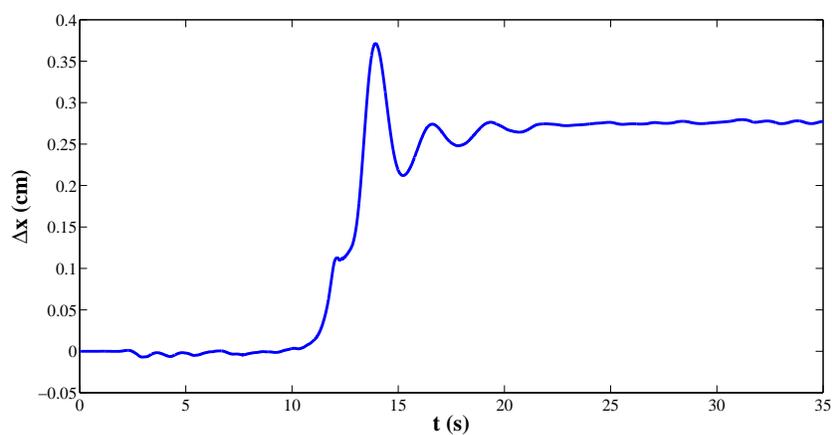}\\
  \bigskip  (c) Angular velocities\\
  \includegraphics[width=0.85\columnwidth]{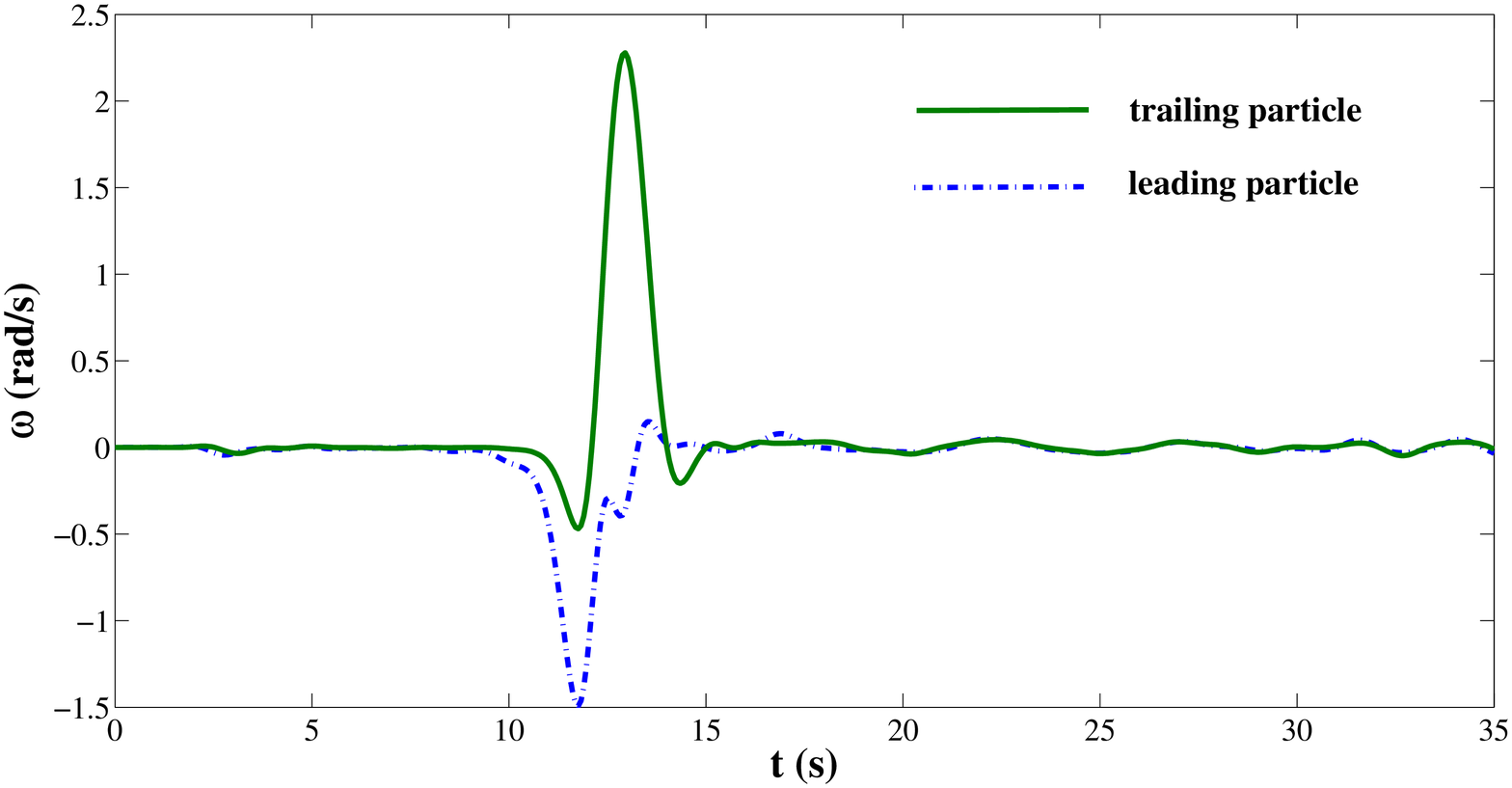}
  \caption{Settling dynamics of two particles at $\Reynolds=14$
    initially aligned vertically along the centerline in a channel of
    width $W=8D$. (a) Vertical separation distance. (b) Horizontal
    separation distance. (c) Angular velocity $\omega$ (positive $=$
    counter-clockwise).}
  \label{fig:cl3} 
\end{figure}

Upon increasing the Reynolds number to $\Reynolds=14$ the dynamics
become much more complex.  The trailing particle catches up with and
subsequently passes the leading particle at time $t\approx
11~\units{s}$, breaking the left-right symmetry.  This tumbling behavior
is evident in Figure~\ref{fig:cl3}(a) where the vertical separation
distance becomes negative.  After the first tumble, the particles
separate horizontally and move to locations symmetrically opposite each
other relative to the centerline and separated by a horizontal distance
of roughly $0.27~\units{cm}$.  They continue to fall vertically at
approximately the same $x$ locations, and for the next 20 seconds they
exchange leading and trailing positions via a small periodic variation
in the vertical velocity whose amplitude decreases in time.  By time
$t\approx 35~\units{s}$, the particles have essentially reached a steady
state in which they are falling at constant velocity and maintaining a
constant separation distance (refer to Figure~\ref{fig:cl3}(b)).

The angular velocity plot in Figure~\ref{fig:cl3}(c) shows that both
particles experience a significant rotation during the tumbling phase
that is several orders of magnitude larger than the small ``wobbling''
motion observed in the $\Reynolds=3$ case.  In fact, the growth of this
rotational motion appears to be connected with the breaking of the
horizontal symmetry that initiates the tumbling motion.  By time
$t=15~\units{s}$, the rotational motion has subsided.

Because of the symmetry in both the initial conditions and the governing
equations, one would expect that the numerical solution should remain
symmetric for all time, regardless of Reynolds number.  The most likely
source of asymmetry that initiates the tumbling behavior observed in the
higher $\Reynolds$ simulation is numerical error -- these errors are
sufficiently damped out when $\Reynolds=3$, but remain large enough to
initiate tumbling at $\Reynolds=14$.  This conjecture is borne out by
the simulations in section~\ref{sec:off_vert}, where two particles are
aligned vertically and released off-center.  We have nonetheless shown
the results for this symmetric initial condition since it is commonly
studied in other simulations~\cite{feng-hu-joseph-1994}.

\begin{figure}[htbp]
  \centering
  (a) Vertical separation distance\\
  \includegraphics[width=0.95\columnwidth]{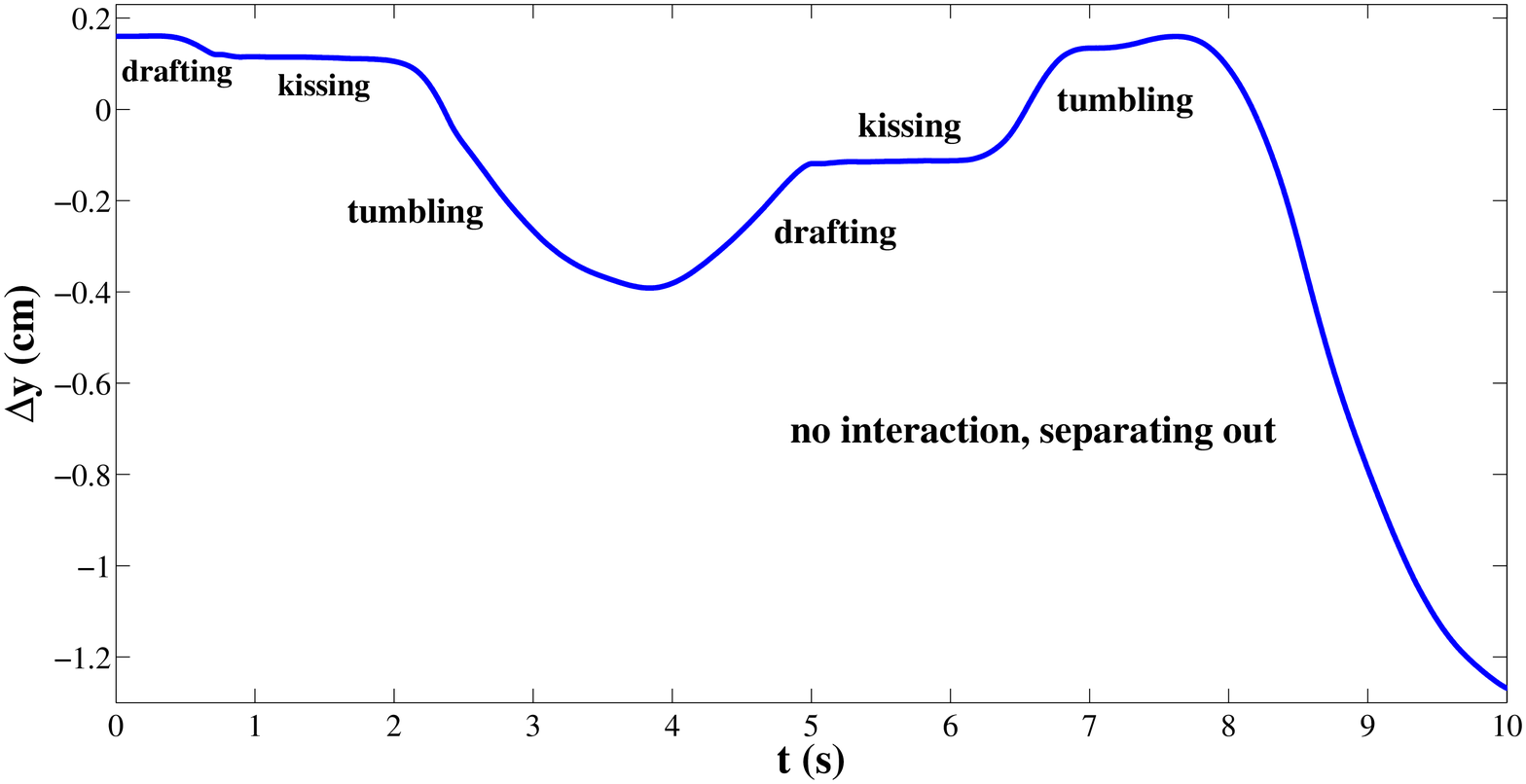}\\
  \bigskip  (b) Angular velocities\\
  \includegraphics[width=0.95\columnwidth]{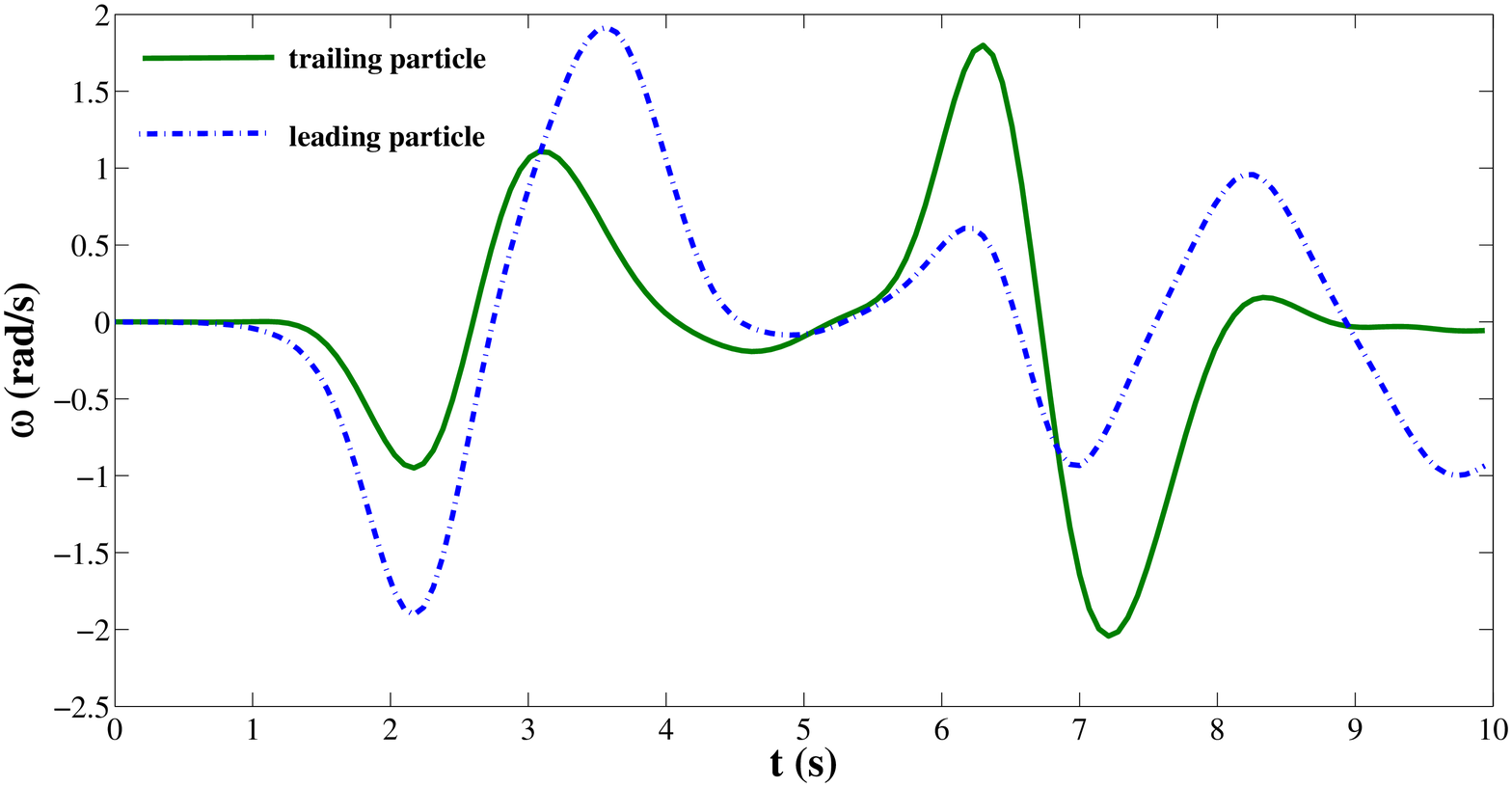}
  \caption{Settling dynamics of two particles at $\Reynolds=80$
    initially aligned vertically along the centerline in a channel of
    width $W=8D$. (a) Vertical separation distance. (b) Angular velocity
    $\omega$ (positive $=$ counter-clockwise).}
  \label{fig:cl_80}
\end{figure}

\begin{figure}[htbp]
  \centering
  \begin{tabular}{ccc}
    \includegraphics[trim=6.0in 0cm 5.0in 0cm,clip=true,width=0.28\columnwidth]{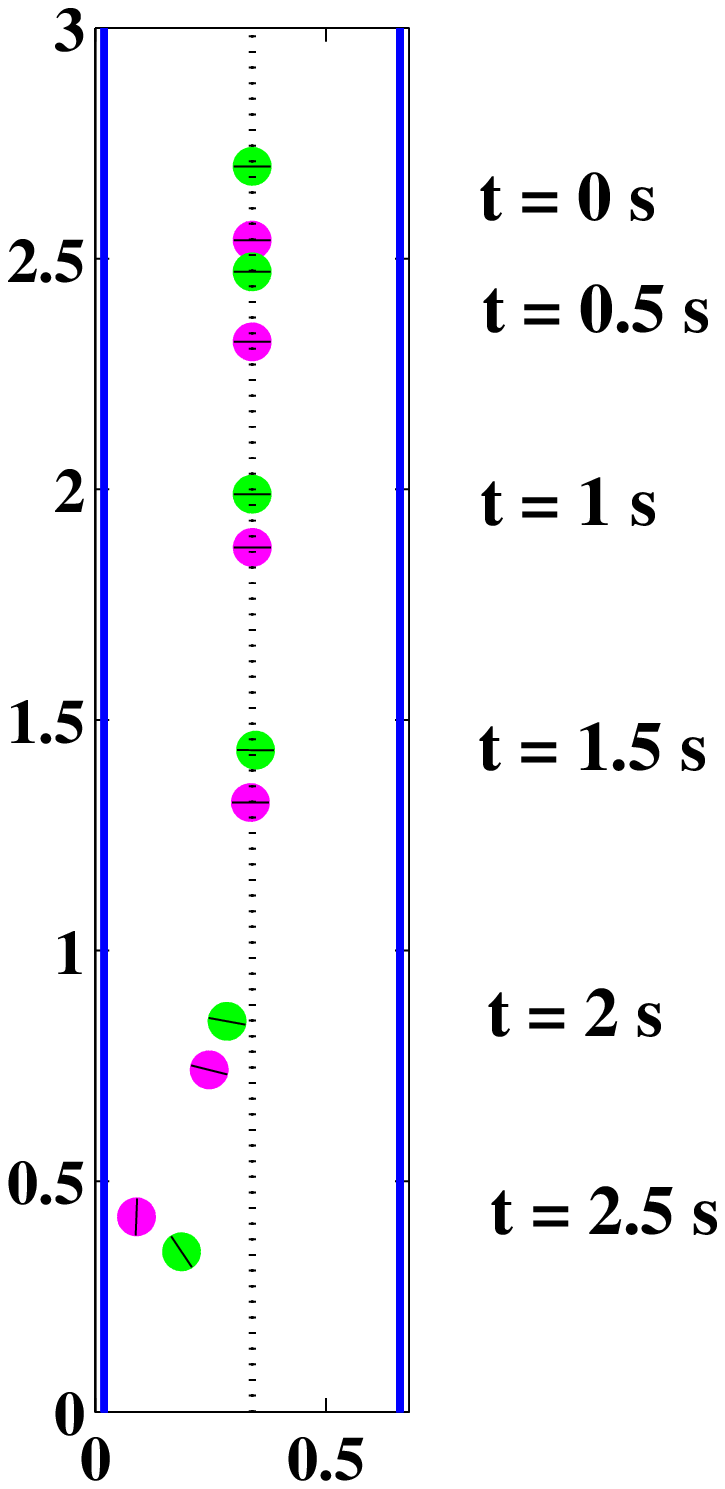}
    & \quad &
    \includegraphics[trim=6.0in 0cm 5.0in 0cm,clip=true,width=0.28\columnwidth]{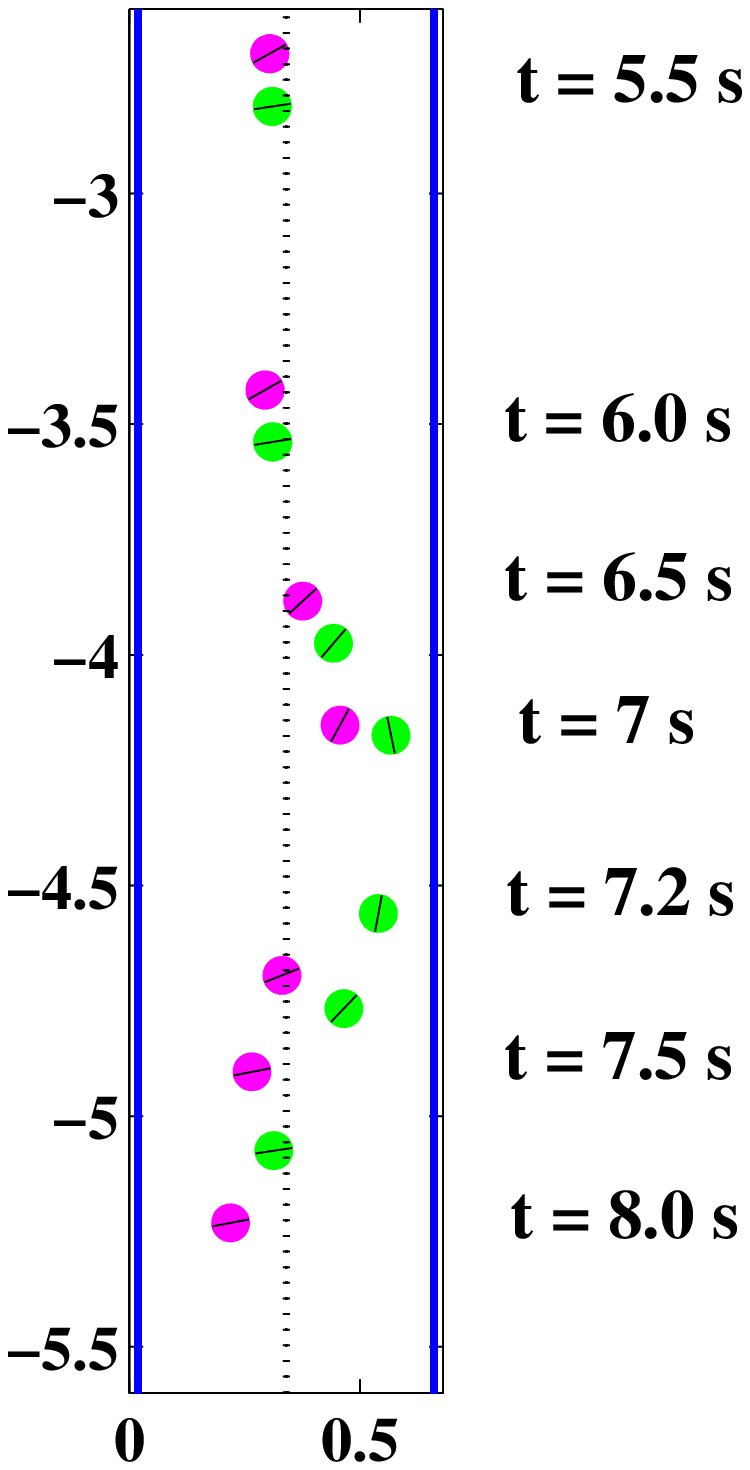}\\
    \includegraphics[trim=6.0in 0cm 5.0in 0cm,clip=true,width=0.28\columnwidth]{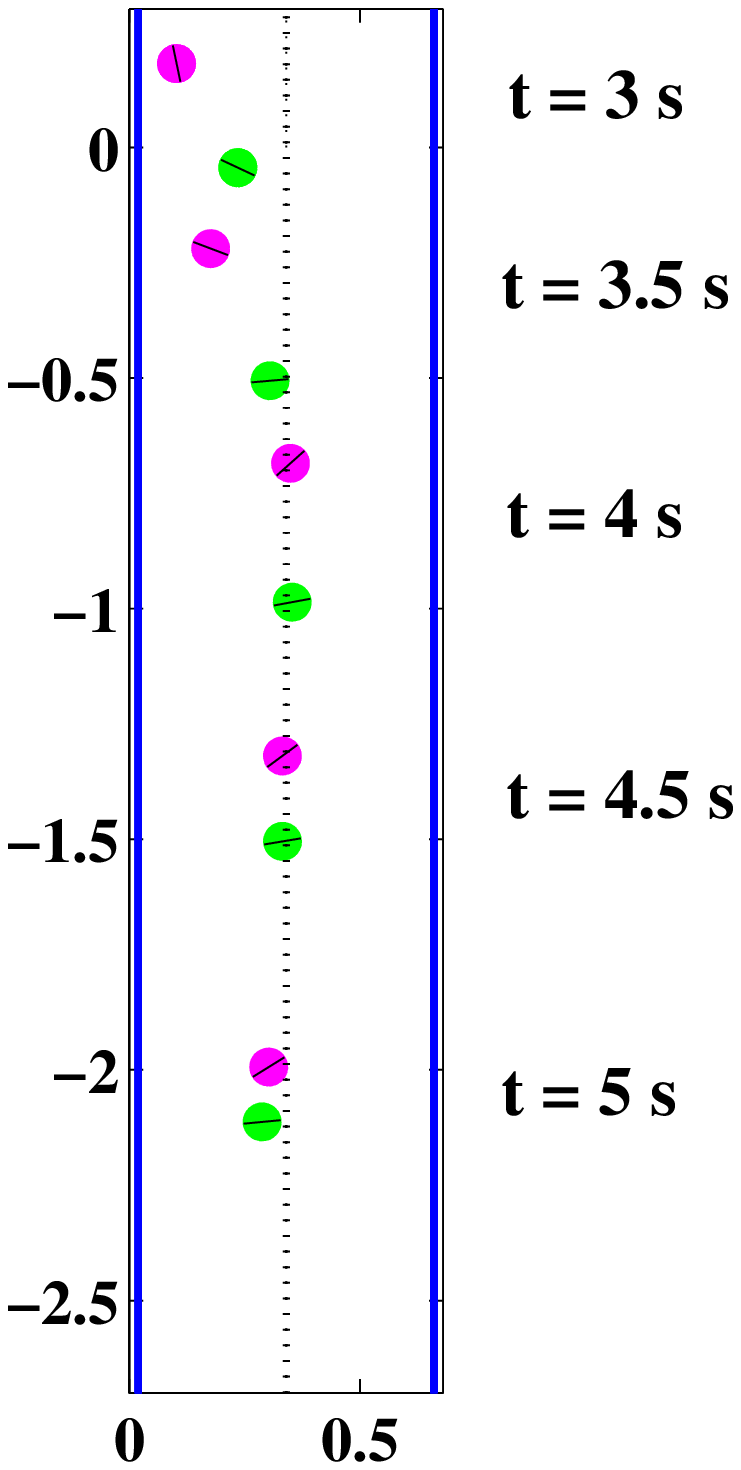}
    & &
    \includegraphics[trim=6.0in 0cm 5.0in 0cm,clip=true,width=0.28\columnwidth]{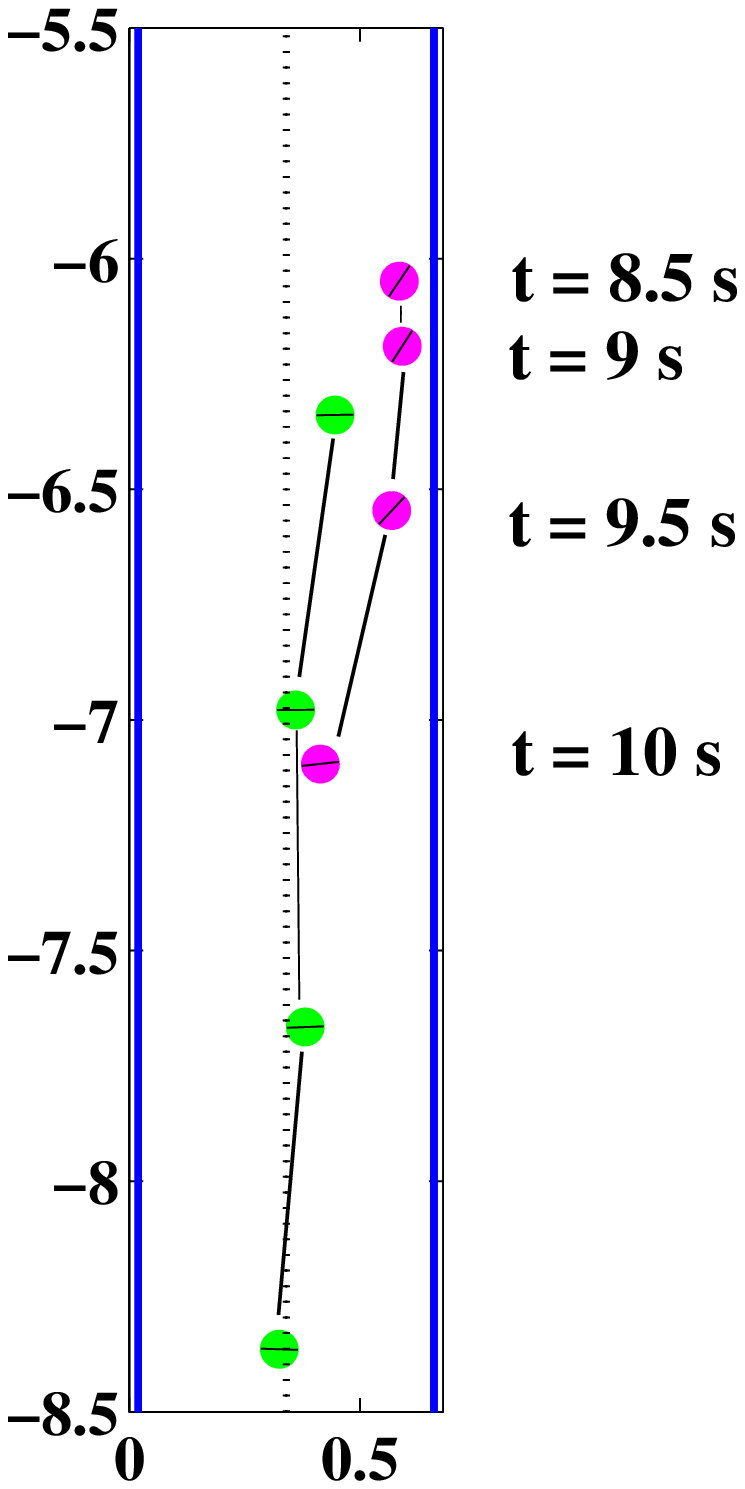}
  \end{tabular}
  \caption{Snaphots of particle interactions in the channel of width
    $W=8D$ with $\Reynolds=80$, for the case when the particles are
    initially aligned vertically along the channel centerline.}
  \label{fig:cl5_snap}
\end{figure}

As the Reynolds number is increased yet further to $\Reynolds=80$, we
observe in Figure~\ref{fig:cl_80} another qualitative change in solution
behavior that is most easily seen in the sequence of snapshots collected
in Figure~\ref{fig:cl5_snap}.  The two particles begin with a DKT
exchange such as that observed for $\Reynolds=14$, however this occurs
as the two particles drift together toward the left channel wall
(instead of toward the channel centerline).  Following that, the
particles drift reverse direction toward the right wall and undergo a
second DKT exchange, after which the particle that was initially
trailing ends up in the lead.  These two DKT exchanges are accompanied
by a back-and-forth rotational motion of each particle that has an
amplitude similar in size to the $\Reynolds=14$ case (refer to
Figure~\ref{fig:cl_80}(b)).  Once again, it appears to be growth in the
small ``wobble'' in the particle angular velocity plot that initiates
tumbling.  We also observe that when a particle nears the left wall it
experiences a clockwise rotation, while the direction of rotation is
reversed near the right wall -- this is consistent with physical
intuition, which suggests that wall drag arising from the no-slip
condition at the channel wall should cause a rolling-type motion as the
portion of the particle closest to the wall slows down.

FHJ~\cite{feng-hu-joseph-1994} have performed a similar computation at
$\Reynolds=70$ (for the same symmetric initial conditions) that exhibits
results consistent with ours up to time $t\approx 5.4~\units{s}$.
However, they terminate their computation at this point and there is no
indication in their paper of the subsequent dynamics.  We have computed
beyond this time and find that after the second tumble, the particles
separate vertically as they migrate toward the center of the channel.
By time $t=10~\units{s}$ they settle at roughly the same speed and no
longer interact in any significant way. Our computations also exhibit
much the same qualitative behavior as the experiments reported in
\cite{fortes-joseph-lundgren-1987,joseph-etal-1987} and numerical
simulations from \cite{glowinski-etal-2001,uhlmann-2005}.

\begin{figure}[htbp]
  \centering
  \begin{tabular}{cc}
    Drafting/kissing phase & Tumbling phase\\
    \includegraphics[trim=3.5in 0cm 3.5in 0cm,clip=true,width=0.45\columnwidth]{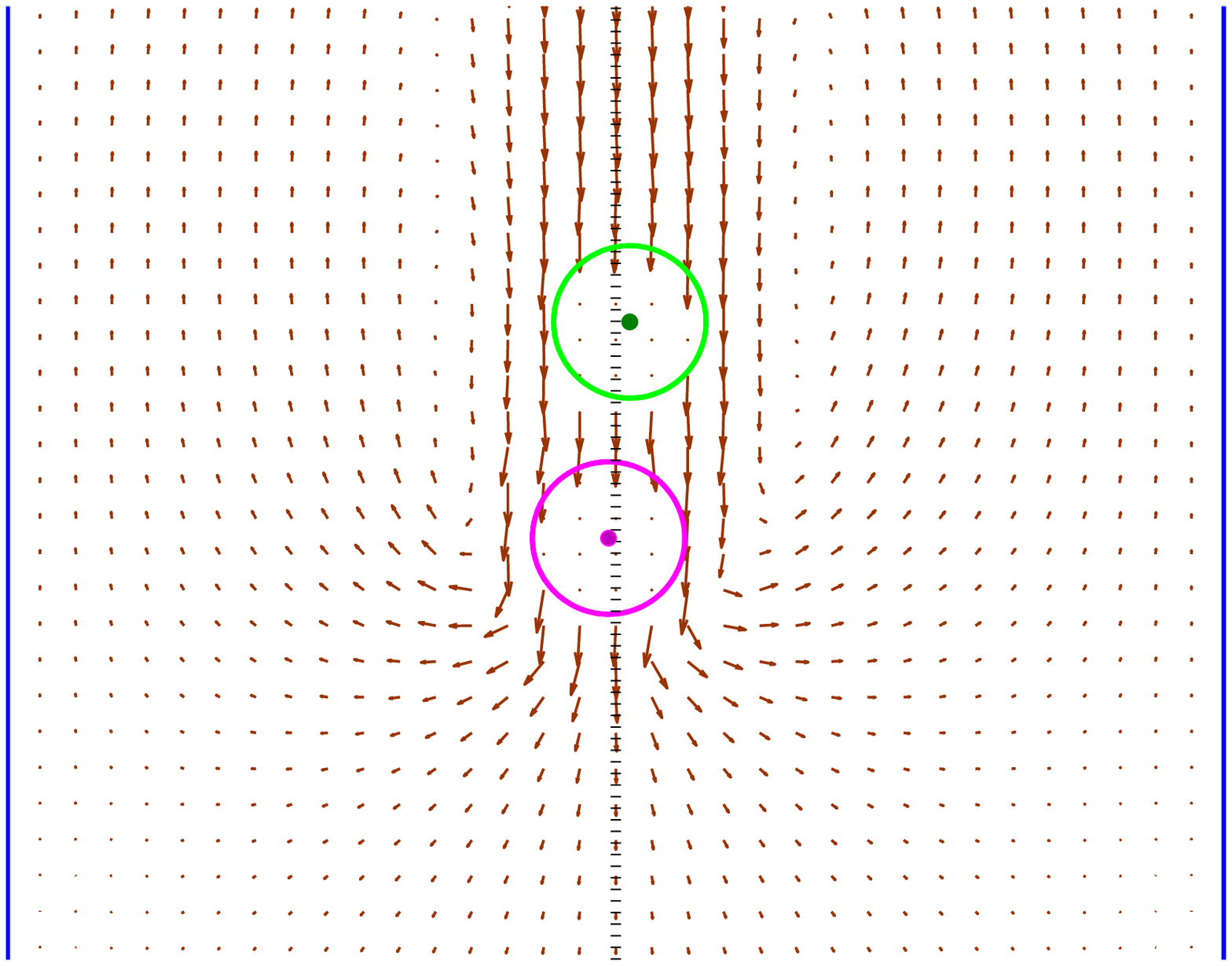}
    &
    \includegraphics[trim=3.5in 0cm 3.5in 0cm,clip=true,width=0.45\columnwidth]{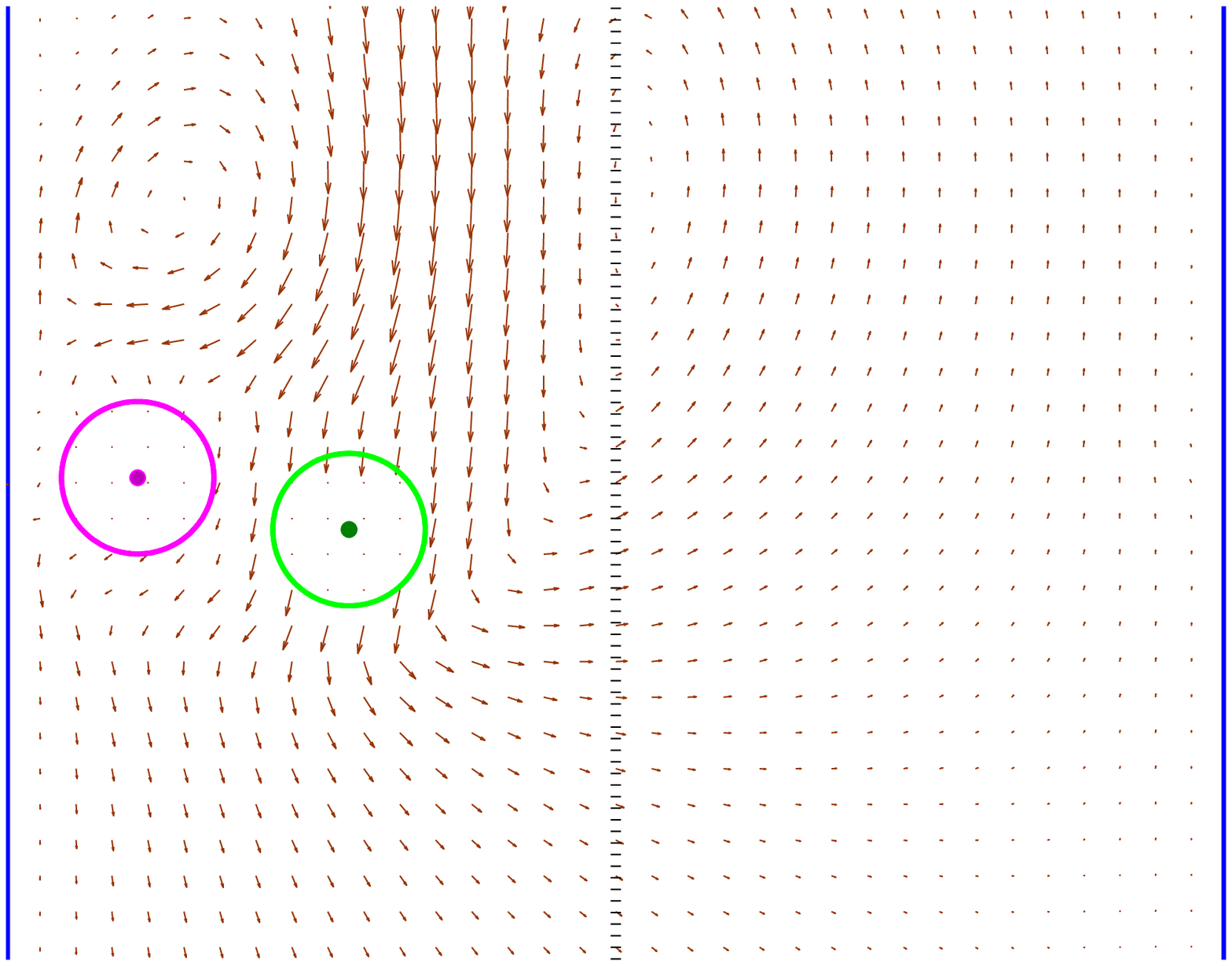}
  \end{tabular}
  \caption{Velocity vector plots for the drafting/kissing (left) and tumbling
    phase (right) for the case when the particles are initially aligned
    vertically along the channel centerline.  Parameters: $W=8D$ and
    $\Reynolds=80$.}
  \label{fig:cl5_vectors}
\end{figure}

\subsection{Two vertically-aligned particles, released off-center}
\label{sec:off_vert}

In this section, we consider an asymmetric initial layout where the two
particles are aligned vertically (again separated by a distance $2D$)
but instead have their centers of mass displaced to the left of center
at location $x=W/4$.  This initial geometry is depicted in
Figure~\ref{fig:init_profile}(b).

\begin{figure}[htbp]
  \centering
  \includegraphics[width=0.95\columnwidth]{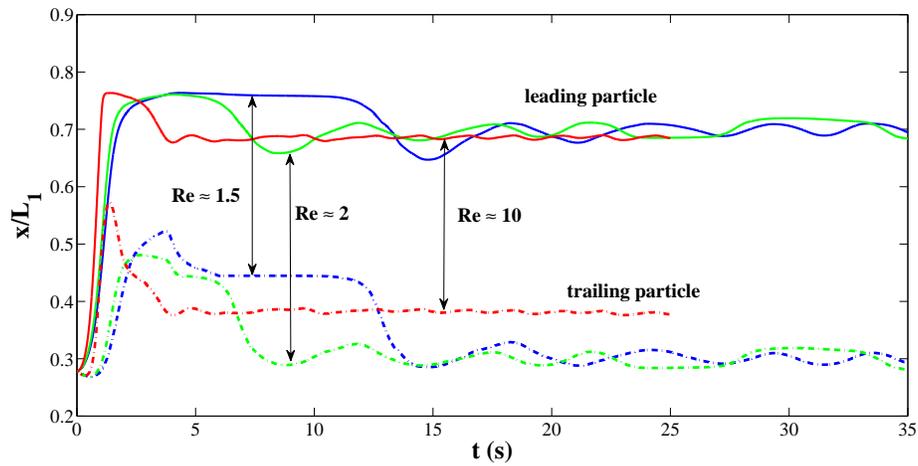}\\
  \caption{Horizontal particle locations for two particles initially
    aligned vertically but off-center.  Parameter values: $W=4D$ and
    $\Reynolds=1.5$, 2, 10.}
  \label{4d_all_hori}
\end{figure}

\begin{figure}[htbp]
  \centering
  \begin{tabular}{ccc}
    \includegraphics[trim=6.0in 0cm 5.0in 0cm,clip=true,width=0.30\columnwidth]{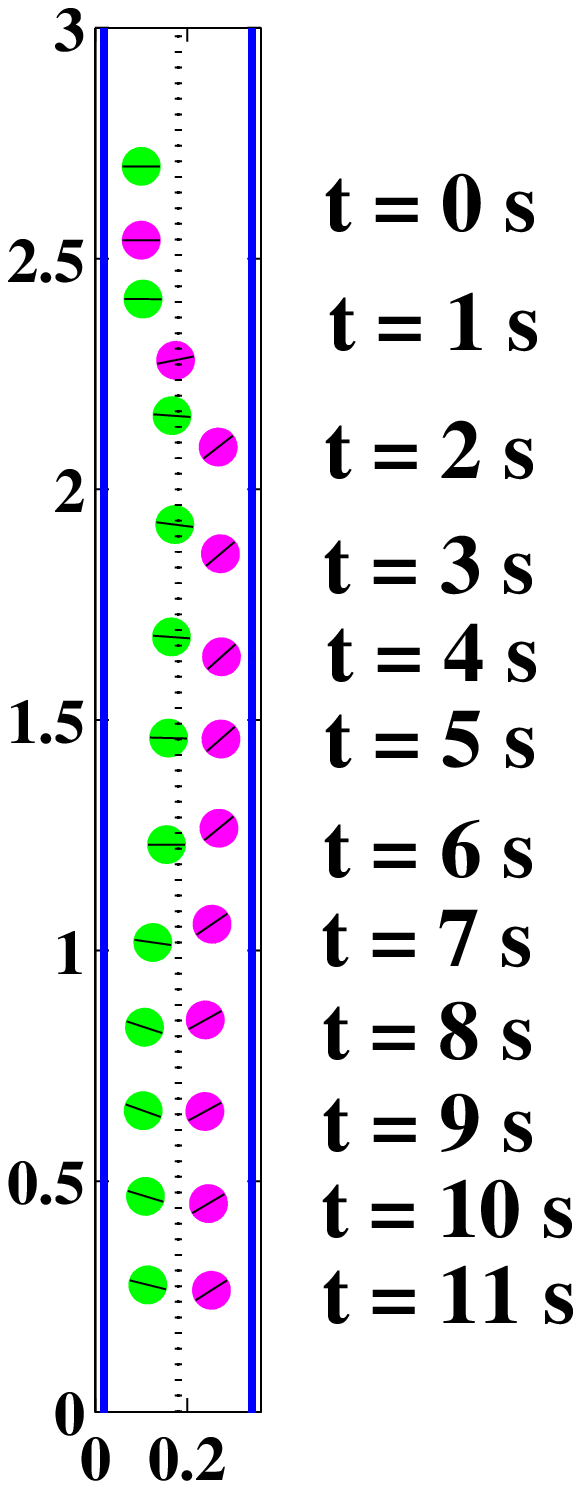}
    & \quad &
    \includegraphics[trim=6.0in 0cm 5.0in 0cm,clip=true,width=0.30\columnwidth]{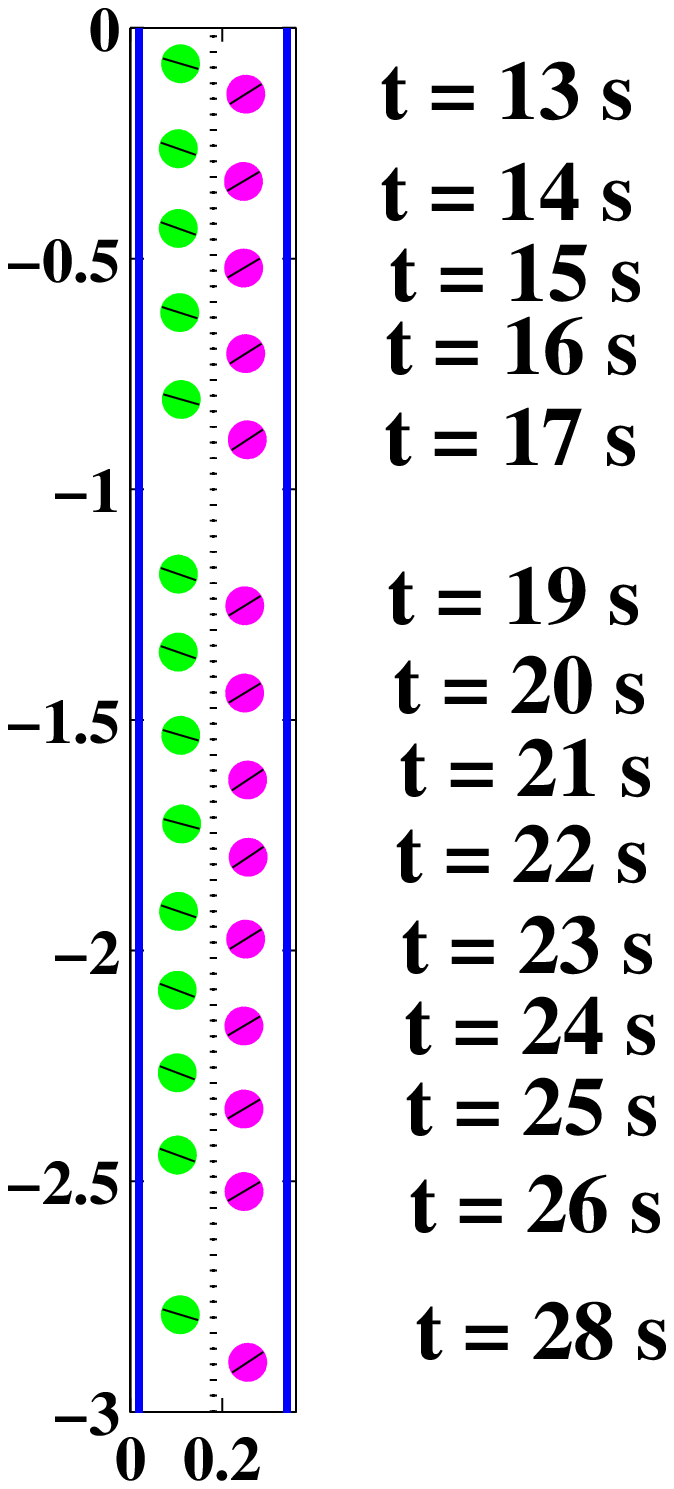}
  \end{tabular}
  \caption{Snapshots of particle interactions for the case when the
    particles are initially aligned vertically, but off-center.
    Parameter values: $W=4D$ and $\Reynolds=2$.}
  \label{4d_re2}
\end{figure}

Results are first reported for a channel of width $W=4D$ and four values
of Reynolds number: $\Reynolds=1.5$, 2, 10 and 47.  For the three
smallest values of Reynolds number, we observe the behavior pictured in
Figures~\ref{4d_all_hori} and~\ref{4d_re2}.  Initially, both particles
drift toward the right, with the trailing particle moving toward the
centerline while the leading particle moves to a point roughly mid-way
between the centerline and the right wall.  At the same time, the
trailing particle speeds up in the wake of the leading particle so that
they approach the same height.  After this initial realignment, the two
lowest Reynolds numbers ($\Reynolds=1.5$, 2) undergo another slight
adjustment in the horizontal locations so that the two particles are
located symmetrically about the centerline; interestingly, the two
particles in the $\Reynolds=10$ case remain in a slightly asymmetric
layout relative to the centerline.  After that, the particles fall with
roughly constant speed and without changing $x$-locations in the
channel.  The snapshots in Figure~\ref{4d_re2} for the $\Reynolds=2$
case show that the particles do not enter either kissing or tumbling
phases.

\begin{figure}[htbp]
  \centering
  \includegraphics[width=0.95\columnwidth]{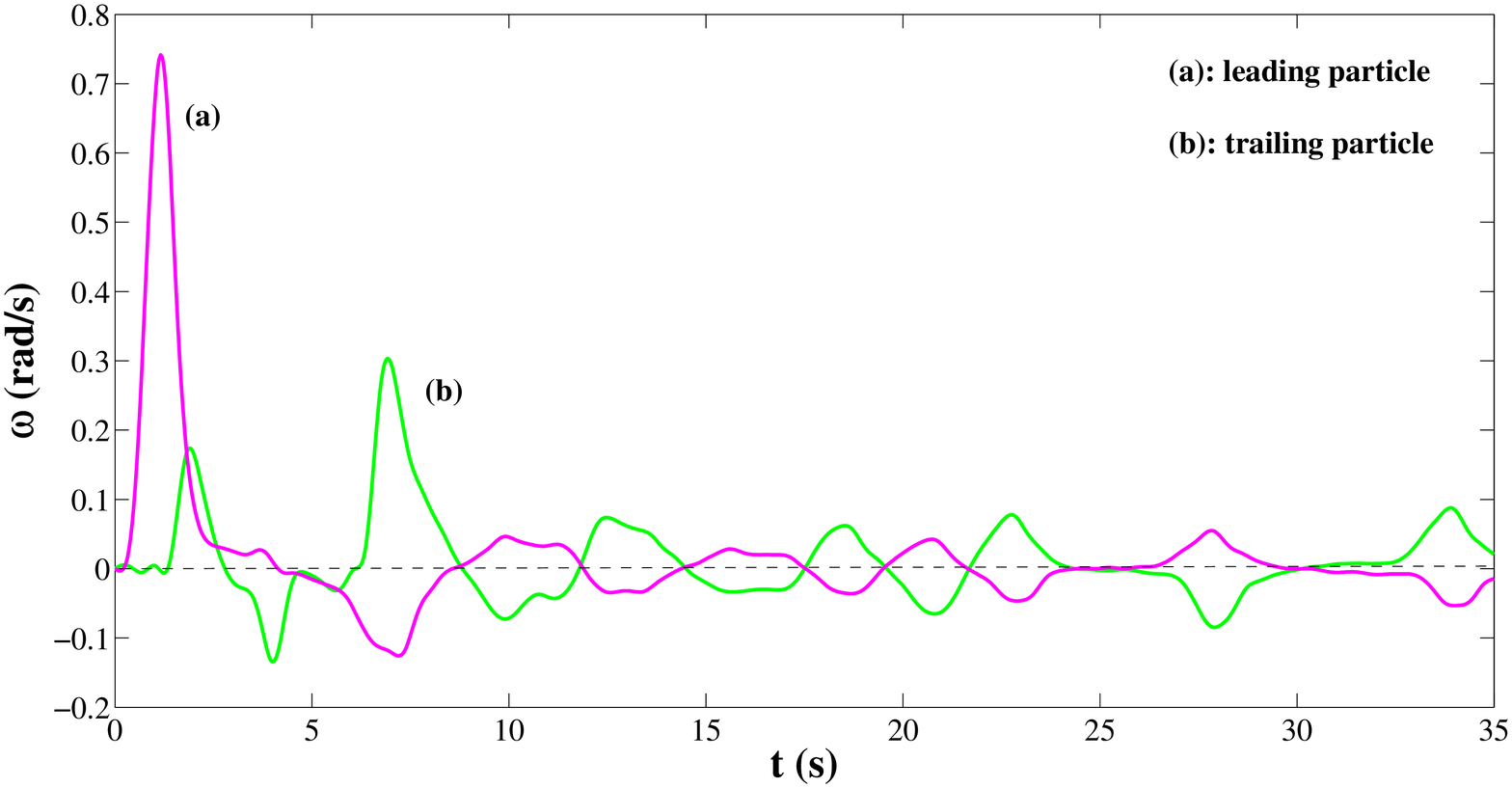}
  \caption{Angular velocity $\omega$ for two particles initially aligned
    vertically and off-center. Parameter values: $W=4D$ and
    $\Reynolds=2$.}
  \label{omega_re_2}
\end{figure}

Both particles experience a distinct rotational motion as shown in
Figure~\ref{omega_re_2} for $\Reynolds=2$, but the magnitude of the
angular velocity is not as large as was observed during the tumbling
phase for the simulations in the previous section.  Corresponding
results for a channel of width $W=8D$ do not show any significant
difference in qualitative behavior.
  
\begin{figure}[htbp]
  \centering
  (a) Horizontal positions\\
  \includegraphics[width=0.95\columnwidth]{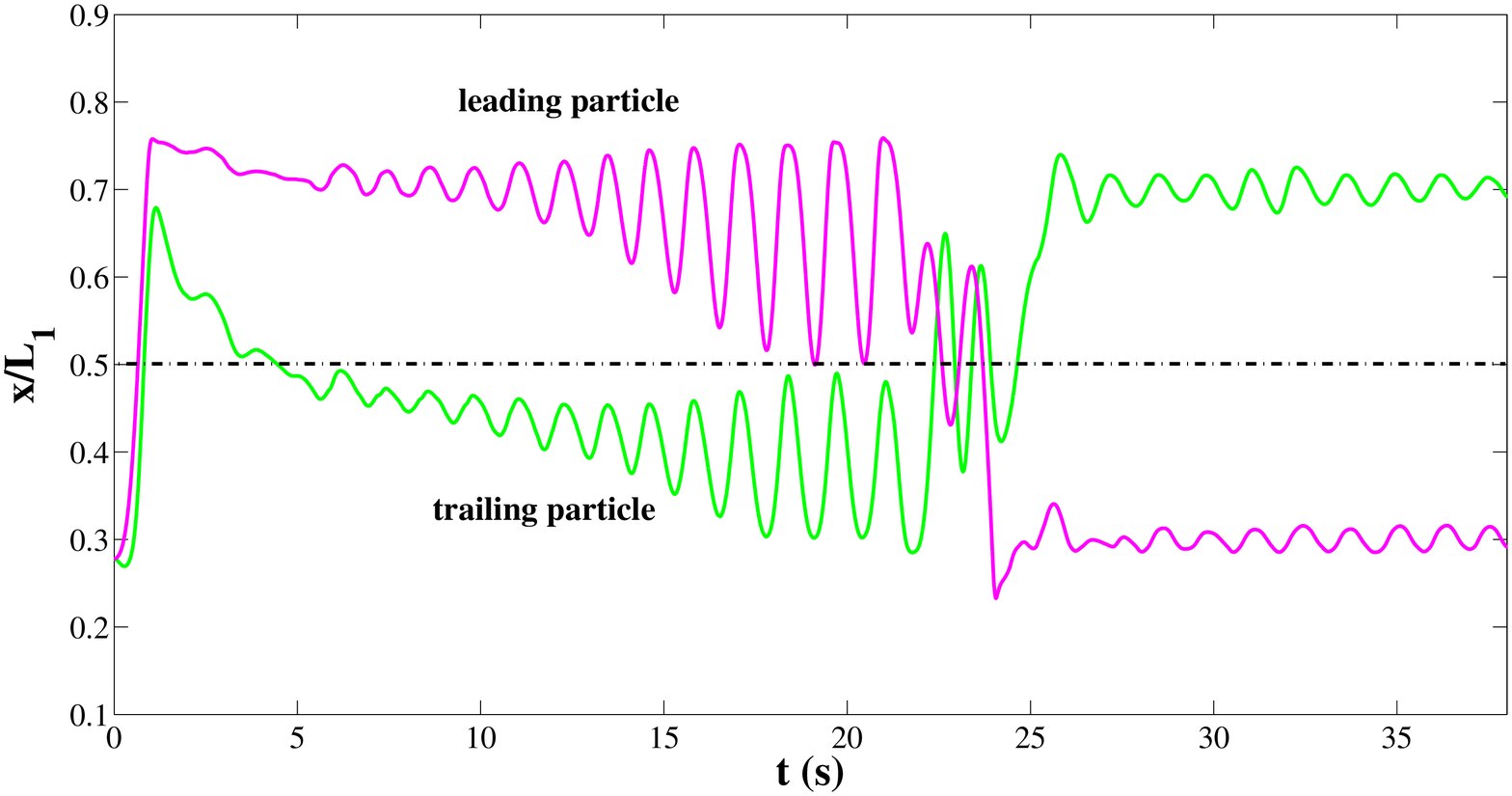}\\
  \bigskip  (b) Angular velocities\\
  \includegraphics[width=0.95\columnwidth]{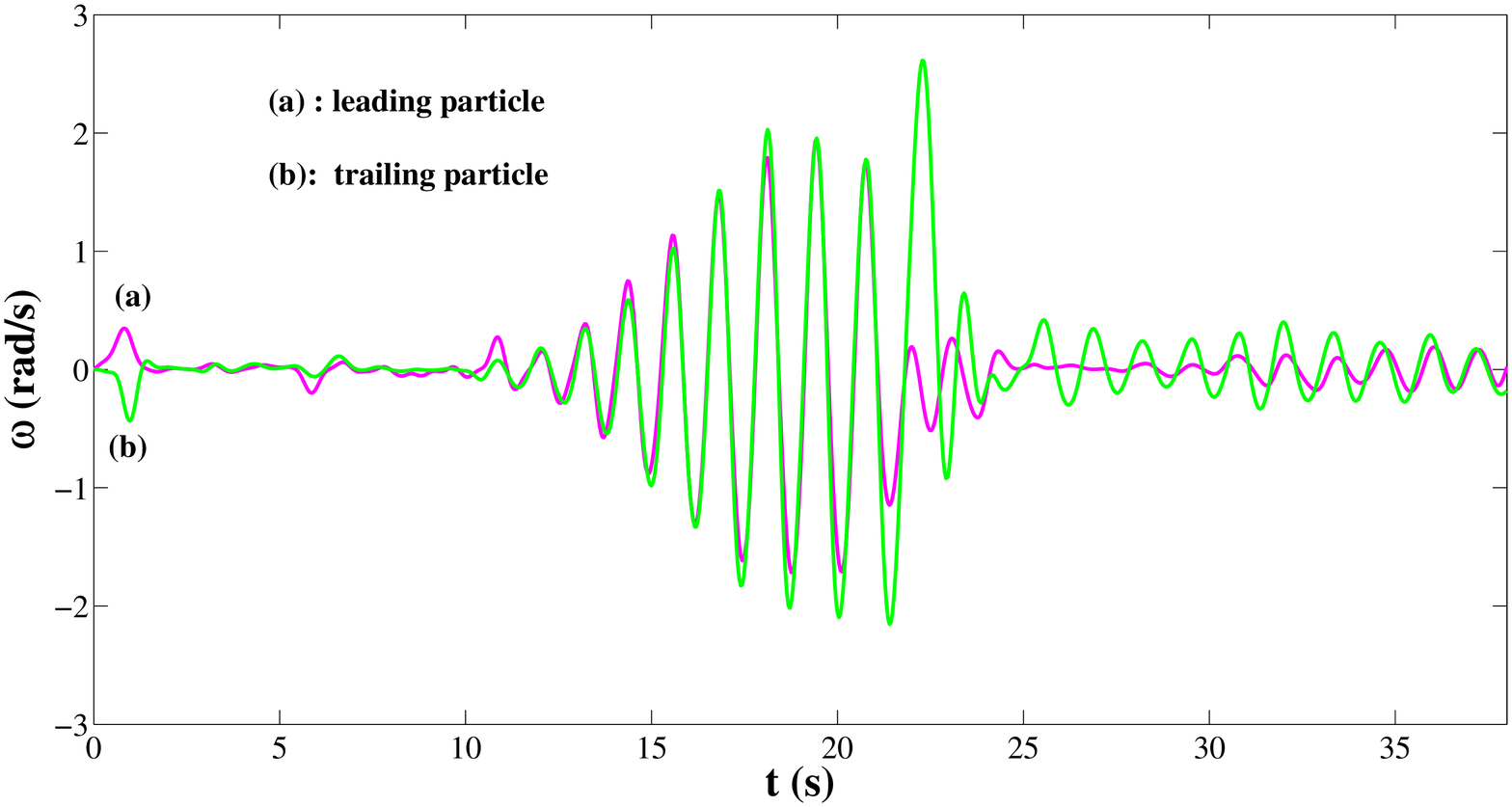}
  \caption{Settling dynamics of two particles initially aligned
    vertically and off-center, with parameters $W=4D$ and
    $\Reynolds=47$.  (a) Horizontal positions.  (b) Angular velocity
    $\omega$ (positive $=$ counter-clockwise).} 
  \label{4d_re47_hori}
\end{figure}

\begin{figure}[htbp]
  \centering
  \includegraphics[trim=6.0in 0cm 5in 0cm,clip=true,width=0.22\columnwidth]{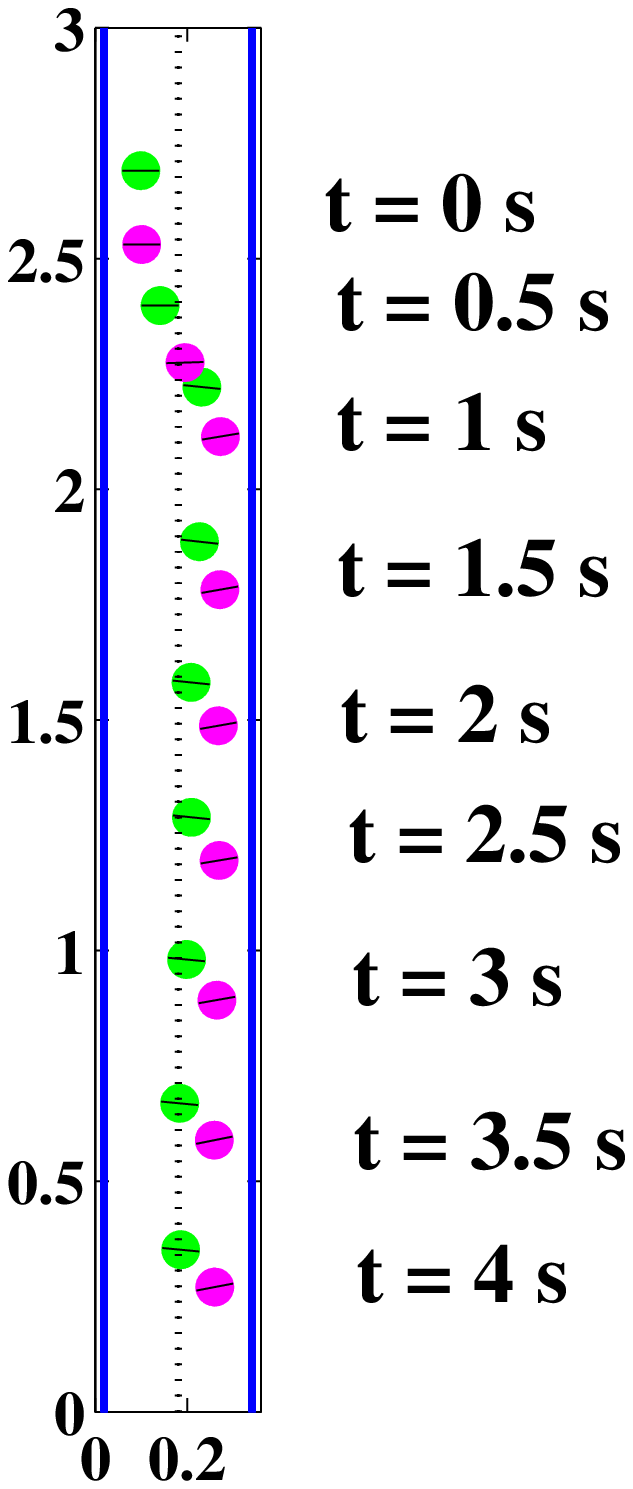}
  \includegraphics[trim=6.0in 0cm 5in 0cm,clip=true,width=0.22\columnwidth]{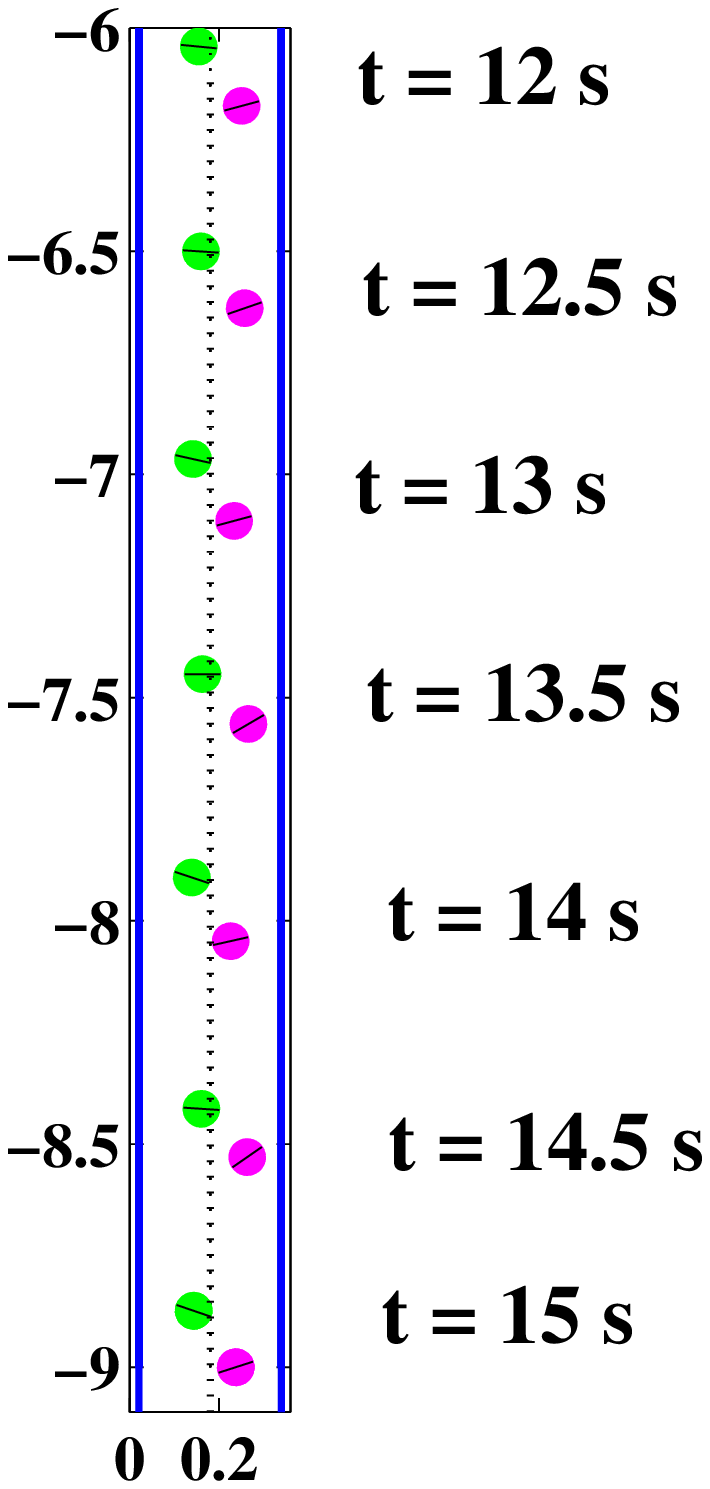}
  \includegraphics[trim=6.0in 0cm 5in 0cm,clip=true,width=0.22\columnwidth]{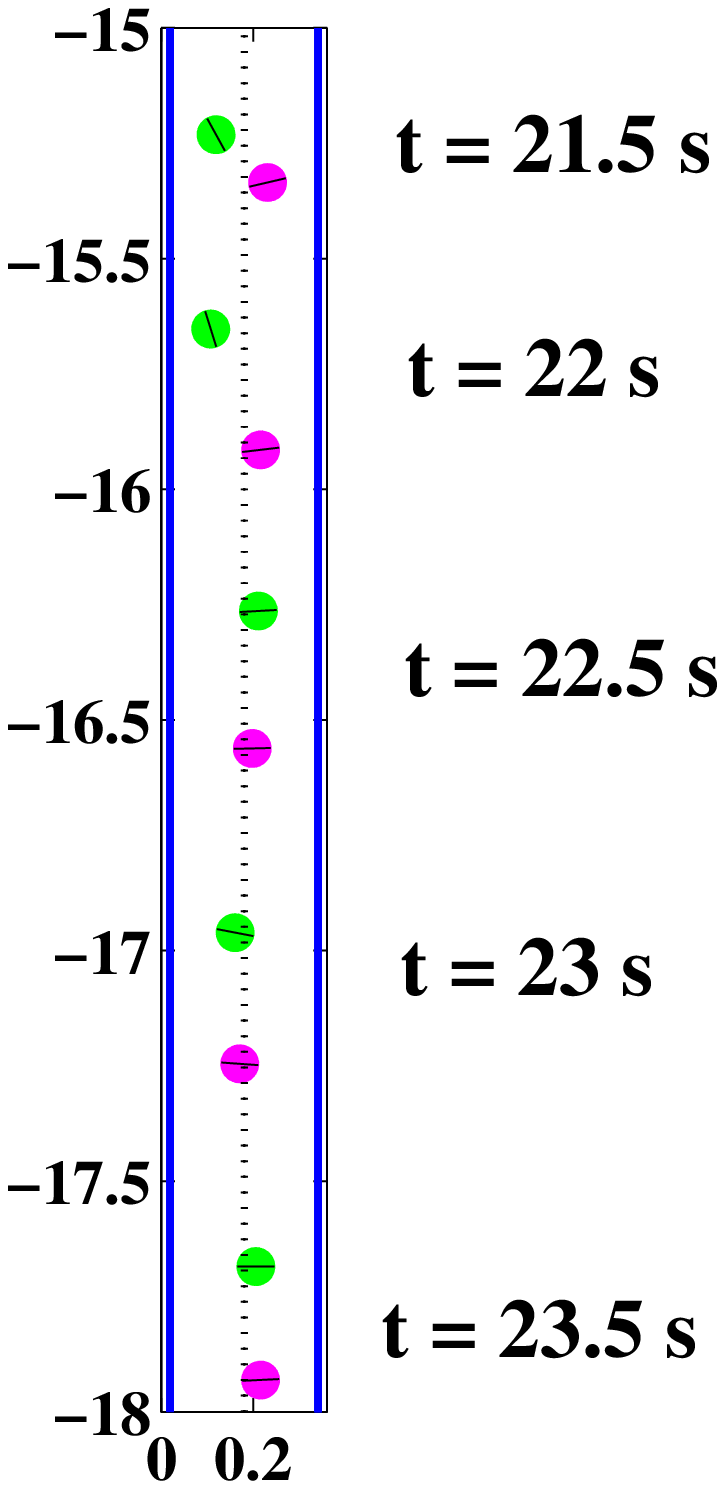}\\
  \includegraphics[trim=6.0in 0cm 5in 0cm,clip=true,width=0.22\columnwidth]{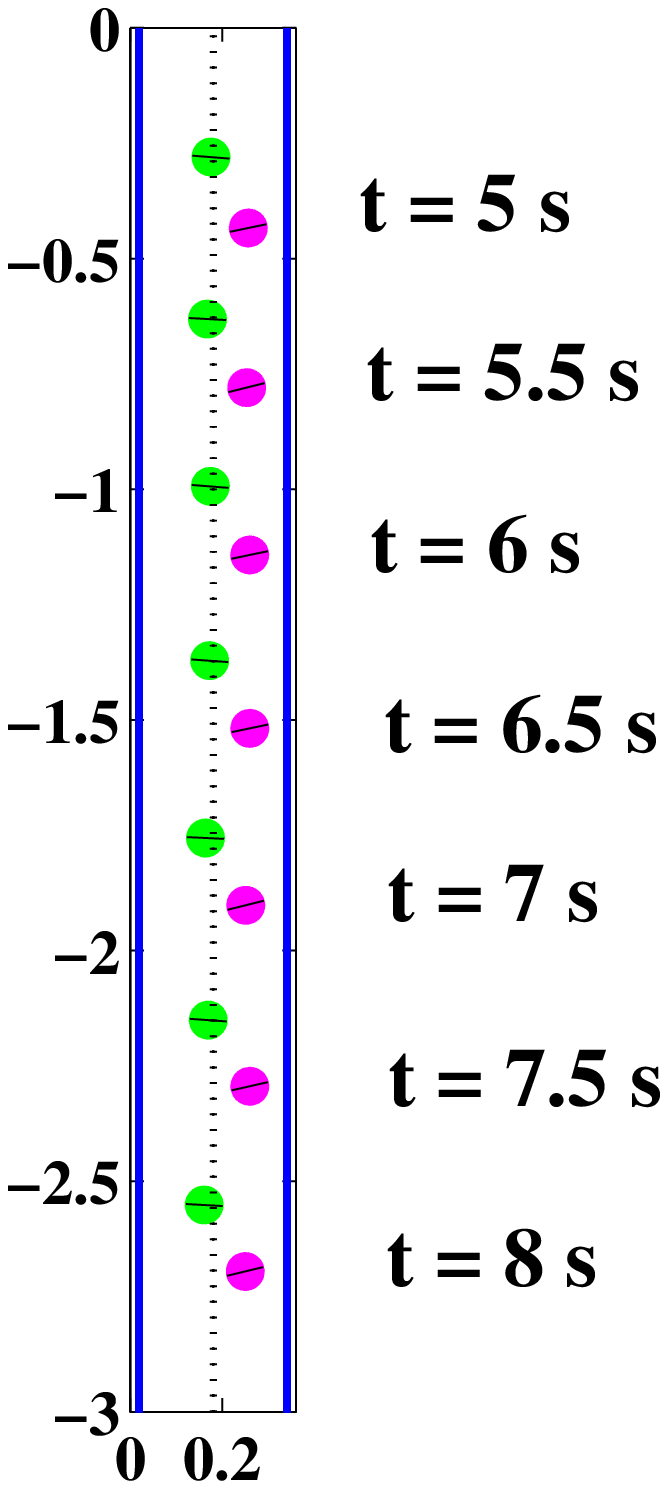}
  \includegraphics[trim=6.0in 0cm 5in 0cm,clip=true,width=0.22\columnwidth]{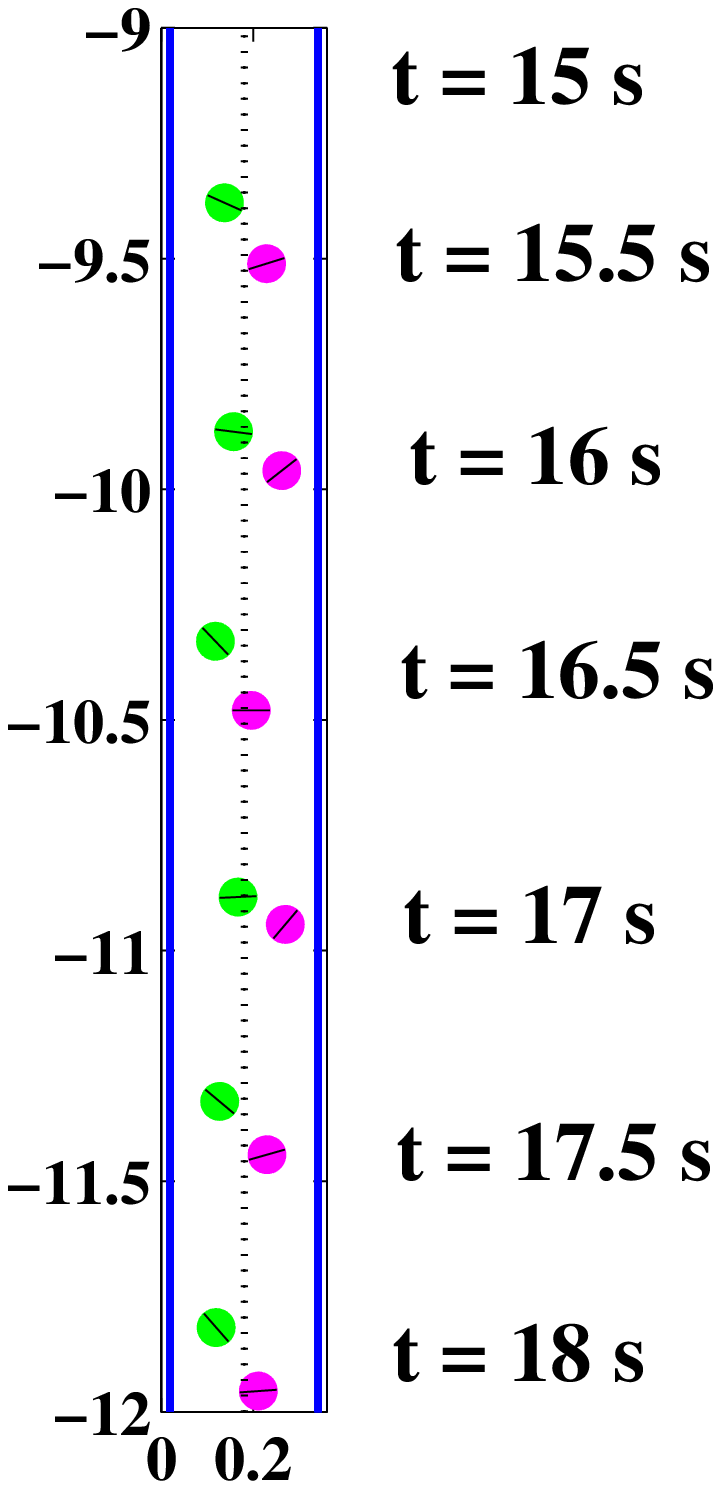}
  \includegraphics[trim=6.0in 0cm 5in 0cm,clip=true,width=0.22\columnwidth]{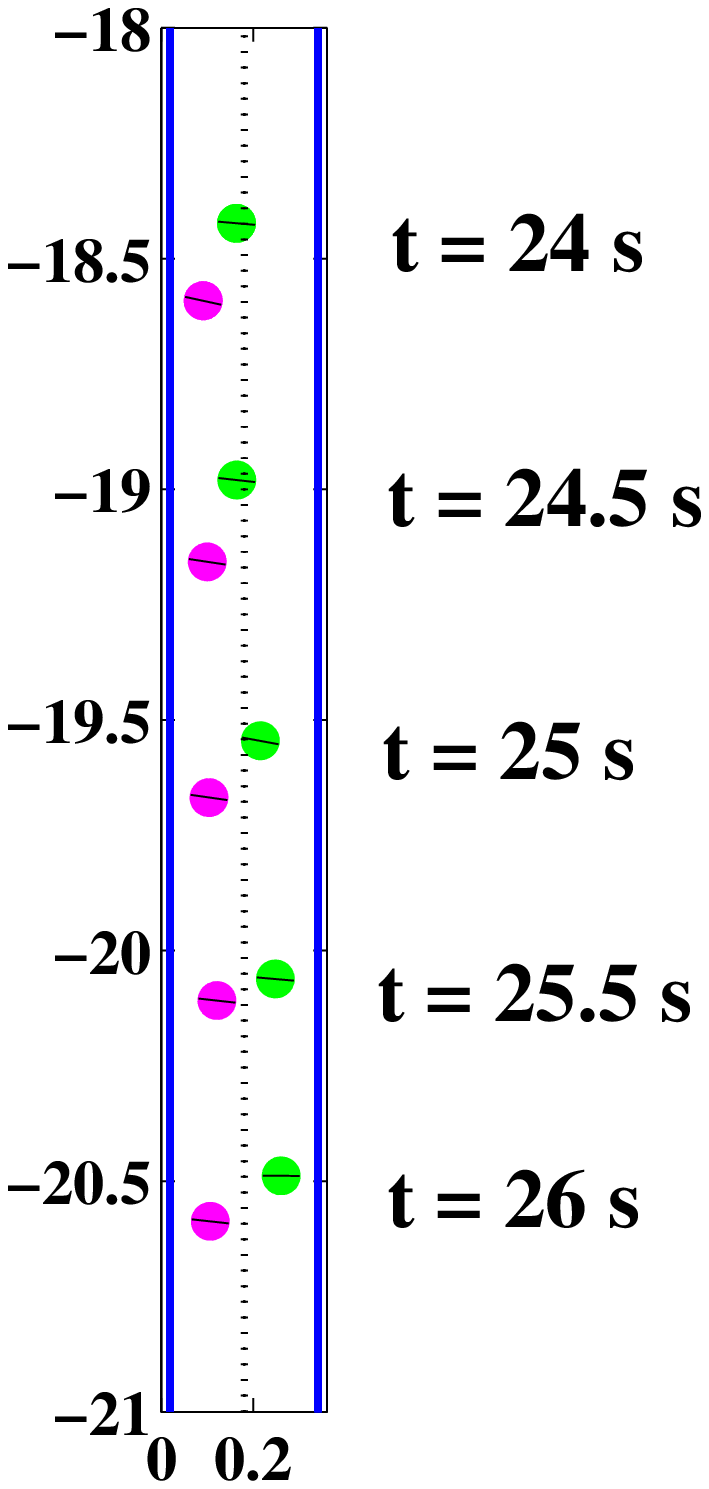}\\
  \includegraphics[trim=6.0in 0cm 5in 0cm,clip=true,width=0.22\columnwidth]{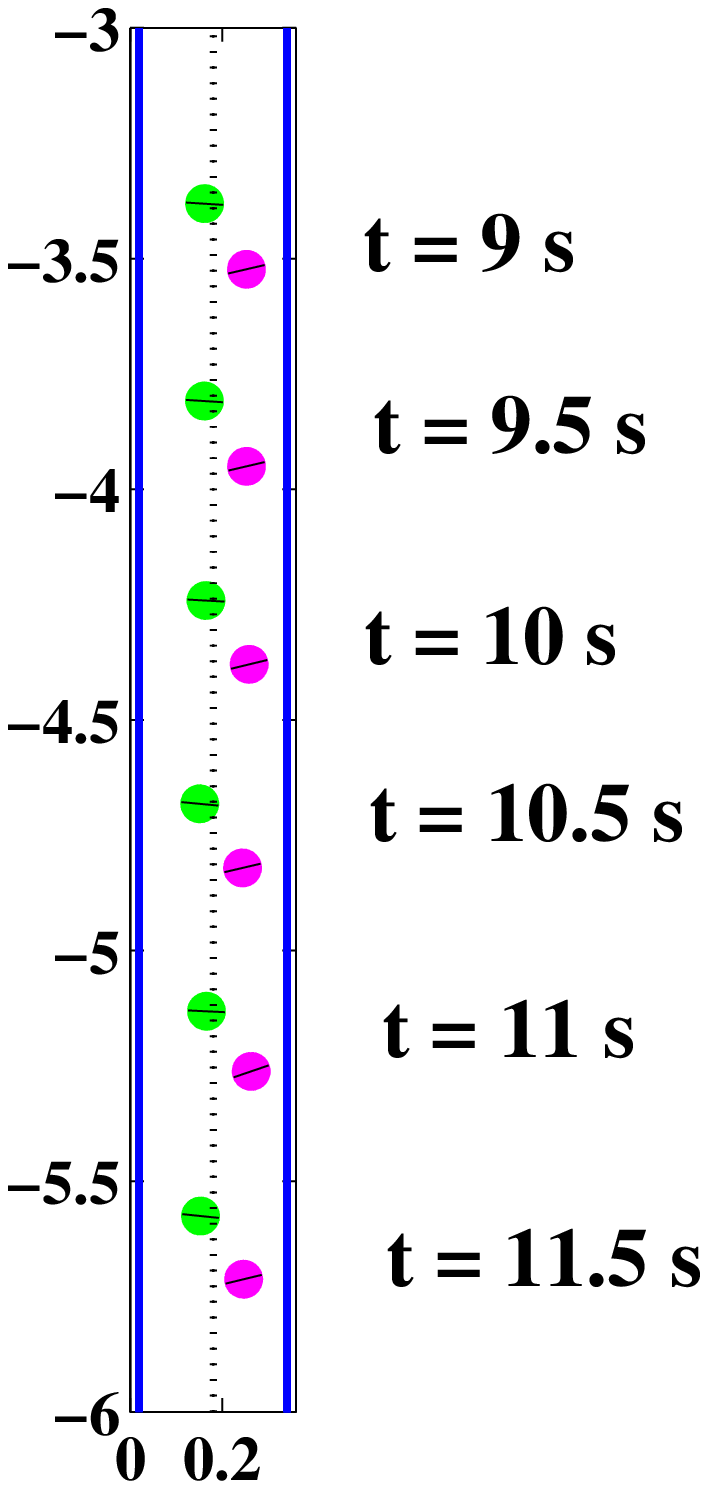}
  \includegraphics[trim=6.0in 0cm 5in 0cm,clip=true,width=0.22\columnwidth]{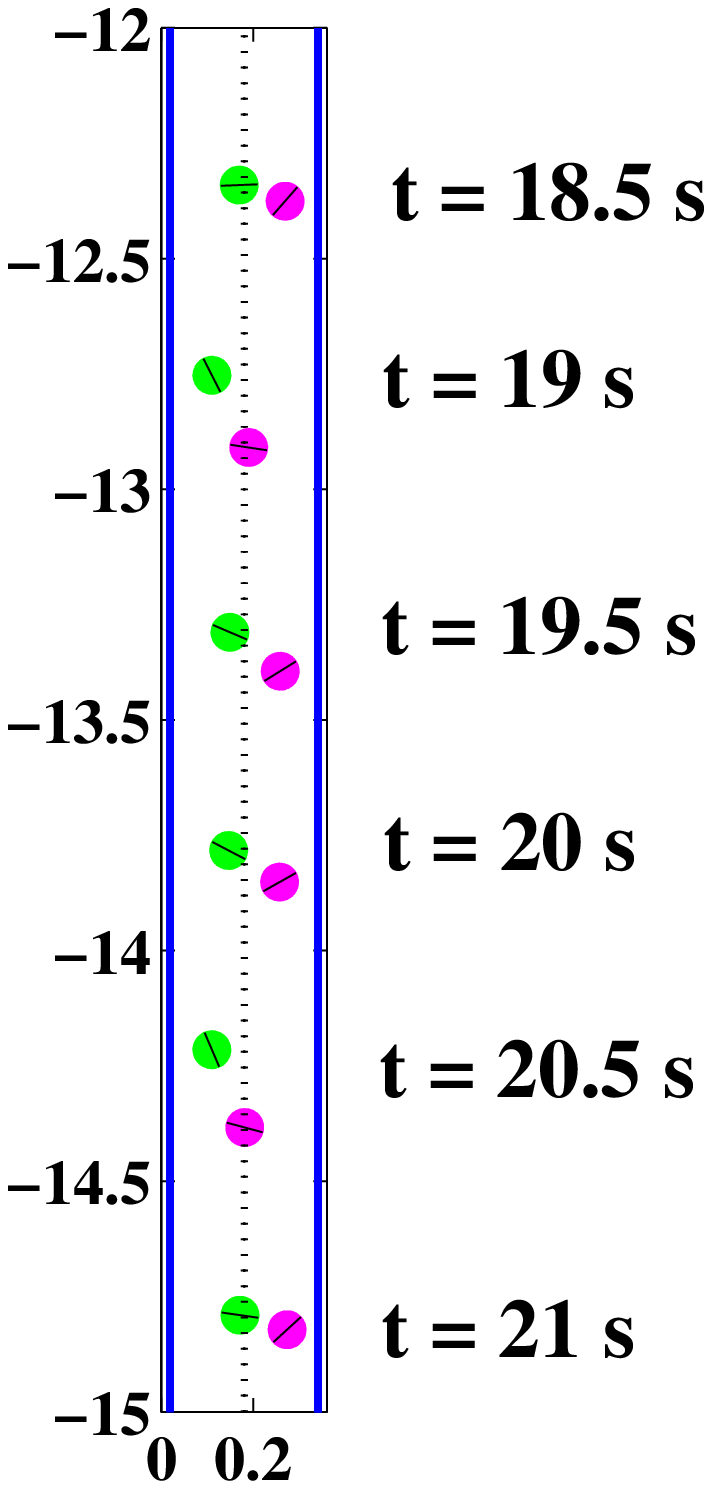}
  \includegraphics[trim=6.0in 0cm 5in 0cm,clip=true,width=0.22\columnwidth]{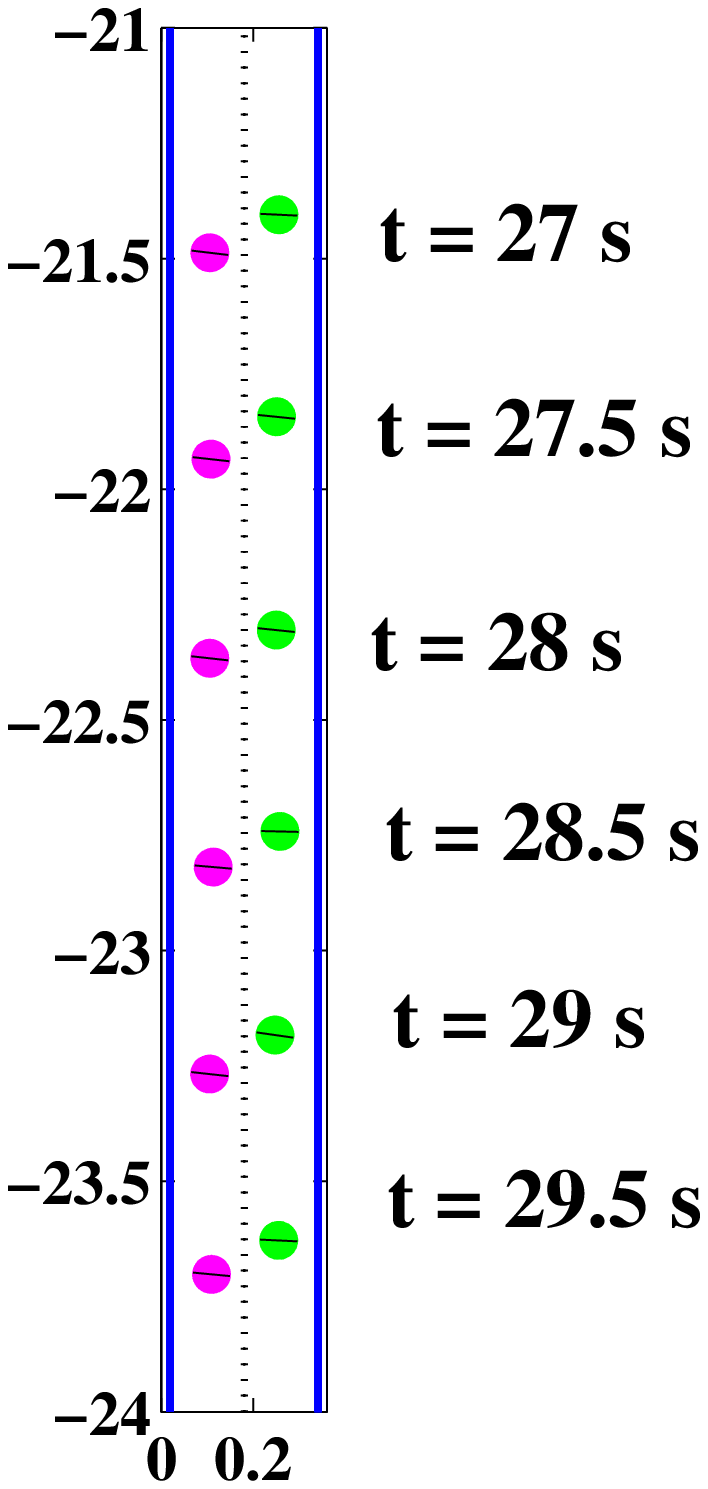}
  \caption{Snapshots of particle interactions for two particles
    initially aligned vertically, but off-center.  Read from top to
    bottom, then left to right.  Parameter values: $W=4D$ and
    $\Reynolds=47$.}
  \label{multi_4d_re_47}
\end{figure}

A very different behavior is observed for the highest value of Reynolds
number ($\Reynolds=47$) as seen in Figures~\ref{4d_re47_hori}
and~\ref{multi_4d_re_47}.  Up to time $t\approx 7 \;\units{s}$ the
dynamics are similar to the lower $\Reynolds$ cases in that the
particles migrate to the right with the leading particle approaching
closest to the wall.  However, at this time the particles separate
horizontally and a marked back-and-forth oscillation appears that grows
in magnitude between times $t\approx 7$ and $23~\units{s}$, until the
oscillating particles overlap with each other near the centerline.  At
$t\approx 23\;\units{s}$, the particles undergo a strong interaction in
which they swap horizontal locations and move to positions symmetrically
opposite to each other on either side of the channel.  At this stage,
they have reached a steady state with roughly constant vertical
velocity.  The growing horizontal oscillations in the time interval
$[7,23]$ are accompanied by a synchronized rotation of both particles
(in opposite directions) that grows rapidly and then also dies out after
$t\approx 23~\units{s}$ (refer to Figure~\ref{4d_re47_hori}(b)).

\subsection{Two horizontally-aligned particles}
\label{sec:hori}

We next simulate the motion of two particles initially aligned
horizontally in a channel of width $W=8D$.  We consider two sets of
initial conditions pictured in Figures~\ref{fig:init_profile}(c) and
(d), first where the particles are located symmetrically with respect to
the channel centerline, and the second an asymmetric arrangement that is
shifted to the left.

\begin{figure}[htbp]
  \centering
  \includegraphics[angle=270,width=0.80\columnwidth]{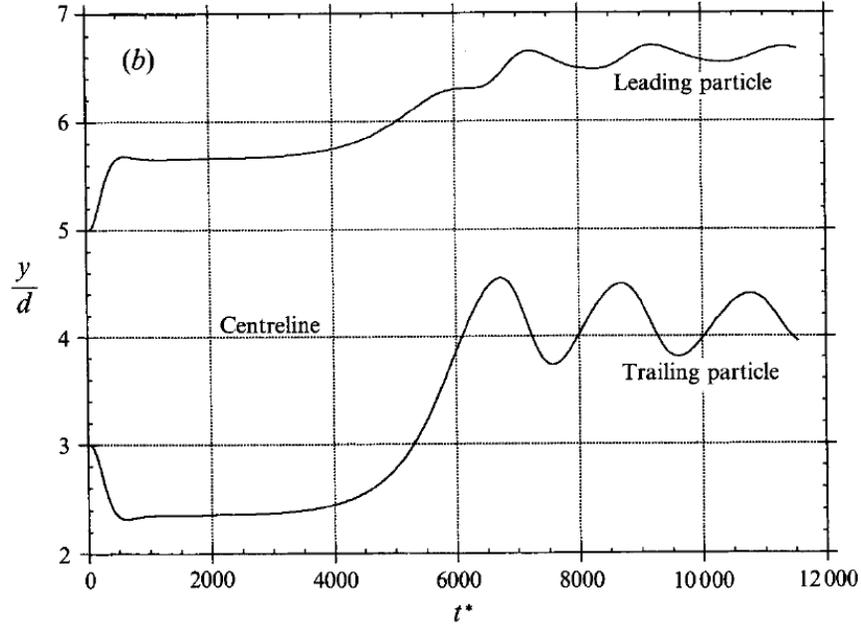}
  \caption{Horizontal locations of the two particles, as simulated by
    FHJ.  Parameters: $W=8D$, $\Reynolds=1.52$.  Reproduced from
    \cite[Figure~32(b)]{feng-hu-joseph-1994}, with permission.}
  \label{feng_original}
\end{figure}

Our main aim here is to determine to what extend our results are able to
reproduce the finite element simulations of
FHJ~\cite{feng-hu-joseph-1994} using $W=8D$ and $\Reynolds=1.52$.  They
computed particle dynamics such as that pictured in
Figure~\ref{feng_original} that can be separated into three distinct
phases: 
\begin{enumerate}
  \renewcommand{\theenumi}{\roman{enumi}}
\item a first phase that consists of a rapid re-adjustment up to time
  $t^{*}\approx 500$ (measured in dimensionless time units, with to
  $t^{*}=t\sqrt{g/D}$\,) during which the particles separate
  horizontally to locations that are equally-spaced from the left and
  right walls.
\item a second phase where the particles maintain their horizontal
  positions and fall together with the same vertical speed until $t^{*}
  \approx 4000$.
\item a third phase in which the particles shift together to the right
  into a new equilibrium state where the left-most particle oscillates
  about the centerline, while the right-most particle is much closer to
  the right wall and also oscillates side-to-side but with smaller
  amplitude.
\end{enumerate}
We remark that FHJ's simulations were intended to reproduce the
experiments of Jayaweera and Mason~\cite{jayaweera-mason-1965}, wherein
two long thin cylinders were settling in a large tank, with the same
initial conditions and $\Reynolds$ between 0.1 and 1.0.  Jayaweera and
Mason's discussion of their experimental results makes mention of the
first two phases but not phase iii.

\begin{figure}[htbp]
  \centering
  \includegraphics[width=0.95\columnwidth]{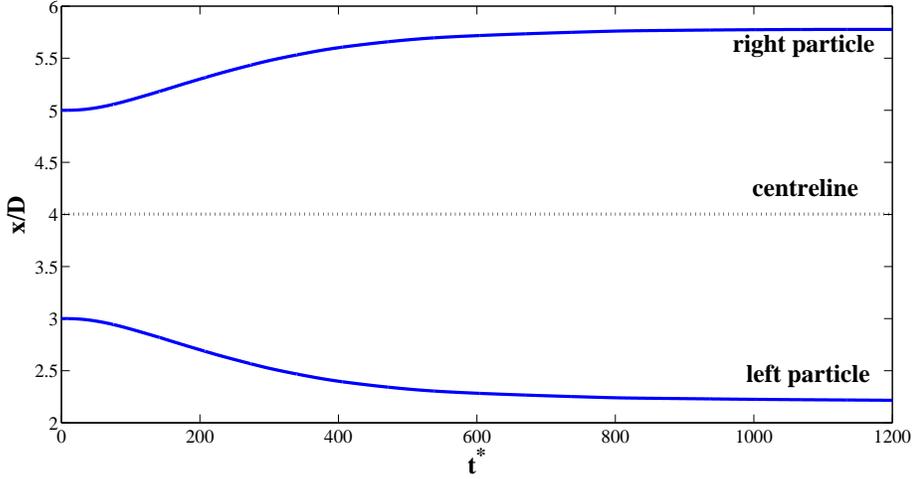}\\
  \caption{Horizontal particle positions for two particles initially
    located on the same horizontal line, and symmetric about the
    centerline.  Parameter values: $W=8D$ and $\Reynolds=1.6$.}
  \label{fig:hori_dist_re1p6}
\end{figure}

We begin with the symmetric case where the two particles have initial
horizontal positions
\begin{alignat}{3}
  x &= \frac{L_x}{2} - \frac{W}{8} \;\; \text{and} \;\;
  x =\frac{L_x}{2} + \frac{W}{8},
  \label{eq:horiz_symmetric}
\end{alignat}
and Reynolds number $\Reynolds=1.6$ which is very close to FHJ's value.
The horizontal locations of the simulated particles are pictured in
Figure~\ref{fig:hori_dist_re1p6}, with the plot axes rescaled to use the
same dimensionless variables as FHJ in Figure~\ref{feng_original}.  Our
solution exhibits a steady state at long times that corresponds to
positions $x/D\approx 2.25$ and $5.75$ located symmetrically across the
centerline; these positions are very close to those obtained in FHJ's
phase ii.  Furthermore, the rapid transient in phase i ends at a
dimensionless time of roughly $t^*=500$, which is also very close to
FHJ's value.  These two results suggest that our numerics are consistent
with FHJ during phases i and ii and that we are capturing this portion
of the motion properly.

However, we do not capture the same phase iii behavior since our two
particles never deviate from their steady state locations for
$t^*\gtrsim 4000$ when FHJ's phase iii begins.  Similar dynamics to
FHJ's phase iii have also been computed by Aidun and
Ding~\cite{aidun-ding-2003} with a lattice-Boltzmann method, and they
ascribe this periodic behavior to the appearance of a solution
bifurcation.  It is likely that this bifurcation is sensitive not only
to solution parameters but also to the presence of numerical error.
Therefore, we suspect that the first order accuracy of our IB method may
be preventing the numerics from capturing this transition to a periodic
state at longer times.

\begin{figure}[htbp]
  \centering
  \includegraphics[width=0.95\columnwidth]{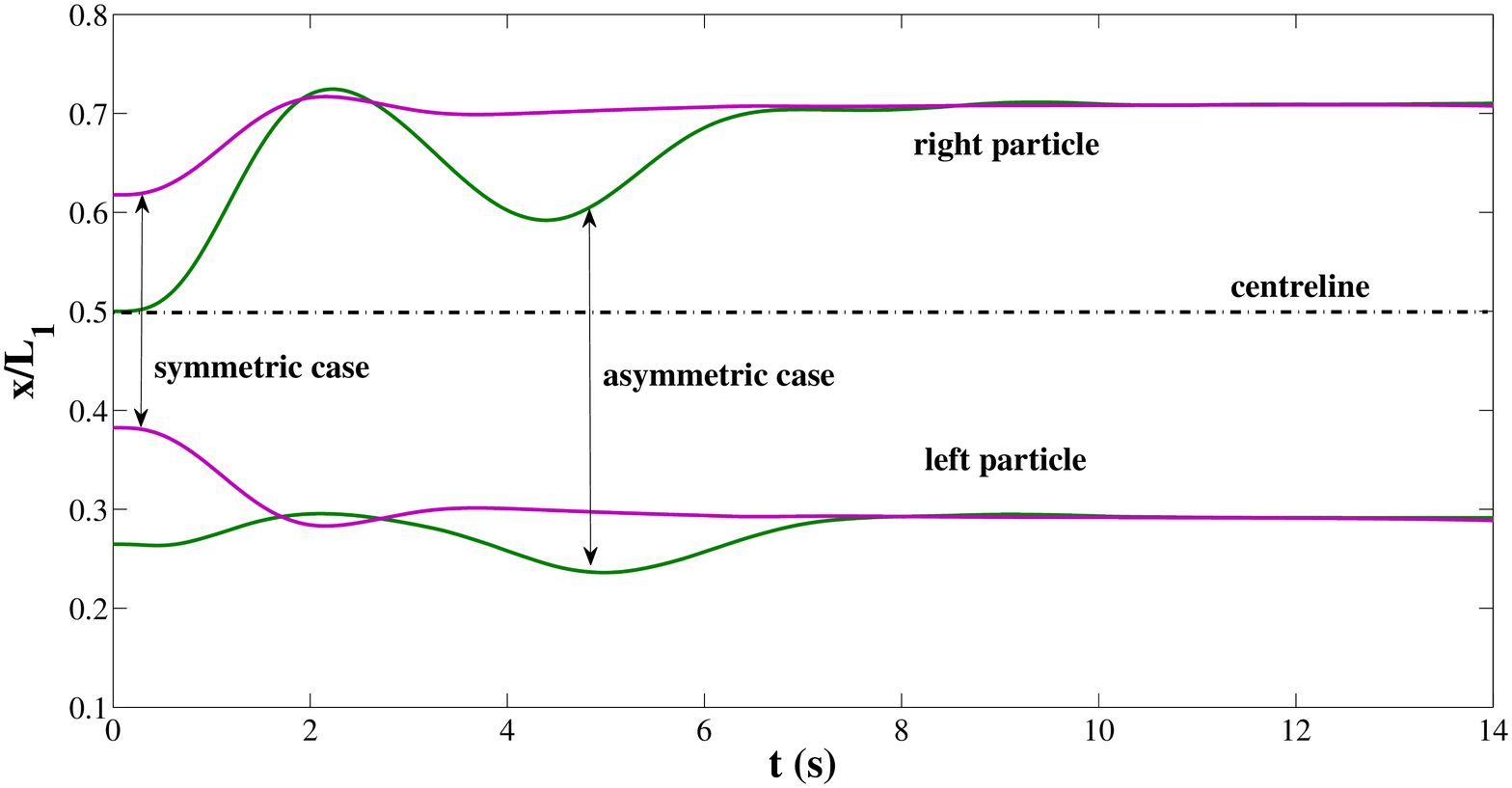}\\
  \caption{Horizontal particle positions for two particles initially
    located on the same horizontal line.  Both symmetrical and
    asymmetric initial conditions are pictured.  Parameter values:
    $W=8D$ and $\Reynolds=4.4$.}
  \label{hori_dist}
\end{figure}

\begin{figure}[htbp]
  \centering
  \includegraphics[trim=6in 0cm 4.8in 0cm,clip=true,width=0.28\columnwidth]{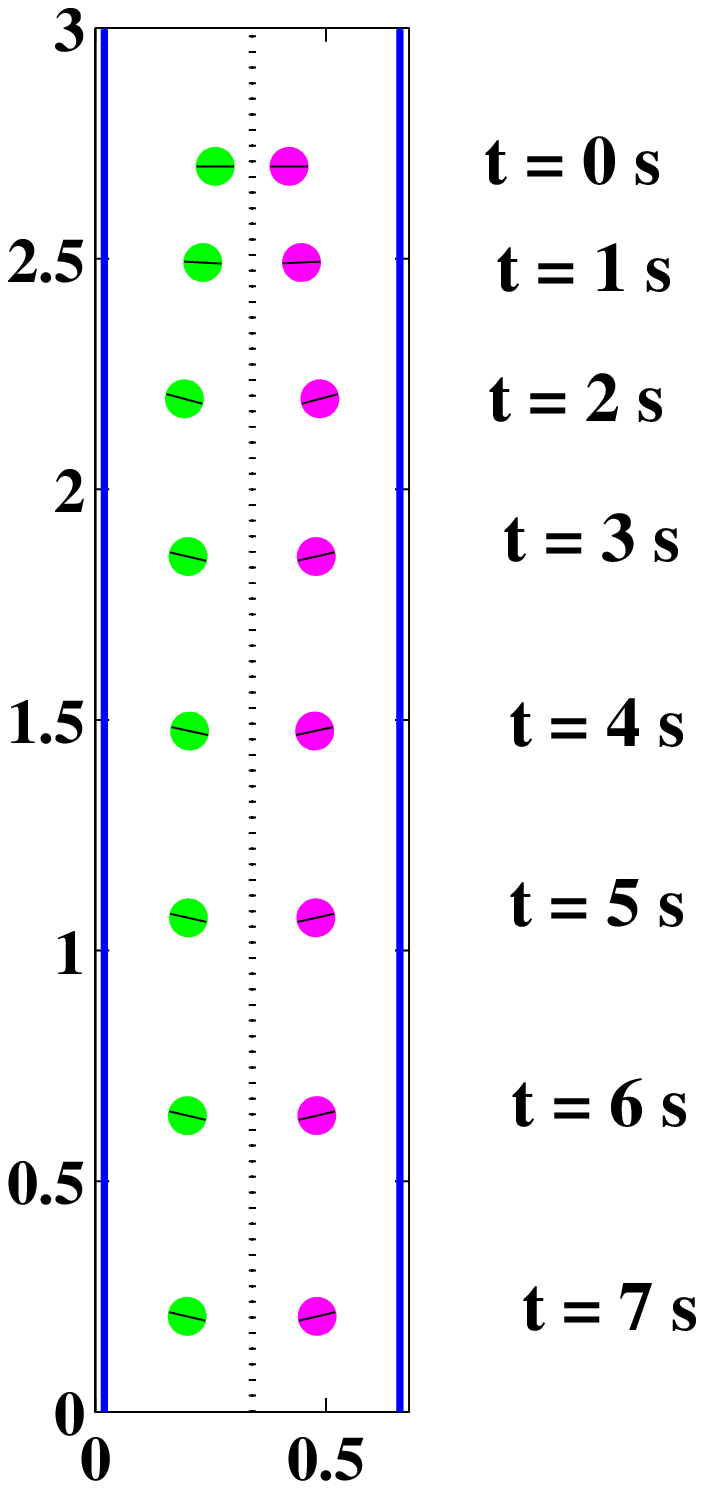}
  \qquad
  \includegraphics[trim=6in 0cm 4.8in 0cm,clip=true,width=0.28\columnwidth]{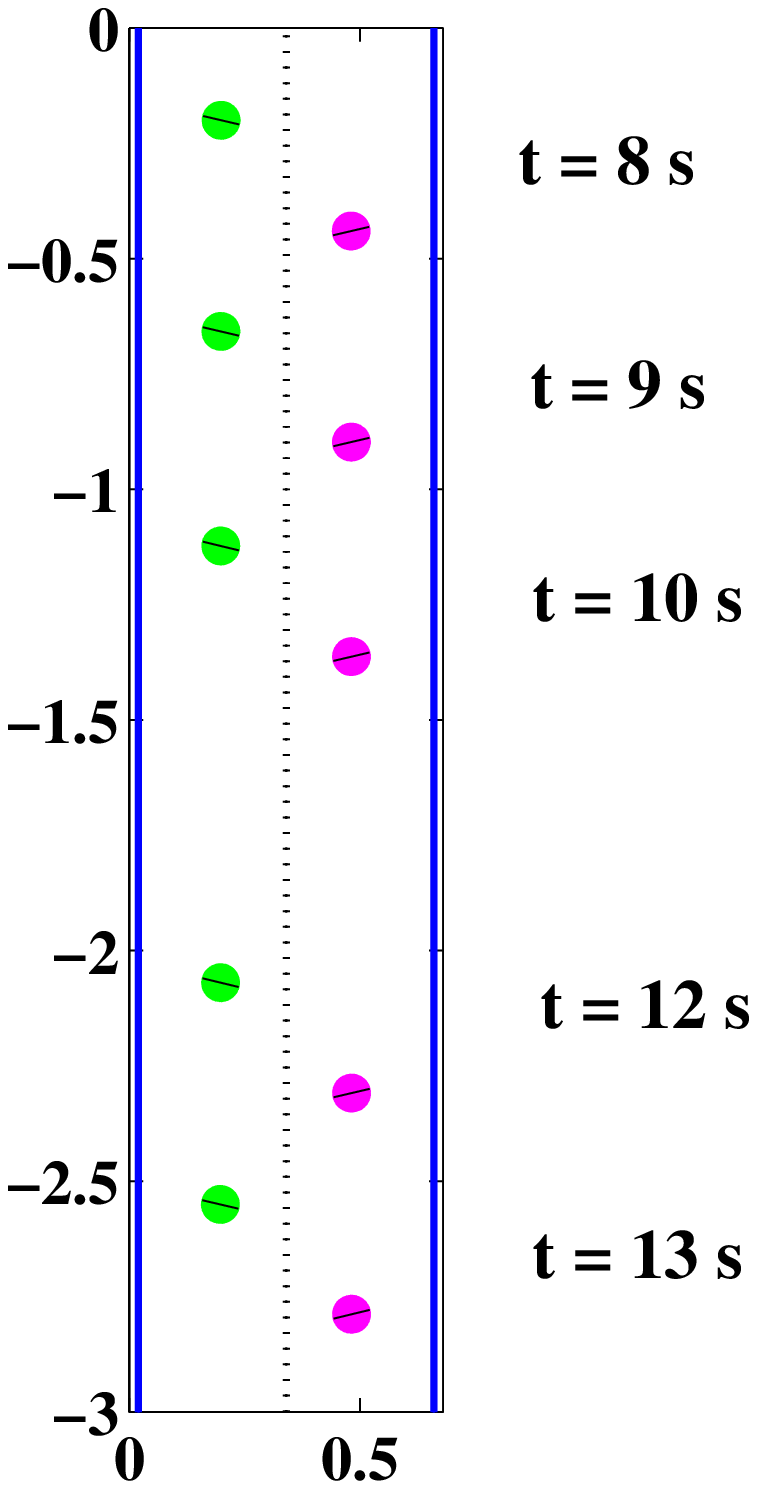}
  \caption{Snapshots of particle interactions for two particles
    initially aligned horizontally and centered.  Parameter values:
    $W=8D$ and $\Reynolds=4.4$.}
  \label{symm_snap}
\end{figure}

We repeated the previous calculation by increasing the Reynolds number
to $\Reynolds=4.4$ and our results are pictured in
Figures~\ref{hori_dist} and~\ref{symm_snap} which exhibit similar
dynamics to the lower $\Reynolds$ case.
In particular, we still observe no transition to phase iii behavior even
at this higher Reynolds number.  These results give us some confidence
that our IB simulations are reproducing physically-relevant behavior
corresponding to phases i and ii, but a more detailed numerical study is
required in order to determine the source of the discrepancy between our
method and FHJ's approach.

\begin{figure}[htbp]
  \centering
  \includegraphics[trim=6in 0cm 4.8in 0cm,clip=true,width=0.28\columnwidth]{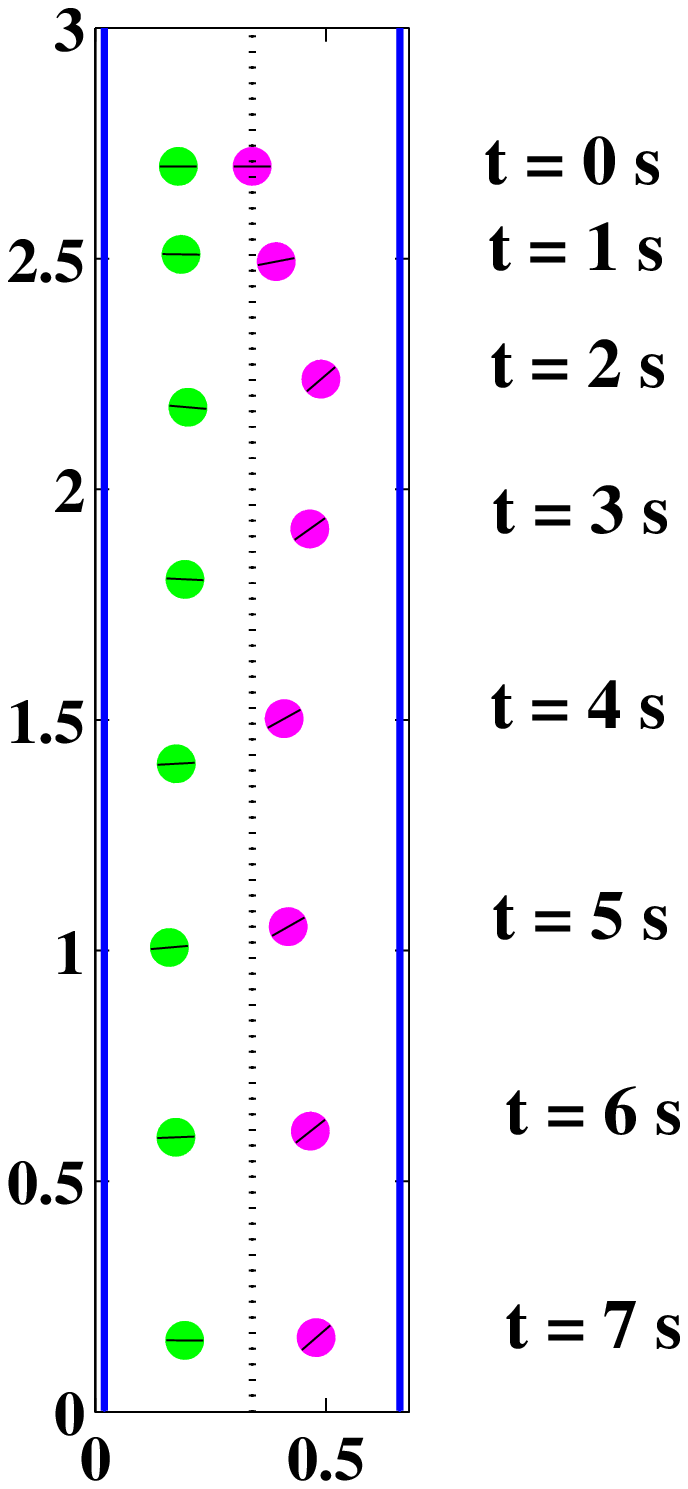}
  \qquad
  \includegraphics[trim=6in 0cm 4.8in 0cm,clip=true,width=0.28\columnwidth]{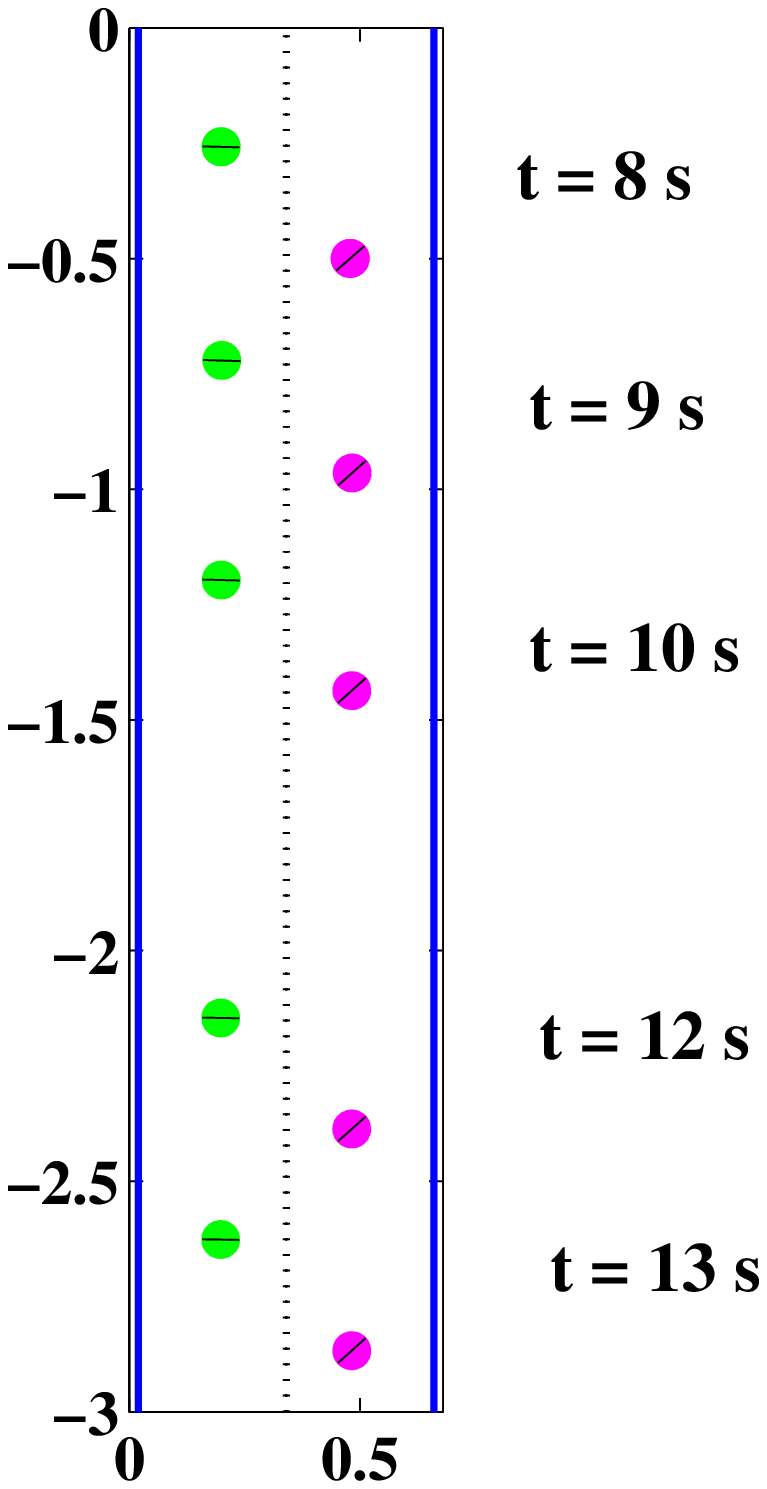}
  \caption{Snapshots of interactions for two particles initially aligned
    horizontally and off-center.  Parameter values: $W=8D$ and
    $\Reynolds=4.4$.}
  \label{asymm_snap}
\end{figure}

The perfectly symmetric initial conditions used above are somewhat
artificial, and will never actually occur in a real flow.  Hence, we have
also simulated an asymmetric initial placement of the particles given by
\begin{alignat}{3}
  x &= \frac{L_x}{2} - \frac{W}{4} \;\; \text{and} \;\;
  x =\frac{L_x}{2}, 
  \label{eq:horiz_asymmetric}
\end{alignat}
in which the initial particle locations from \en{eq:horiz_symmetric} are
shifted a distance $W/8$ to the left as pictured in
Figure~\ref{fig:init_profile}(d).  Otherwise, the channel width $W=8D$
and Reynolds number $\Reynolds=4.4$ remain the same as in the symmetric
case.  The numerical results are shown in Figures~\ref{hori_dist}
and~\ref{asymm_snap}, where the particles exhibit similar dynamics to
the symmetric case and approach the same long-term equilibrium solution.
The only difference can be seen in the transient motion where the
particles undergo one additional oscillation in the horizontal locations
en~route to steady state.

\section{Conclusions}
\label{sec:conclude}

The main aim of this paper is to demonstrate the ability of the immersed
boundary method to simulate realistic dynamics of solid particles
settling under gravity within a Newtonian incompressible fluid.  The
solid particles are modelled as a network of stiff springs, while the
added mass of the particles is incorporated using an extra gravitational
forcing term that is spread onto fluid points via a regularized delta
function.  Numerical simulations of a single particle show good
agreement with the most accurate empirical formula for wall-corrected
settling velocity due to Fax\'en.  Furthermore, two-particle simulations
reproduce qualitatively features of the dynamics seen in both
experiments and numerical simulations.

This study is by no means a comprehensive comparison to other results
from the extensive literature on particle sedimentation, but rather sets
the stage for such a study in future.  In particular, we plan to perform
a more detailed comparison with other published results, focusing first
on our idealized cylindrical particles.  By implementing improvements to
the numerical algorithm that increase accuracy of the solution
approximation (such as in~\cite{lai-peskin-2000}) we hope to be able to
explain the discrepancy we observed between our results and those of
Feng, Hu and Joseph~\cite{feng-hu-joseph-1994}.  After that, the natural
next step would be to extend our numerical method to 3D in order to
permit simulations spherical particle interactions in a more realistic
geometry.

We emphasize that this study is a ``proof-of-concept'' that the immersed
boundary method may be applied to simulating the sedimentation of
particles that are denser than the suspending fluid.  We make no claim
to improve on or to compete with other numerical methods that are
specially-tailored to deal with rigid, non-deformable particles.
Instead, our ultimate goal is to solve sedimentation problems involving
irregularly-shaped and highly deformable particles, which to our
knowledge has not been sufficiently well studied in the literature.
Such particle systems arise in the study of suspensions of red blood
cells, wood pulp fibers, vesicles, bubbles, etc.  Making use of the
uniform triangulated meshes from the DistMesh package will allow us to
deal with more general particle shapes.  Furthermore, we plan to take
advantage of recent developments in massively parallel immersed boundary
algorithms by Wiens and Stockie~\cite{wiens-stockie-2013}, which should
prove instrumental in allowing efficient 2D and 3D immersed boundary
simulations to be performed for non-dilute suspensions containing large
numbers of particles.

\section*{Acknowledgments}

This work was supported by grants from the Natural Sciences and
Engineering Research Council of Canada and the Mprime Network of Centres
of Excellence.

\providecommand{\noopsort}[1]{}


\begin{thebibliography}{10}

\bibitem{aidun-ding-2003}
C.~K. Aidun and E.-J. Ding.
\newblock Dynamics of particle sedimentation in a vertical channel:
  {P}eriod-doubling bifurcation and chaotic state.
\newblock {\em Phys. Fluids}, 15(6):1612--1621, 2003.

\bibitem{alpkvist-klapper-2007}
E.~Alpkvist and I.~Klapper.
\newblock Description of mechanical response including detachment using a novel
  particle model of biofilm/flow interaction.
\newblock {\em Water Sci. Tech.}, 55:265--273, 2007.

\bibitem{batchelor-1967}
G.~K. Batchelor.
\newblock {\em An Introduction to Fluid Dynamics}.
\newblock Cambridge University Press, 1967.

\bibitem{benrichou-etal-2005}
A.~{Ben Richou}, A.~Ambari, M.~Lebey, and J.~K. Naciri.
\newblock Drag force on a circular cylinder midway between two parallel plates
  at {$Re \ll 1$}. {P}art 2: moving uniformly (numerical and experimental).
\newblock {\em Chem. Eng. Sci.}, 60(10):2535--2543, 2005.

\bibitem{benrichou-etal-2004}
A.~{Ben Richou}, A.~Ambari, and J.~K. Naciri.
\newblock Drag force on a circular cylinder midway between two parallel plates
  at {$Re \ll 1$}. {P}art 1: {P}oiseuille flow (numerical).
\newblock {\em Chem. Eng. Sci.}, 59(15):3215--3222, 2004.

\bibitem{boffi-gastaldi-heltai-2007}
D.~Boffi, L.~Gastaldi, and L.~Heltai.
\newblock On the {CFL} condition for the finite element immersed boundary
  method.
\newblock {\em Comput. Struct.}, 85:775--783, 2007.

\bibitem{brenner-1966}
H.~Brenner.
\newblock Hydrodynamic resistance of particles at small {R}eynolds number.
\newblock {\em Adv. Chem. Eng.}, 6:287--438, 1966.

\bibitem{breugem-2012}
W.-P. Breugem.
\newblock A second-order accurate immersed boundary method for fully resolved
  simulations of particle-laden flows.
\newblock {\em J. Comput. Phys.}, 231(13):4469--4498, 2012.

\bibitem{bringley-thesis-2008}
T.~T. Bringley.
\newblock {\em Analysis of the Immersed Boundary Method for {S}tokes Flow}.
\newblock PhD thesis, Department of Mathematics, New York University, May 2008.

\bibitem{burger-wendland-2001}
R.~B\"urger and W.~L. Wendland.
\newblock Sedimentation and suspension flows: {H}istorical perspective and some
  recent developments.
\newblock {\em J. Eng. Math.}, 41:101--116, 2001.

\bibitem{champmartin-ambari-2007}
S.~Champmartin and A.~Ambari.
\newblock Kinematics of a symmetrically confined cylindrical particle in a
  ``{S}tokes-type'' regime.
\newblock {\em Phys. Fluids}, 19:073303, 2007.

\bibitem{chhabra-etal-2003}
R.~P. Chhabra, S.~Agarwal, and K.~Chaudhary.
\newblock A note on wall effect on the terminal falling velocity of a sphere in
  quiescent {N}ewtonian media in cylindrical tubes.
\newblock {\em Powder Tech.}, 129:53--58, 2003.

\bibitem{davis-acrivos-1985}
R.~H. Davis and A.~Acrivos.
\newblock Sedimentation of noncolloidal particles at low {R}eynolds numbers.
\newblock {\em Annu. Rev. Fluid Mech.}, 17:91--118, 1985.

\bibitem{difelice-1999}
R.~{Di Felice}.
\newblock The sedimentation velocity of dilute suspensions of nearly monosized
  spheres.
\newblock {\em Int. J. Multiphase Flow}, 25:559--574, 1999.

\bibitem{dupuis-etal-2008}
A.~Dupuis, P.~Chatelain, and P.~Koumoutsakos.
\newblock An immersed boundary-lattice-{B}oltzmann method for the simulation of
  the flow past an impulsively started cylinder.
\newblock {\em J. Comput. Phys.}, 227(9):4486--4498, 2008.

\bibitem{faxen-1946}
O.~H. Fax{\'e}n.
\newblock Forces exerted on a rigid cylinder in a viscous fluid between two
  parallel fixed planes.
\newblock {\em Proceedings of the Royal Swedish Academy of Sciences},
  187:1--13, 1946.

\bibitem{feng-hu-joseph-1994}
J.~Feng, H.~H. Hu, and D.~D. Joseph.
\newblock Direct simulation of initial value problems for the motion of solid
  bodies in a {N}ewtonian fluid. {P}art 1. {S}edimentation.
\newblock {\em J. Fluid Mech.}, 261:95--134, 1994.

\bibitem{fortes-joseph-lundgren-1987}
A.~F. Fortes, D.~D. Joseph, and T.~S. Lundgren.
\newblock Nonlinear mechanics of fluidization of beds of spherical particles.
\newblock {\em J. Fluid Mech.}, 177:467--483, 1987.

\bibitem{ghosh-2013}
S.~Ghosh.
\newblock {\em The immersed boundary method for simulating gravitational
  settling and fluid shear-induced deformation of elastic structures}.
\newblock PhD thesis, Department of Mathematics, Simon Fraser University,
  Burnaby, Canada, Mar. 2013.

\bibitem{glowinski-etal-2001}
R.~Glowinski, T.~W. Pan, T.~I. Hesla, D.~D. Joseph, and J.~P\'eriaux.
\newblock A fictitious domain approach to the direct numerical simulation of
  incompressible viscous flow past moving rigid bodies: {A}pplication to
  particulate flow.
\newblock {\em J. Comput. Phys.}, 169(2):363--426, 2001.

\bibitem{griffith-etal-2007}
B.~E. Griffith, R.~D. Hornung, D.~M. McQueen, and C.~S. Peskin.
\newblock An adaptive, formally second order accurate version of the immersed
  boundary method.
\newblock {\em J. Comput. Phys.}, 223(1):10--49, 2007.

\bibitem{guazzelli-morris-2012}
{\'{E}}.~Guazzelli and J.~F. Morris.
\newblock {\em A Physical Introduction to Suspension Dynamics}.
\newblock Cambridge Texts in Applied Mathematics. Cambridge University Press,
  2012.

\bibitem{haeri-shrimpton-2012}
S.~Haeri and J.~S. Shrimpton.
\newblock On the application of immersed boundary, fictitious domain and
  body-conformal mesh methods to many particle multiphase flows.
\newblock {\em Int. J. Multiphase Flow}, 40:38--55, 2012.

\bibitem{happel-brenner-1983}
J.~Happel and H.~Brenner.
\newblock {\em Low {R}eynolds number hydrodynamics, with special applications
  to particulate media}.
\newblock Mechanics of fluids and transport processes. Martinus Nijhoff
  Publishers, 1983.

\bibitem{hernandez-ortiz-phdthesis-2004}
J.~P. Hern\'andez-Ortiz.
\newblock {\em Boundary Integral Equations for Viscous Flows -- non-{N}ewtonian
  Behavior and Solid Inclusions}.
\newblock PhD thesis, University of Wisconsin-Madison, Department of Mechanical
  Engineering, 2004.

\bibitem{hopkins-fauci-2002}
M.~M. Hopkins and L.~J. Fauci.
\newblock A computational model of the collective fluid dynamics of motile
  microorganisms.
\newblock {\em J. Fluid Mech.}, 455:149--174, 2002.

\bibitem{hou-shi-2008a}
T.~Y. Hou and Z.~Shi.
\newblock Removing the stiffness of elastic force from the immersed boundary
  method for the {2D} {S}tokes equations.
\newblock {\em J. Comput. Phys.}, 227:9138--9169, 2008.

\bibitem{hu-1996}
H.~H. Hu.
\newblock Direct simulation of flows of solid-liquid mixtures.
\newblock {\em Int. J. Multiphase Flow}, 22(2):335--352, 1996.

\bibitem{hu-joseph-fortes-1997}
H.~H. Hu, D.~D. Joseph, and A.~F. Fortes.
\newblock Experiments and direct simulation of fluid particle motions.
\newblock {\em Int. Vid. J. Eng. Res.}, 2:17--24, 1997.

\bibitem{jayaweera-mason-1965}
K.~O. L.~F. Jayaweera and B.~J. Mason.
\newblock The behaviour of freely falling cylinders and cones in a viscous
  fluid.
\newblock {\em J. Fluid Mech.}, 22(4):709--720, 1965.

\bibitem{joseph-etal-1987}
D.~D. Joseph, A.~Fortes, T.~S. Lundgren, and P.~Singh.
\newblock Nonlinear mechanics of fluidization of spheres, cylinders and disks
  in water.
\newblock In {\em Advances in Multiphase Flow and Related Problems}, pages
  101--122. SIAM, Philadelphia, PA, 1987.

\bibitem{ladd-1994II}
A.~J.~C. Ladd.
\newblock Numerical simulations of particulate suspensions via a discretized
  {B}oltzmann equation. {P}art {II}. {N}umerical results.
\newblock {\em J. Fluid Mech.}, 271:311--339, 1994.

\bibitem{lai-thesis-1998}
M.-C. Lai.
\newblock {\em Simulations of the flow past an array of circular cylinders as a
  test of the immersed boundary method}.
\newblock PhD thesis, New York University, Sept. 1998.

\bibitem{lai-peskin-2000}
M.-C. Lai and C.~S. Peskin.
\newblock An immersed boundary method with formal second-order accuracy and
  reduced numerical viscosity.
\newblock {\em J. Comput. Phys.}, 160(2):705--719, 2000.

\bibitem{mittal-iaccarino-2005}
R.~Mittal and G.~Iaccarino.
\newblock Immersed boundary methods.
\newblock {\em Annu. Rev. Fluid Mech.}, 37:239--261, 2005.

\bibitem{munster-mierka-turek-2012}
R.~M\"unster, O.~Mierka, and S.~Turek.
\newblock Finite element-fictitious boundary methods {(FEM-FBM)} for {3D}
  particulate flow.
\newblock {\em Int. J. Numer. Meth. Fluids}, 69(2):294--313, 2012.

\bibitem{persson-strang-2004}
P.~O. Persson and G.~Strang.
\newblock A simple mesh generator in {MATLAB}.
\newblock {\em SIAM Review}, 46(2):329--345, 2004.

\bibitem{peskin-2002}
C.~S. Peskin.
\newblock The immersed boundary method.
\newblock {\em Acta Numer.}, 11:1--39, 2002.

\bibitem{phanthien-fan-2002}
N.~Phan-Thien and X.-J. Fan.
\newblock Viscoelastic mobility problem using a boundary element method.
\newblock {\em J. Non-Newton. Fluid Mech.}, 105(2-3):131--152, 2002.

\bibitem{pianet-arquis-2008}
G.~Pianet and E.~Arquis.
\newblock Simulation of particles in fluid: a two-dimensional benchmark for a
  cylinder settling in a wall-bounded box.
\newblock {\em Euro. J. Mech. B Fluids}, 27:309--321, 2008.

\bibitem{numrecipes}
W.~H. Press, S.~A. Teukolsky, W.~T. Vetterling, and B.~P. Flannery.
\newblock {\em Numerical Recipes in C: The Art of Scientific Computing}.
\newblock Cambridge University Press, second edition, 1992.

\bibitem{prosperetti-tryggvason-2007}
A.~Prosperetti and G.~Tryggvason, editors.
\newblock {\em Computational Methods for Multiphase Flow}.
\newblock Cambridge University Press, 2007.

\bibitem{qi-1999}
D.~Qi.
\newblock Lattice-{B}oltzmann simulations of particles in
  non-zero-{R}eynolds-number flows.
\newblock {\em J. Fluid Mech.}, 385:41--62, 1999.

\bibitem{richardson-zaki-1954b}
J.~F. Richardson and W.~N. Zaki.
\newblock Sedimentation and fluidisation: {P}art {I}.
\newblock {\em Trans. Inst. Chem. Eng.}, 32:35--53, 1954.

\bibitem{stockie-thesis-1997}
J.~M. Stockie.
\newblock {\em Analysis and Computation of Immersed Boundaries, with
  Application to Pulp Fibres}.
\newblock PhD thesis, Institute of Applied Mathematics, University of British
  Columbia, Vancouver, Canada, 1997.
\newblock Available at \url{http://circle.ubc.ca/handle/2429/7346}.

\bibitem{stockie-2009}
J.~M. Stockie.
\newblock Modelling and simulation of porous immersed boundaries.
\newblock {\em Comput. Struct.}, 87(11-12):701--709, 2009.

\bibitem{stokes-1966}
G.~G. Stokes.
\newblock Section {IV}. {D}etermination of the motion of a fluid about a sphere
  which moves uniformly with a small velocity.
\newblock In {\em Mathematical and Physical Papers}, volume III of {\em The
  Sources of Science, No. 33}, pages 55--67. Cambridge University Press,
  Teddington, UK, second edition, 1966.
\newblock First published in 1901.

\bibitem{sucker-brauer-1975}
D.~Sucker and H.~Brauer.
\newblock Fluid\-dy\-na\-mik bei quer angestr{\"o}mten {Z}ylindern.
\newblock {\em Heat Mass Transfer}, 8(3):149--158, 1975.

\bibitem{takaisi-1955}
Y.~Takaisi.
\newblock The drag on a circular cylinder moving with low speeds in a viscous
  liquid between two parallel walls.
\newblock {\em J. Phys. Soc. Japan}, 10:685--693, 1955.

\bibitem{tu-peskin-1992}
C.~Tu and C.~S. Peskin.
\newblock Stability and instability in the computation of flows with moving
  immersed boundaries: {A} comparison of three methods.
\newblock {\em SIAM J. Sci. Stat. Comput.}, 13(6):1361--1376, 1992.

\bibitem{uhlmann-2005}
M.~Uhlmann.
\newblock An immersed boundary method with direct forcing for the simulation of
  particulate flows.
\newblock {\em J. Comput. Phys.}, 209:448--476, 2005.

\bibitem{vasseur-cox-1977}
P.~Vasseur and R.~G. Cox.
\newblock The lateral migration of spherical particles sedimenting in a
  stagnant bounded fluid.
\newblock {\em J. Fluid Mech.}, 80(3):561--591, 1977.

\bibitem{wang-layton-2009}
J.~Wang and A.~Layton.
\newblock Numerical simulations of fiber sedimentation in {N}avier-{S}tokes
  flow.
\newblock {\em Commun. Comput. Phys.}, 5(1):61--83, 2009.

\bibitem{wang-fan-luo-2008}
Z.~Wang, J.~Fan, and K.~Luo.
\newblock Combined multi-direct forcing and immersed boundary method for
  simulating flows with moving particles.
\newblock {\em Int. J. Multiphase Flow}, 34:283--302, 2008.

\bibitem{white-1946}
C.~M. White.
\newblock The drag of cylinders in fluids at slow speeds.
\newblock {\em Proc. Roy. Soc. A}, 186:472--479, 1946.

\bibitem{wiens-stockie-2013}
J.~K. Wiens and J.~M. Stockie.
\newblock An efficient parallel immersed boundary algorithm using a
  pseudo-compressible fluid solver.
\newblock {\em J. Comput. Phys.}, May 2013.
\newblock Submitted, arXiv:1305.3976.

\bibitem{zhu-peskin-2002}
L.~Zhu and C.~S. Peskin.
\newblock Simulation of a flapping flexible filament in a flowing soap film by
  the immersed boundary method.
\newblock {\em J. Comput. Phys.}, 179(2):452--468, 2002.

\end{thebibliography}
\end{document}